\documentclass[pdflatex,sn-mathphys-num]{sn-jnl}% Math and Physical Sciences Numbered Reference Style
\usepackage{graphicx}%
\usepackage{multirow}%
\usepackage{amsmath,amssymb,amsfonts}%
\usepackage{amsthm}%
\usepackage{mathrsfs}%
\usepackage[title]{appendix}%
\usepackage{xcolor}%
\usepackage{textcomp}%
\usepackage{manyfoot}%
\usepackage{booktabs}%
\usepackage{algorithm}%
\usepackage{algorithmicx}%
\usepackage{algpseudocode}%
\usepackage{listings}%
\usepackage{natbib}
\usepackage{bm}
\usepackage{graphicx}
\usepackage{subcaption}
\usepackage{tabularx}
\usepackage{multirow}
\usepackage{adjustbox}
\usepackage{comment}
\usepackage{booktabs}
\usepackage{caption}
\usepackage{array}
\usepackage{xcolor}
\usepackage{todonotes}
\usepackage{booktabs}
\usepackage{makecell}
\newcounter{xxx}
\theoremstyle{thmstyleone}%
\theoremstyle{thmstyletwo}%

\theoremstyle{thmstylethree}%

\begin{document}

\title[Brain Connectivity]{Bayesian Brain Edge-Based Connectivity (BBeC): a Bayesian model for brain edge-based connectivity  inference}


\author[1]{\fnm{Zijing} \sur{Li}}
\author[1]{\fnm{Chenhao} \sur{Zeng}}

\author*[1]{\fnm{Shufei} \sur{Ge }}\email{geshf@shanghaitech.edu.cn}

\author[~]{\sur{for the Alzheimer’s Disease Neuroimaging Initiative$\dagger$}}
%.


\affil[1]{\orgdiv{Institute of Mathematical Sciences}, \orgname{ShanghaiTech University\let\thefootnote\relax\footnote{$\dagger$ Data used in preparation of this article were obtained from the Alzheimer’s Disease Neuroimaging Initiative (ADNI) database (adni.loni.usc.edu). As such, the investigators within the ADNI contributed to the design and implementation of ADNI and/or provided data but did not participate in analysis or writing of this report. A complete listing of ADNI investigators can be found at: \url{http://adni.loni.usc.edu/wp-content/uploads/how_to_apply/ADNI_Acknowledgement_List.pdf}}}, \orgaddress{\street{393 Middle Huaxia Road}, \city{Shanghai}, \postcode{201210}, %\state{Shanghai}, 
\country{China}}}


\abstract{
\textbf{Background:} Brain connectivity analysis based on magnetic resonance imaging is crucial for understanding neurological mechanisms. However, edge-based connectivity inference faces significant challenges, particularly the curse of dimensionality when estimating high-dimensional covariance matrices. Existing methods often struggle to account for the unknown latent topological structure among brain edges, leading to inaccurate parameter estimation and unstable inference.

\textbf{Methods:} To address these issues, this study proposes a Bayesian hierarchical model based on a finite-dimensional Dirichlet distribution. Unlike non-parametric approaches, our method utilizes a finite-dimensional Dirichlet distribution to model the latent topological structure of brain networks, ensuring constant parameter dimensionality and improving algorithmic stability. We reformulate the covariance matrix structure to guarantee positive definiteness and employ a Metropolis-Hastings algorithm to simultaneously infer network topology and correlation parameters. Furthermore, to alleviate the computational burden of parameter inference in large-scale networks, we optimized the calculation process of the likelihood function to reduce the algorithm's time complexity. Our implementation is available at \url{https://github.com/mimi6501/BBeC}.

\textbf{Results:} Simulations validated the recovery of both network topology and correlation parameters across various settings. Furthermore, we quantitatively compared the proposed framework with the Graphical Lasso and a Dirichlet process-based non-parametric Bayesian model. Experimental results show our model offers flexible parameter tuning while outperforming baselines in  estimation accuracy and convergence stability. Sensitivity analysis reveals the diagonal adjustment parameter $\lambda$ has minimal impact on model accuracy, and parameter sampling order has negligible impact on final inference. When applied to the Alzheimer's Disease Neuroimaging Initiative dataset, the model successfully identified structural subnetworks. The identified clusters were not only validated by composite anatomical metrics but also consistent with established findings in the literature, collectively demonstrating the model's reliability. The estimated covariance matrix also revealed that intragroup connection strength is stronger than intergroup connection strength.

\textbf{Conclusions:} This study introduces a Bayesian framework for inferring brain network topology and high-dimensional covariance structures. The model configuration effectively reduces parameter dimensionality while ensuring the positive definiteness of covariance matrices. As a result, it offers an efficient and reliable tool for investigating intrinsic brain connectivity in large-scale neuroimaging studies.
}


\maketitle

\section{Introduction}\label{sec1}

The concept of human brain connectivity has sparked extensive research and witnessed explosive growth since its introduction~\cite{sporns2005human}. Research on brain connectivity is fundamental to advancing our understanding of human brain disorders, with magnetic resonance imaging (MRI) emerging as the predominant methodological approach in this field. In this domain, functional MRI (fMRI) is widely used to map whole brain functional architecture at high spatial resolution by inferring functional connectivity from temporally correlated blood oxygen level dependent signals~\cite{voineskos2024functional}. Complementing this, structural MRI (sMRI) offers a parallel approach to investigate the brain's physical connectivity and organizational patterns. By modeling the statistical interdependencies of morphological properties, such as cortical thickness or volume, across different brain regions, it is possible to construct structural covariance networks~\cite{mongay2025structural}. These networks are thought to reflect long-term, coordinated neurodevelopmental and pathological processes, potentially arising from shared genetic or environmental influences, or long-range axonal connections~\cite{evans2013networks}. This approach allows connectivity data to be systematically represented as network graphs, where anatomically defined brain areas constitute nodes, and the correlations between their structural properties form weighted edges—collectively establishing a biologically grounded topological architecture~\cite{bullmore2009complex}. Research indicates that alterations in the topological architecture of these structural brain connections are linked to numerous neurological disorders, including Parkinson's and Alzheimer's disease~\cite{ji2020convolutional}. Consequently, the analysis of structural connectivity provides a powerful paradigm for investigating human brain organization, enabling the identification of potential biomarkers for neurodegenerative disorders.

In brain connectivity studies, researchers employ phenotypic measures derived from MRI data to estimate interregional correlations, subsequently applying statistical analyses to identify connectivity patterns that covary with pathological and behavioral phenotypes. For instance, Durante et al. developed a Bayesian framework to detect both global network topology alterations and local nodal changes, thereby revealing between-group differences in structural brain organization~\cite{durante2018bayesian}. However, these statistical analyses ignore the correlations between connectivity edges, which may lead to inaccurate parameter estimation. Current edge-based analytical strategies primarily adopt three distinct paradigms. First, pairwise edge-level analysis operationalizes connectivity through correlation coefficients computed between regional time series. This approach necessitates rigorous multiple comparison correction, typically employing false discovery rate (FDR) procedures, to account for elevated false positive rates inherent in mass univariate testing~\cite{simpson2013permutation,bullmore2009complex}. Alternatively, subnetwork-based approaches reconceptualize functional connectivity alterations as modular assemblies of edges. These methods detect group differences by identifying coordinated variations within functionally defined subnetworks, thereby capturing mesoscale organizational changes \cite{chen2015parsimonious}. Finally, whole-network analyses adopt a comprehensive perspective by explicitly modeling higher-order interactions across the complete edge ensemble. This paradigm enables detection of system-wide reorganization patterns through multivariate characterization of network topology \cite{zalesky2010network}. However, due to the complexity of brain connectivity data and internal brain structures, fitting an appropriate model to determine connectivity characteristics poses a significant challenge.

Generally, modeling edge-based brain connectivity patterns requires estimation of a covariance structure characterizing interdependencies among connection edges. However, reliable estimation of the covariance matrix parameters presents some challenges. The correlation between connected edges is likely influenced by the underlying network topology of brain regions. Since each edge connects two nodes, a group of edges may be constrained by a set of nodes with an inherent topological structure. Due to the large number of nodes, the number of edges also surges, leading to a need to estimate a vast number of parameters in the covariance matrix. The dependency structure between edges is related to the network topology of brain regions, so considering this characteristic can accurately estimate model parameters. Previous parameter structure modeling strategies mainly relied on spatial distance for estimation ~\cite{brown2014incorporating,derado2010modeling,li2015spatial}. However, the edge correlations in brain networks are determined by the underlying topological structure and are not necessarily related to spatial distance. Therefore, models based on spatial proximity may not be suitable for accurately analyzing edge correlations. While advanced regularization techniques have been developed for high-dimensional data ~\cite{rothman2009generalized,cui2016sparse,bickel2008regularized}, these methods often fail to incorporate the intrinsic topological organization of neural systems. This oversight may lead to biologically implausible connectivity estimates that disregard the brain's hierarchical network architecture.

Fiecas et al. developed a variance component linear modeling framework to enable statistical inference of functional connectivity networks, with specific application to comparative analysis between neurotypical young adults and those with dyslexia ~\cite{fiecas2017variance}. However, this modeling framework does not account for the topological architecture of brain networks and becomes computationally intractable for edge parameter estimation in high-dimensional scenarios where the number of regions of interest (ROIs) exceeds practical thresholds. Chen et al. developed a non-parametric Bayesian framework for simultaneous estimation of brain network topology and covariance matrix parameters ~\cite{chen2020bayesian}. Their method took the sample covariance matrix of edges as the input and employed a Dirichlet process (DP) to identify the underlying topological structure of brain networks, followed by Markov Chain Monte Carlo (MCMC) sampling for posterior inference of model parameters. Modeling brain network topology with a DP enables automatic selection of the cluster number but introduces parameter dimensionality variations during MCMC iterations.  
The change of parameter dimensionality not only increases the complexity of parameter estimation but may also lead to difficulties in achieving algorithm convergence. To overcome this limitation, in this work, we adopt a finite-dimensional Dirichlet distribution to model the latent topological structure of brain networks, ensuring a constant parameter dimension throughout the MCMC iterations and thereby improving algorithmic stability. At the same time, we reformulate the structure of the covariance matrix to guarantee that it remains positive definite throughout the entire inference process. Furthermore, to alleviate the computational burden of large-scale network inference, this study proposes a computational optimization strategy for the likelihood function. We utilize Cholesky decomposition to reduce the number of floating-point operations and solve systems of linear equations to lower the computional  complexity of calculating the trace term.

The remainder of the manuscript is organized as follows. Section~\ref{sec2} provides a detailed description of the proposed Bayesian model and presents the corresponding mathematical formulations. In Section~\ref{sec3}, we explain how the Metropolis-Hastings (MH) algorithm is employed to perform sampling from the posterior distribution of the model parameters. Additionally, this section covers the optimization of the likelihood function calculation. Section~\ref{sec4} evaluates the performance and effectiveness of the proposed method through simulation studies, compares our model against competing approaches to demonstrate its advantages, and conducts a sensitivity analysis of the diagonal adjustment parameter in the covariance matrix. In Section~\ref{sec5}, we demonstrate an application of the proposed model on a real-world dataset consisting of $632$ subjects and $20$ ROIs, aiming at identifying potential brain connectivity patterns. Finally, Section~\ref{sec6} summarizes the model, covering its advantages, limitations, and potential future extensions. 

\section{Model}\label{sec2}

Statistically speaking, given  $ V $ ROIs, the structural relationships of the $ s $-th individual ($ s \in \{1,\cdots,S\} $) can be represented by a $ V \times V $ matrix, $ \mathbf{M}^s $, where each element denotes the relationship between a pair of ROIs. $ \mathbf{M}^s $ is symmetric for undirected relationships (e.g., covariance/correlation matrices ~\cite{kim2015highly} or undirected connectivity adjacency matrices ~\cite{bullmore2009complex}) and asymmetric for directed relationships ~\cite{goebel2003investigating}. A symmetric $ \mathbf{M}^s $ can represent an undirected graph with $ V $ nodes and $ E=V(V-1)/2 $ unique weighted edges. This matrix can be reshaped into a $ 1 \times E $ vector $ \mathbf{R}^s $, where each element corresponds to a unique ROI pair edge. Conventionally, $ \mathbf{R}^s $ is assumed to follow a multivariate normal distribution ~\cite{anderson1958introduction}.

Assume the centralized and standardized vectors $ \mathbf{R}^s \ (s=1,\ldots, S) $ follow a multivariate normal distribution $ \mathbf{R}^s \sim MVN(\mathbf{0}, \mathbf{\Sigma}) $, where $ \mathbf{\Sigma} \in \mathbb{R}^{E \times E} $ denotes the unknown edge covariance matrix. This matrix can be estimated either directly as a whole or as a structured parameterized matrix ~\citep{marrelec2006partial,friedman2008sparse}. While Bayesian frameworks often employ inverse-Wishart priors for full-matrix estimation ~\citep{gelman2013bayesian}, Monte Carlo methods become inefficient in high dimensions due to local mode trapping. In contrast, parameterized structured matrices offer advantages by reducing inference from a high-dimensional matrix to a small-sized set of parameters. This study aims to efficiently estimate $ \mathbf{\Sigma} $ using sample data $ \{\mathbf{R}^s\}_{s=1}^S $.

In this work, the covariance matrix $\mathbf{\Sigma }$
 is defined as $\mathbf{\Lambda}+\lambda \boldsymbol{I}$, where $\mathbf{\Lambda}\in \mathbb{R}^{E\times E}$
 is a function of the brain network topology $G$ and  correlation parameters $\boldsymbol{\rho} = (\rho_0, \rho_1, \cdots, \rho_K)$.  $G$ is a random measure of the latent $K$ brain networks, $\boldsymbol{I}$ is the $E\times E$ identity matrix, and $\lambda$ usually takes small values  ensuring the invertibility of $\mathbf{\Sigma }$~\cite{golub2013matrix}.   Different from ~\cite{chen2020bayesian}, here $G$ is restricted to follow a finite-dimensional Dirichlet distribution with parameters $(\gamma_1,\ldots,\gamma_K)$, rather than the DP, to avoid the dimensional changes of parameters and label switching issues during the inference~\cite{jasra2005markov, neal2000markov}. The Bayesian model used in this study is as follows.
\begin{align}  
\mathbf{R}^{s}\mid \boldsymbol{\Lambda} &\sim MVN(\mathbf{0},(\boldsymbol{\Lambda}+\lambda \boldsymbol{I})), s=1,\ldots, S,\nonumber \\  
\boldsymbol{\Lambda} &= f(G,\boldsymbol{\rho}), \label{eq2} \\  
G &\sim \mathrm{Dirichlet}(\gamma_1,\ldots,\gamma_K),\nonumber  
\end{align}
where $\mathrm{Dirichlet}(\gamma_1,\ldots,\gamma_K)$ represents a finite-dimensional Dirichlet distribution with parameter $(\gamma_1,\ldots,\gamma_K)$. And   the mapping relationship $f$ between $G$ and $\boldsymbol{\rho}$ takes the following form,
\begin{equation}  
\Lambda_{e_{i,j},e_{i',j'}}=\begin{cases}  
exp\left\{-\rho_k\right\}, &\text{if }\omega_i=\omega_j=\omega_{i'}=\omega_{j'}=C_k, \\  
exp\left\{-\rho_0\right\}, &\text{otherwise.}  
\end{cases}  
\label{eq3}  
\end{equation}
The entry of $\Lambda_{e_{i,j},e_{i',j'}}$ is determined by four vertices ($i,~j,~i',~j' =1,\ldots, V, \text{ and } i \neq j, i' \neq j'$) of the two edges. For $i=1,\ldots,V$, $\omega_i = C_k$ ($k=1,\ldots,K$) serves as an indicator variable for brain region $i$, signifying that brain region $i$ is assigned to the $k$-th class of brain networks. The four brain regions belong to the same class if and only if $\omega_i = \omega_j = \omega_{i'} = \omega_{j'}$.

Additionally, unlike the form in ~\cite{chen2020bayesian}, the covariance matrix here takes the form $(\boldsymbol{\Lambda}+\lambda \boldsymbol{I})$ instead of  $\boldsymbol{\Lambda}$. The adjustment term $\lambda \boldsymbol{I}$ ensures computational stability: during the MCMC sampling for  $\omega_i$ or $\rho_k$,  numerical underflow or overflow  can make $\boldsymbol{\Lambda}$ ill-conditioned or non-invertible, and halting the inference process. Incorporating a diagonal matrix $\lambda \boldsymbol{I}$ maintains  the invertibility of the covariance matrix throughout sampling. However, this adjustment may introduce bias into   parameter estimation. Consequently, the value of $\lambda$ must be sufficiently small to maintain computational stability while minimizing the impact on the accuracy of the estimation results ~\cite{golub2013matrix}.

To ensure the validity of the structured covariance matrix in practice, the following assumptions are made.  Firstly, the correlation between edges across classes is weak or nearly zero or at least is significantly different from that within classes.   Secondly, a positive correlation is assumed for edges within the same class, arising from their shared similarity in patterns. Therefore, if given two edges whose vertices do not belong to the same class, the entry of the log covariance corresponding to two edges will be assigned $\exp\left\{-\rho_0\right\}$, and this value should be small or different from $\exp\left\{-\rho_k\right\}$ ($k=1,\ldots,K)$. For vertices with the same class $k$,  it will be assigned a value of $\exp\left\{-\rho_k\right\}$. This exponential form ensures that the resulting covariance matrix remains positive definite, thus avoiding numerical difficulties in subsequent computations such as determinant evaluation and matrix inversion~\cite{stein1999interpolation}.

We take a discrete distribution $Discrete(\mathbf{\pi})$ to represent that brain region $i$ ($i=1,\ldots,V$) is associated with the brain network ${\omega}_i \in (C_1,\cdots,C_K)$ with probabilities $\mathbf{\pi}=(\pi_1,\cdots,\pi_K)$. We  use the following model to explore the topology of the brain network,
\begin{align}
\omega_i = C_k \mid \boldsymbol{\pi} &\sim \mathrm{Discrete}(\boldsymbol{\pi}), \quad i = 1, \ldots, V \label{eq4} \\
\boldsymbol{\pi} \mid (\gamma_1, \ldots, \gamma_K) &\sim \mathrm{Dirichlet}(\gamma_1, \ldots, \gamma_K), \nonumber
\end{align}
where $K$ is the pre-defined number of brain network categories and is treated as a tunable parameter.

Stacking all subject vectors row-wise ($\mathbf{R}^s$,$s=1,\ldots,S$) forms the full ${S \times E} $ data matrix $ \mathbf{R}$. Given the model described in Equations (\ref{eq2},~\ref{eq3},~\ref{eq4}),  the joint posterior  of  $G$ and $\boldsymbol{\rho}$ with prior distributions $p(\boldsymbol{\rho}),p(\mathbf{\omega}$) takes the following form:
\begin{align}
p(G, \boldsymbol{\rho} \mid \mathbf{R}) &\propto p(\mathbf{R}\mid G, \boldsymbol{\rho})p(G)p(\boldsymbol{\rho}) \label{eq5}
\end{align}
\vspace{-\baselineskip}
\vspace{-\baselineskip}
\begin{align*}
\small
\propto \exp \Bigg\{ &-\frac{S}{2} \log \left( \det(\boldsymbol{\Lambda} + \lambda \mathbf{I}) \right) \\
& - \frac{S}{2} \mathrm{tr} \left( \mathbf{H} (\boldsymbol{\Lambda} + \lambda \mathbf{I})^{-1} \right) \Bigg\} p(G) p(\boldsymbol{\rho}),
\end{align*}
where $\mathbf{H} = \mathbf{R}^T \mathbf{R}/S$. The derivation of Equation (\ref{eq5}) is provided in \nameref{append1}.

We can also derive the full conditional
distribution of $\mathbf{\omega}$ and $\boldsymbol{\rho}$ as follows: 
\begin{equation}
\begin{aligned}
p(\omega_i = C_k \mid \boldsymbol{\omega}_{-i}, \boldsymbol{\rho}, \mathbf{R}) 
&\propto\; \exp\Bigg\{ -\frac{S}{2}\log\left(\det\left(\mathbf{\Lambda} + \lambda \mathbf{I}\right)\right)\\
&\quad -\frac{S}{2}\,\mathrm{tr}\left(\mathbf{H}\left(\mathbf{\Lambda} + \lambda \mathbf{I}\right)^{-1}\right) \Bigg\}
\times \frac{m_{-ik}}{V-1}, i=1,\ldots, V,
\end{aligned}
\label{eq6}
\end{equation}

\vspace{-\baselineskip}

\begin{multline}
p(\rho_k\mid \boldsymbol{\omega},\mathbf{H},\boldsymbol{\rho}_{-k}) \propto \exp\left\{-\frac{S}{2}\log\left(\det\left(\mathbf{\Lambda}+\lambda \mathbf{I}\right)\right) \right. \\
\left. -\frac{S}{2}\mathrm{tr}\left(\mathbf{H}\left(\mathbf{\Lambda}+\lambda \mathbf{I}\right)^{-1}\right)\right\} \times p(\rho_k), k=0,1,\ldots,K, \label{eq7}
\end{multline}
where $\boldsymbol{\omega}_{-i}=\{\omega_i\}_{1\le i \le V}\setminus \{\omega_i\}$, the symbol $\setminus$ denotes set subtraction, and $m_{-ik}=\sum_{j\neq i}I(\omega_j=C_k)$. Similarly, denote $\boldsymbol{\rho}_{-k}=\{\rho_k\}_{0\le k \le K}\setminus \{\rho_k\}$ and $\rho_k$ with $k=0, 1, \ldots, K$.

\medskip 
The priors of $\rho_k$ ($0\le k\le K$) can be selected as a normal distribution where values falling into the high-intensity region are positive, or alternatively a Gamma distribution or exponential distribution. The derivation of the full conditional distributions of $\mathbf{\omega}$ and $\boldsymbol{\rho}$ can be found in  \nameref{append2}. 

\section{Model Inference}\label{sec3}
%\quad 

This section describes the posterior inference workflow for the BBeC model, which consists of two core parts. First, we detail the MH algorithm for parameter sampling, covering the implementation of the finite-dimensional Dirichlet distribution, the specific inference steps, and the formulation of the acceptance rate. Second, to mitigate the computational challenges posed in high-dimensional brain networks, we introduce a computational optimization strategy to improve algorithmic efficiency for large-scale networks.

\subsection{Metropolis Hasting algorithm}

This section provides a detailed description of the implemented MH sampling procedure for inferring the brain network assignment parameters $\mathbf{\omega}$ 
 and the correlation parameters $\boldsymbol{\rho}$. Please note that in our model, the number of brain network categories $K$ is treated as a tuning parameter. In practice, when assigning initial class labels to the $V$ brain regions, we suggest ensuring  exactly $K$ distinct categories are included-consistent with the tuning parameter $K$, and each category contains at least two brain regions. This requirement arises because each category must include at least two edges, and two edges necessitate at least two nodes.  Formally, there exists a vector of category sizes $(n_1, n_2, \ldots, n_K)$ of length $K$, satisfying:
\begin{equation}
n_k \geq 2, \quad \sum_{k=1}^{K} n_k = V.
\label{eq9}
\end{equation}

This requirement also guides the tuning of the parameter $K$. In theory, the initial values of $\boldsymbol{\rho}$ can be set to any real numbers; however, in practice, $\boldsymbol{\rho}$ is initialized to be positive to avoid the explosion of Eq. (\ref{eq3}).

In this work, we employ a \textit{block-wise updating strategy} to enhance sampling efficiency. The procedure iterates in the following two alternating phases.
\begin{enumerate}
    \item \textbf{Label updating}: Using the MH algorithm, we sequentially update the class labels $ \omega_i $ for each brain region.
    \item \textbf{Parameter updating}: With class assignments fixed, the MH algorithm is reapplied to update the correlation parameters $ \boldsymbol{\rho} $.
\end{enumerate}

As highlighted by ~\cite{chen2020bayesian}, such an alternating scheme not only improves sampling performance but also generally demonstrates favourable convergence behaviour.

In the updating of $\omega_i$, we adopt a proposal distribution identical to its prior, implementing it as a degenerate form of the DP~\cite{ishwaran2002exact}. Specifically, we conceptualize the sampling mechanism as drawing from a DP with the concentration parameter set to zero ($\alpha=0$). In this setting, when the base measure is discrete, the DP induces a standard finite-dimensional Dirichlet distribution in terms of the probabilities of the underlying categories~\cite{ferguson1973bayesian,teh2017dirichlet}. And for the update of $\omega_i$,  the acceptance probability of new proposal $\omega_i^*$ is computed as follows:\begin{equation}
{\alpha}_{\omega_i^*} = \min\left(1, \frac{p(\mathbf{R} \mid \omega_i^*, \boldsymbol{\omega}_{-i}, \boldsymbol{\rho})}{p(\mathbf{R} \mid \omega_i, \boldsymbol{\omega}_{-i}, \boldsymbol{\rho})}\right).
\label{eq:omega_acceptance}
\end{equation}

For each parameter $\rho_k$, $k = 0, \ldots, K$, the update is performed by generating a candidate value $\rho_k^*$  from the proposal distribution $q(\cdot \mid \rho_k)$,
$$
\rho_k^* \mid \rho_k \sim q(\cdot \mid \rho_k), \quad k = 0, \ldots, K.
$$
The choice of the proposal distribution $q(\cdot)$ is flexible. Typical prior selections include a range of distributions, such as the Normal distribution (often employed in random-walk formulations), as well as Gamma or exponential families, which offer flexibility in modeling different underlying processes. The acceptance probability for the candidate value $\rho_k^*$, denoted as $\alpha_{\rho_k}$, is given by the generalized MH criterion:\begin{equation}
\alpha_{\rho_k^*} = \min\left(1, \frac{p(\mathbf{R} \mid \rho_k^*, \boldsymbol{\omega}, \boldsymbol{\rho}_{-k}) p(\rho_k^*)}{p(\mathbf{R} \mid \rho_k, \boldsymbol{\omega}, \boldsymbol{\rho}_{-k}) p(\rho_k)} \times \frac{q(\rho_k \mid \rho_k^*)}{q(\rho_k^* \mid \rho_k)}\right).
\label{eq:rho_acceptance}
\end{equation}

This formula can be simplified under certain conditions. For instance, if the proposal distribution is chosen to be the parameter's prior, i.e., $q(\rho_k^* \mid \rho_k) = p(\rho_k^*)$, the acceptance probability reduces to the ratio of the likelihood functions. Alternatively, if the proposal distribution is symmetric, such that $q(\rho_k \mid \rho_k^*) = q(\rho_k^* \mid \rho_k)$, the acceptance probability becomes the ratio of the posterior distributions.

The procedure of the MH algorithm aforementioned is described in Algorithm~\ref{MH_algorithm}.

%\XXX{Add the description of the reversed order of the updating parameters with specifying the results are valiable in Appendix, and add the results to Appendix }

\floatname{algorithm}{Algorithm}  
\renewcommand{\algorithmicrequire}{\textbf{Input:}}  
\renewcommand{\algorithmicensure}{\textbf{Output:}}  
 
\renewcommand{\thealgorithm}{1}  
    \begin{algorithm}[!htbp]
        \caption{Metropolis-Hastings Algorithm}  
        \label{MH_algorithm}
        \begin{algorithmic}[1] 
            \Require $\mathbf{R}\text{, number of iterations: } M\text{, regions: } V \text{, }\text{number of individuals: }S
            \text{, and } \lambda$.
            \Ensure The network of brain regions, denoted as $\omega_i$, along with $\rho_k$.
            \State Assign initial values to $\mathbf{\omega}, \boldsymbol{\rho}$
            \State $\theta^{(0)} \leftarrow \mathbf{\omega}$, $\boldsymbol{\rho}^{(0)}\leftarrow\boldsymbol{\rho}$
            \For{$t=1 \text{\, to\, } M$} //M is iteration
                \For{$i=1 \text{\, to\, } V$} //Classify each region.
                    \State $\omega_i^* \leftarrow \text{CRP}\text{ with probability } \frac{m_{-ik}}{V - 1}$
                    \State $\theta^{(t)} \leftarrow$ Replace the network of the i-th region in $\theta^{(t-1)}$ with $\omega_i^*$.
                    \State $p(\theta^{(t)}) \leftarrow$ Calculate the likelihood using the parameters $\left\{\theta^{(t)}, \rho_k^{(t-1)}\right\}$ according to Eq.(\ref{eq6}).
                    \State $p(\theta^{(t-1)}) \leftarrow$ Calculate the likelihood function using the parameters $\left\{\theta^{(t-1)},  \rho_k^{(t-1)}\right\}$ according to Eq.(\ref{eq6}).
                    \State $\mathbf{\alpha}_{\omega_i}\leftarrow$Eq.(\ref{eq:omega_acceptance}) 
                    \If{$\mathbf{\alpha}_{\omega_i} > $ uniform(0,1)}
                        \State Record the classification of the current $\omega_i$
                    \Else
                        \State $\theta^{(t)}\leftarrow\theta^{(t-1)}$
                    \EndIf
                \EndFor
                \State $\mathbf{K} \leftarrow \text{Unique}(\theta^{(t)})$
                \For{$i \text{\, in \,} \mathbf{K}$}
                    \State $\rho_i^{(t)} \leftarrow q\left(\cdot \mid \rho_i^{(t-1)}\right)$.
                    \State $p^{*} \leftarrow$ Calculate the likelihood function using the parameters $\left\{\theta^{(t)}, \rho_i^{(t)}, \rho_{-i}\right\}$ according to Eq.(\ref{eq7}).
                    \State $p \leftarrow$ Calculate the likelihood using the parameters $\left\{\theta^{(t)}, \rho_i^{(t-1)}, \rho_{-i}\right\}$ according to Eq.(\ref{eq7}). 
                    \State $\mathbf{\alpha}_{\rho_k}\leftarrow$ Eq.(\ref{eq:rho_acceptance})  
                    \If{$\mathbf{\alpha}_{\rho_k} > $ uniform(0,1)}
                        \State Record the current $\rho_i$
                    \Else
                        \State $\rho_i^{(t)}\leftarrow\rho_i^{(t-1)}$
                    \EndIf
                \EndFor
            \EndFor
        \end{algorithmic}
    \end{algorithm}

\subsection{Computational Optimization}

As the number of nodes $V$ increases, the number of edges $E$ in the brain network grows quadratically as $O(V^2)$. This leads to a significant expansion in the dimensionality ($E \times E$) of the edge-based covariance matrix, imposing a heavy computational burden on the posterior inference process. Therefore, it is essential to optimize the likelihood calculation to improve overall efficiency.

In standard numerical computing, the log-determinant of a matrix is typically computed via eigenvalue decomposition, while the inverse matrix in the trace term is obtained through direct inversion methods such as Gauss-Jordan elimination or LU decomposition. For an $E \times E$ matrix, the computational complexity of these general algorithms is $O(E^3)$. Given that these high-dimensional operations must be repeated in every iteration of the MH algorithm, conventional methods may result in prohibitive execution times for large-scale networks. Since the edge covariance matrix in our model is guranteed to be symmetric positive definite, we can leverage this property to implement more efficient computational strategies.

The primary bottlenecks in the likelihood calculation are the log-determinant of the high-dimensional covariance matrix and the trace term involving its inverse. Let the edge covariance matrix be $\mathbf{A} = \boldsymbol{\Lambda} + \lambda \mathbf{I}$. Since $\mathbf{A}$ is symmetric positive definite, we can perform Cholesky decomposition such that $\mathbf{A} = \mathbf{L}\mathbf{L}^T$, where $\mathbf{L}$ is a lower triangular matrix.

By utilizing the property that the determinant of a triangular matrix equals the product of its diagonal elements, the log-determinant of $\mathbf{A}$ can be simplified as follows:
\begin{equation*}
    \log(\det(\mathbf{A})) = \log\left(\det(\mathbf{L})\det(\mathbf{L}^T)\right) = 2 \sum_{i=1}^E \log(L_{ii}).
\end{equation*}
After obtaining $\mathbf{L}$, this step only requires summing the logarithms of the diagonal elements. Although the computational complexity of Cholesky decomposition remains $O(E^3)$, its constant factor is much smaller than that of eigenvalue decomposition, leading to a faster computation in practice.

For the second bottleneck, the trace term $\mathrm{tr}(\mathbf{H}\mathbf{A}^{-1})$, recall that $\mathbf{H} = \frac{1}{S}\mathbf{X}^T\mathbf{X}$, where $S$ is the sample size and $\mathbf{X}$ is the $S \times E$ data matrix. Using the cyclic property of the trace, the expression can be rewritten as:
\begin{equation*}
    \mathrm{tr}(\mathbf{H}\mathbf{A}^{-1}) = \mathrm{tr}\left(\frac{1}{S}\mathbf{X}^T\mathbf{X}\mathbf{A}^{-1}\right) = \frac{1}{S}\mathrm{tr}(\mathbf{X}\mathbf{A}^{-1}\mathbf{X}^T).
\end{equation*}
To avoid direct matrix inversion, we define an intermediate matrix $\mathbf{Y} = \mathbf{A}^{-1}\mathbf{X}^T$, which is equivalent to solving the linear system $\mathbf{A}\mathbf{Y} = \mathbf{X}^T$. Substituting the Cholesky decomposition, we solve $\mathbf{L}\mathbf{L}^T\mathbf{Y} = \mathbf{X}^T$ in two steps: first, solving $\mathbf{L}\mathbf{Z} = \mathbf{X}^T$ via forward substitution, and then solving $\mathbf{L}^T\mathbf{Y} = \mathbf{Z}$ via backward substitution.

Since the sample size $S$ is typically much smaller than the number of edges $E$, the complexity of this step is $O(E^2 S)$, which is significantly more efficient than the $O(E^3)$ complexity of direct inversion. Finally, the trace calculation is reduced to:
\begin{equation*}
    \mathrm{tr}(\mathbf{X}\mathbf{Y}) = \sum_{i=1}^S \sum_{j=1}^E X_{ij} Y_{ji}.
\end{equation*}
This element-wise summation avoids the storage of large intermediate matrices, saving memory and further increasing computational speed. These optimized values are then used to calculate the MH acceptance probability.

\section{Simulation Studies}\label{sec4}

\subsection{Data Generation Settings} 
%\quad 

 In this section, we simulated data under different scenarios to evaluate the performance of our proposed method. We generated datasets containing $ S $ subjects, $ V $ brain regions, involving $ E = V \times (V-1)/2 $ edges. Specifically, given a predefined brain network topology $ G $ and  a set of correlation parameters $ \boldsymbol{\rho} $, an $ E \times E $ covariance matrix $ \mathbf{\Lambda} $ was constructed as defined in Equation (\ref{eq3}) in Section~\ref{sec2}.  Then, for each subject $ s $ ($s=1,\ldots,S$), data vectors $ \mathbf{R}^s $ were generated as specified in Section~\ref{sec2}; the full data matrix $ \mathbf{R}$ and sample covariance matrix $ \mathbf{H}$
 were then calculated following description in Section~\ref{sec2}. The complete data generation procedure is outlined in Algorithm ~\ref{Gen_algorithm}.

\floatname{algorithm}{Algorithm} 
\renewcommand{\algorithmicrequire}{\textbf{Input:}} 
\renewcommand{\algorithmicensure}{\textbf{Output:}} 
 
\renewcommand{\thealgorithm}{2} 
    \begin{algorithm}
        \caption{Data Generating Algorithm} 
        \label{Gen_algorithm}
        \begin{algorithmic}[1] 
            \Require $\omega_i$, $\rho_k$, $S$, $\lambda$, $V$
            \Ensure Data: $\mathbf{H}$, $\mathbf{R}$, Matrix $\boldsymbol{\Lambda}$ 
            \State Assign a category $\omega_i$ to each node $i$, $i=1,\ldots, V$.
            \State $E = V\times(V-1)/2$ 
            \State Construct the $\boldsymbol{\Lambda}$ matrix according to Eq.(\ref{eq3}).  
            \State Draw $S$ samples from a multivariate normal distribution $N(\mathbf{0}, (\boldsymbol{\Lambda} + \lambda \boldsymbol{I}))$ to form the matrix $\mathbf{R}$.
            \State $\mathbf{H} \leftarrow \mathbf{R}^T \mathbf{R}$/$S$
            
        \end{algorithmic}
    \end{algorithm}

In the simulation studies, to mimic the real scenarios, we fixed the number of subjects at $S = 632$ and generated datasets with varying numbers of brain regions, $V = 10, 20, 30, 40 $,  representing simple, moderate, and complex scenarios respectively (with the number of edges ranging from $45$ to $780$). This experimental design allowed  comprehensive  evaluation of   the proposed model's performance.  Each $V$ was paired with a  group structure $G$ and a set of correlation parameters $\rho$.  The partitioning parameter $ G $ defined the sizes of distinct subnetworks (e.g., $ G = (3,3,4) $ indicates three subnetworks with $3$, $3$, and $4$ brain regions, respectively). The correlation parameter vector $\boldsymbol{\rho}$ specified the connection strength within each subnetwork (e.g., $\exp(-\boldsymbol{\rho}) = [0.2, 0.4, 0.9] $ means the connection strengths for the first, second, and third subnetworks are $0.2$, $0.4$, and $0.9$, respectively).   The primary objective was to assess whether the proposed Bayesian model can identify the underlying network topology embedded in the covariance matrix and correctly infer the dependency relationships among neural connections when using the sample covariance matrix as input.

For $V=10$, we configured the group structure as $G=(3,3,4)$ with correlation parameters  $\exp(-\boldsymbol{\rho})=[0.2, 0.4, 0.9]$. For $V=20$, the group structure was $G=(4,4,4,4,4)$ paired with $\exp(-\boldsymbol{\rho})=[0.2, 0.4, 0.6, 0.8, 0.9]$. For $V=30$, we used $G=(5,5,5,5,5,5)$ and $\exp(-\boldsymbol{\rho})=[0.001, 0.2, 0.4, 0.5, 0.6, 0.8]$. 
This setting incorporated a very low correlation value ($0.001$) to test the model's ability to infer weak connectivity between groups.   Finally, for $V=40$, the group structure was $G=(5,5,5,5,5,5,5,5)$ 
 with correlation parameters $\exp(-\boldsymbol{\rho})=[0.001, 0.1, 0.2, 0.3, 0.4, 0.5, 0.6, 0.7]$.

 For each $ V $, the group structure $ G $ and correlation parameter $ \rho $ were held constant,  two distinct topologies were considered (differing in their group assignments $ \boldsymbol{\omega} $), denoted as \textit{structure 1} (S1) and \textit{structure 2} (S2). These topologies primarily differed in community member arrangement.  S1 represented a well-organized clustered topology where nodes were ordered by group membership (e.g., $ \boldsymbol{\omega} = (\underbrace{0, \ldots, 0}_{G_1}, \underbrace{1, \ldots, 1}_{G_2}, \ldots, \underbrace{K-1, \ldots, K-1}_{G_K}) $). In contrast, S2 represented a random topology where members of the same community were scattered (e.g., $ \boldsymbol{\omega} = (2, K-1, 1, \ldots, 0) $).   The purpose of this comparative design is to evaluate whether the model's capabilities are influenced by the organizational scheme of the nodes. Specifically, we aimed to verify that the model can identify the topological structure, regardless of whether the brain regions are arranged in a clustered fashion or a random order. The specific node assignments $\boldsymbol{\omega}$ employed for both S1 and S2 across different network sizes are detailed in Table ~\ref{tab:sim_structures}.
 
\begin{table}[htbp]
  \centering
  \caption{Configurations of $\boldsymbol{\omega}$ 
 for S1 and S2 across different network sizes.}
  \label{tab:sim_structures}
  \renewcommand{\arraystretch}{1.3} 
  \small  
  \begin{tabularx}{\textwidth}{c c >{\raggedright\arraybackslash}X} 
    \toprule
    \textbf{V} & \textbf{Structure} & \textbf{Node Assignment Vector} $\boldsymbol{\omega}$ \\
    \midrule
    \multirow{2}{*}{10} 
      & S1 & $(1, 1, 1, 2, 2, 2, 3, 3, 3, 3)$ \\
      & S2 & $(1, 2, 3, 1, 2, 3, 1, 2, 3, 3)$ \\
    \midrule
    \multirow{2}{*}{20} 
      & S1 & $(1, 1, 1, 1, 2, 2, 2, 2, 3, 3, 3, 3, 4, 4, 4, 4, 5, 5, 5, 5)$ \\
      & S2 & $(1, 5, 2, 4, 1, 3, 4, 5, 3, 2, 5, 1, 4, 3, 2, 5, 1, 4, 2, 3)$ \\
    \midrule
    \multirow{2}{*}{30} 
      & S1 & $(1, 1, 1, 1, 1, 2, 2, 2, 2, 2, 3, 3, 3, 3, 3, 4, 4, 4, 4, 4, 5, 5, 5, 5, 5, 6, 6, 6, 6, 6)$ \\
      & S2 & $(3, 5, 2, 4, 1, 6, 1, 5, 3, 2, 4, 6, 1, 3, 5, 2, 4, 6, 1, 3, 5, 2, 4, 6, 1, 3, 5, 2, 4, 6)$ \\
    \midrule
    \multirow{2}{*}{40} 
      & S1 & $(1, 1, 1, 1, 1, 2, 2, 2, 2, 2, 3, 3, 3, 3, 3, 4, 4, 4, 4, 4, \newline 5, 5, 5, 5, 5, 6, 6, 6, 6, 6, 7, 7, 7, 7, 7, 8, 8, 8, 8, 8)$ \\
      & S2 & $(6, 1, 7, 7, 4, 8, 2, 5, 8, 5, 2, 3, 2, 4, 3, 1, 6, 7, 4, 3, \newline 5, 4, 7, 1, 6, 2, 8, 1, 4, 1, 8, 6, 6, 3, 2, 5, 7, 5, 8, 3)$ \\
    \bottomrule
  \end{tabularx}
\end{table}

To evaluate the effect of the regularization parameter $\lambda$, a sensitivity analysis was performed using the S1 topological structure. This analysis helps quantify the impact the choices of $\lambda$ on the structural features identified by the model and evaluate the sensitivity of results to $\lambda$ selection. In the data generation process, the ground truth value of $\lambda$ was set to $0$. During the inference stage, $\lambda$ was fixed at various orders of magnitude: $10^{-10}$, $10^{-8}$, $10^{-6}$, $10^{-4}$, $10^{-3}$, and $10^{-2}$.

%This analysis helps quantify the impact of hyperparameter choices on the structural features identified by the model and assesses the sensitivity of results to hyperparameter selection.

\subsection{Evaluation Metrics}

To evaluate the discrepancy between estimated and true covariance matrices, we employed two key metrics: (1) the normalized Frobenius norm $||\mathbf{\Lambda} - \hat{\mathbf{\Lambda}}||_F / E$, where $\mathbf{\Lambda}$ denotes the true connectivity matrix and $\hat{\mathbf{\Lambda}}$ denotes the estimated counterpart ~\cite{chavarria2024froimpro}; and (2) the normalized error $||\mathbf{\Lambda} - \hat{\mathbf{\Lambda}}||_1/(E\times E)$.

To quantitatively evaluate the node clustering performance, the adjusted rand index (ARI) was adopted~\cite{hubert1985comparing}. The ARI measures the similarity between the estimated clusters and the ground-truth topological structure, where a higher value indicates better clustering performance. The index reaches its maximum value of $1$ when the estimated results perfectly match the true labels.

Furthermore, we evaluated the performance of our method in identifying the underlying brain network topology by assessing its sensitivity and specificity. To this end, we directly compared the presence and absence of non-zero entries in the estimated covariance matrix $\hat{\mathbf{\Lambda}}$ against the ground-truth matrix $\mathbf{\Lambda}$. The true/false positive and negative rates are defined by comparing the non-zero entries of the true covariance matrix and its estimate. Specifically: a \textit{true positive} (TP) occurs if both $\mathbf{\Lambda}_{ij}$ and $\hat{\mathbf{\Lambda}}_{ij}$ are non-zero; a \textit{false positive} (FP) if  $\mathbf{\Lambda}_{ij}$ is zero but $\hat{\mathbf{\Lambda}}_{ij}$ is non-zero; a \textit{true negative} (TN) if both $\mathbf{\Lambda}_{ij}$ and $\hat{\mathbf{\Lambda}}_{ij}$ are zero; and a \textit{false negative} (FN) if 
 $\mathbf{\Lambda}_{ij}$ is non-zero but $\hat{\mathbf{\Lambda}}_{ij}$ is zero. Sensitivity is calculated as $ \text{TP}/(\text{TP}+\text{FN}) $, reflecting the model's ability to detect true brain connections, while specificity is defined as $ \text{TN}/(\text{TN}+\text{FP}) $, representing the capacity to suppress false positives.

Finally, the split-$\hat{R}$ statistic~\cite{riutort2020bayesian} was introduced to assess the convergence of the MCMC algorithm. This metric evaluates convergence by comparing the between-chain variance and within-chain variance of the split sequences. Specifically, after discarding a certain proportion of burn-in samples, the remaining sequence of length $N$ is divided into two halves of length $n$. Let $W$ denote the average within-chain variance and $B$ represent the between-chain variance. The split-$\hat{R}$ is calculated as follows:
\begin{equation*}
\hat{R} = \sqrt{\frac{\frac{n-1}{n}W + \frac{B}{n}}{W}}.
\end{equation*}
Generally, a split-$\hat{R}$ value between $1.0$ and $1.1$ indicates that the model has reached a satisfactory state of convergence~\cite{vehtari2021rank}.

\subsection{Tuning K} 
Since $K$ was treated as a tuning parameter, we evaluated the model's performance across different values of $K$ using the Watanabe-Akaike information criterion (WAIC) ~\cite{millar2018conditional}, defined as:
$$
\text{WAIC} = -2 \sum_{i=1}^{n} \log \tilde{p}(y_i | \mathbf{y}) + 2\text{Var},
$$
where $\text{Var} = \sum_{i=1}^{n} \text{Var}_i$ denotes the sum of the posterior variances of the pointwise log-likelihoods, with
$$
Var_i = \text{Var}_{\mathbf{\theta}|\mathbf{y}} (\log p(y_i | \mathbf{\theta})) \,.
$$
In this context, $\mathbf{y} = (y_1, ..., y_n)$ represents the vector of observed data, $\mathbf{\theta}$ denotes the vector of model parameters, and $\tilde{p}(\cdot | \mathbf{y})$ generically denotes the posterior predictive density.
A lower WAIC value indicates a better model fit. For the $ V = 10 $ case  $ G = (3, 3, 4) $ (true subnetwork count was $3$), we test $ K = 2, 3, 4 $ for both S1 and S2. For $ V = 20 $ and $ V = 30 $, we test $ K = 4, 5, 6 $ and $ K = 5, 6, 7 $, respectively, with both structures sharing the same $ K $ range. For $ V = 40 $,  due to structural  differences between S1 and S2, we test $ K = 7, 8, 9 $ for S1 and $ K = 8, 9, 10 $ for S2. The WAIC values across $ K $ settings were summarized in Table ~\ref{tab:waic_comparison} and ~\ref{tab:waic_comparison_extend}, guiding the selection of optimal clustering numbers.

For each combination of settings,   the value of $K$ was determined by identifying the  lowest WAIC presented in  Tables ~\ref{tab:waic_comparison}, ~\ref{tab:waic_comparison_extend} and we defined $\hat{K}$ to represent the distinct number of labels.

\subsection{Results} 
We executed the MH algorithm with $2,500$ iterations. For the initialization of $\mathbf{\omega}$, the brain regions were randomly assigned to categories subject to the size constraints specified in Eq.~(\ref{eq9}). Regarding the initialization of $\boldsymbol{\rho}$, the parameters $\rho_0, \rho_1, \ldots, \rho_K$ were randomly generated from the interval $[0, 0.5]$. For  $\rho_k$, we selected a normal prior distribution with a mean of $0.25$ and a standard deviation of $0.01$. The proposal distribution was configured as a random walk normal distribution with a step size (i.e., standard deviation) of $0.01$.

When $V=10$, for S1,  the best value of the tuning parameter was $K=3$, and the estimated parameters  $\exp(-\hat{\boldsymbol{\rho}}) = [0.1963, 0.4619, 0.8748]$ closely matched the ground truth, with the estimated topology fully aligned to the true configuration (Fig.~\ref{fig:brain_analysis}(a)). For S2, the best value of $K$ was $4$ (with one group vanishing during the evolution of the Markov chain, i.e., $\hat{K}=3$) and the estimated parameters $\exp(-\hat{\boldsymbol{\rho}}) = [0.1952, 0.4622, 0.8733]$ also aligned closely  with the true values, demonstrating consistent performance (Fig.~\ref{fig:brain_analysis}(b)).

When $V=20$, the optimal tuning value of $K$ for model inference was $6$ in both S1 and S2 scenarios, aligning perfectly with the true $K$ value. This yielded estimated parameters of $\exp(-\hat{\boldsymbol{\rho}})=[0.2084, 0.4616, 0.6176, 0.8482, 0.9028]$ and $\exp(-\hat{\boldsymbol{\rho}})=[0.2084, 0.4579, 0.6189, 0.8438, 0.9001]$, respectively, both estimated vectors closely aligned with the ground truth (Fig.~\ref{fig:brain_analysis}(c)-(d)). The quantitative
results in Table \ref{tab:comparison} indicated that the proposed model maintained robust performance
in structural identification, even as the number of nodes increased.

When $V=30$, the best value of S1 was $K=7$  (with one group disappeared as the Markov chain evolved, i.e., $\hat{K}=6$) , and the estimated parameters $\exp(-\hat{\boldsymbol{\rho}})=[0.00106, 0.1964, 0.4965, 0.5075, 0.6264, 0.7539]$ were close to the true values, with the topology fully aligned to the ground truth (Fig.~\ref{fig:large_scale_networks}(a)). In contrast, for  S2, the best value of $K$ was $6$, yielding an estimated $\exp(-\hat{\boldsymbol{\rho}})=[0.00106, 0.1962, 0.49606, 0.51202, 0.62204, 0.7579]$, which also matched the true configuration well (Fig.~\ref{fig:large_scale_networks}(b)).

In a larger-scale experiment with $V=40$ nodes, involving $780$ edges. For S1, the best value for $K$ was $9$ (one group vanished as the Markov chain evolved, i.e., $\hat{K}=8$), and the estimated parameters $\exp(-\hat{\boldsymbol{\rho}}) = [0.00096, 0.09729, 0.2029, 0.2993, 0.4839, 0.5651, 0.6684, 0.7234]$ were close to the true values, with the estimated topology perfectly aligned to the ground truth (Fig.~\ref{fig:large_scale_networks}(c)). For S2, the best value for $K$ was $10$ (two groups vanished as the Markov chain evolved, i.e., $\hat{K}=8$), yielding estimated parameters $\exp(-\hat{\boldsymbol{\rho}}) = [0.0009, 0.09728, 0.2038, 0.3004, 0.4862, 0.56601, 0.6688, 0.7233]$, which also matched the true configuration well (Fig.~\ref{fig:large_scale_networks}(d)). These results indicated that the proposed model remains effective even when the network was more complex. The quantitative results in Table~\ref{tab:comparison_extended}  indicated that the proposed model maintained robust performance in structural identification, even as the number of nodes increased.

In the simulation experiments with $V = 10, 20, 30, 40$, we observed that the convergence of the model was highly sensitive to the choice of $ K $. When the specified $ K $ was smaller than the true number of clusters, the MH algorithm failed to converge, resulting in excessively large WAIC values. When $ K $ matched the true cluster count, convergence was achieved for certain topological configurations (e.g., S1), but remained challenging for others (e.g., S2). For $ V = 40 $, where higher-dimensional parameter estimation and increased topological complexity were involved, setting $ K $ to values slightly larger than the true cluster count significantly improved convergence behavior. Notably, when $ K $ was overestimated, the converged $\hat{K}$ automatically aligned with the true cluster number, demonstrating the model's robustness regarding the initial selection of cluster numbers. These findings suggested that setting $ K $ slightly larger than the expected true value is advisable for ensuring algorithmic stability.

\begin{table}[t]
\centering
\caption{Model assessments using WAIC for varying $K$ when $V=10$ and $V=20$. log(-WAIC) values are reported for different specified cluster counts ($K$).  A hyphen (-) indicates non-convergence of the model. $\hat{K}$ denotes the actual number of clusters identified by the model.}
\label{tab:waic_comparison}

\setlength{\tabcolsep}{18pt} 

\begin{tabular}{ccc}
\toprule
\multicolumn{3}{c}{\textbf{$V = 10$}} \\
\multicolumn{3}{c}{\small $G=(3, 3, 4)$} \\
\multicolumn{3}{c}{\small $\exp(-\boldsymbol{\rho})=[0.2,0.4,0.9]$} \\
\midrule
\textbf{K} & S1 & S2 \\
\midrule
2 & - & - \\
3 & 12.917506 & - \\
4 & 12.9175060 $(\hat{K}=3)$ & 12.9175066 $(\hat{K}=3)$ \\
\midrule
\multicolumn{3}{c}{} \\[-1ex]  
\toprule
\multicolumn{3}{c}{\textbf{$V = 20$}} \\
\multicolumn{3}{c}{\small $G=(4,4,4,4,4)$} \\
\multicolumn{3}{c}{\small $\exp(-\boldsymbol{\rho})=[0.2,0.4,0.6,0.8,0.9]$} \\
\midrule
\textbf{K} & S1 & S2 \\
\midrule
4 & - & - \\
5 & 14.408328 & - \\
6 & 14.4083294 $(\hat{K}=5)$ & 14.4083291 $(\hat{K}=5)$ \\
\bottomrule
\end{tabular}
\end{table}

\begin{table}[t]
\centering
\caption{
Model assessments using WAIC for varying $K$ when $V=30$ and $V=40$. log(-WAIC) values are reported for different specified cluster counts ($K$).  A hyphen (-) indicates non-convergence of the model. $\hat{K}$ denotes the actual number of clusters identified by the model.}
\label{tab:waic_comparison_extend}

\setlength{\tabcolsep}{10pt} 

\begin{tabular}{cccc}
\toprule
\multicolumn{4}{c}{\textbf{$V = 30$}} \\
\multicolumn{4}{c}{\small$G=(5,5,5,5,5,5)$} \\
\multicolumn{4}{c}{\small $\exp(-\boldsymbol{\rho})=[0.001,0.2,0.4,0.5,0.6,0.8]$}\\
\midrule
\textbf{K} & S1 & & S2 \\
\midrule
5 & - & & - \\
6 & 15.2534880 & & 15.2534881 \\
7 & 15.2534882 $(\hat{K}=6)$ & & 15.2534880 $(\hat{K}=6)$ \\

\midrule
\multicolumn{4}{c}{} \\[-1ex]  

\toprule 
\multicolumn{4}{c}{\textbf{$V = 40$}} \\
\multicolumn{4}{c}{\small $G=(5,5,5,5,5,5,5,5)$} \\
\multicolumn{4}{c}{\small $\exp(-\boldsymbol{\rho})=[0.001,0.1,0.2,0.3,0.4,0.5,0.6,0.7]$} \\
\midrule
\textbf{K} & S1 & \textbf{K} & S2 \\
\midrule
7 & - & 8 & 15.841786 \\
8 & - & 9 & - \\
9 & 15.8417868 $(\hat{K}=8)$ & 10 & 15.8417867 $(\hat{K}=8)$ \\
\bottomrule
\end{tabular}
\end{table}

\begin{table*}[t]
\caption{Network estimation accuracy for small networks ($V$=$10$, $V$=$20$) concerning metrics normalized error, error of Frobenius norm, sensitivity, specificity, and ARI.}
\label{tab:comparison}
\tabcolsep=0pt
\begin{tabular*}{\textwidth}{@{\extracolsep{\fill}}lcccc@{\extracolsep{\fill}}}
\toprule
& \multicolumn{2}{c}{$V=10$} & \multicolumn{2}{c}{$V=20$} \\
\cmidrule(lr){2-3} \cmidrule(lr){4-5}
& \multicolumn{2}{c}{$G=(3,3,4)$} & \multicolumn{2}{c}{$G=(4,4,4,4,4)$} \\
& \multicolumn{2}{c}{$\exp(-\boldsymbol{\rho})=[0.2,0.4,0.9]$} & \multicolumn{2}{c}{$\exp(-\boldsymbol{\rho})=[0.2,0.4,0.6,0.8,0.9]$} \\
\cmidrule(lr){2-3} \cmidrule(lr){4-5} 
& S1 & S2 
& S1 & S2 \\
\midrule
normalized error & 0.0013 & 0.0003 & 0.0002 & 0.0003 \\
ARI & 1.00 & 1.00 & 1.00 & 1.00 \\
sensitivity & 100\% & 100\% & 100\% & 100\% \\
specificity & 100\% & 100\% & 100\% & 100\% \\
Frobenius norm & 0.00532 & 0.00547 & 0.00254 & 0.00237 \\
\bottomrule
\end{tabular*}
\end{table*}

\begin{table*}[t]
\caption{Network estimation accuracy for large networks ($V$=$30$, $V$=$40$) concerning metrics normalized error, error of Frobenius norm, sensitivity, specificity, and ARI.}
\label{tab:comparison_extended}
\setlength{\tabcolsep}{0pt}
\begin{tabular*}{\textwidth}{@{\extracolsep{\fill}}lcccc@{}}
\toprule
& \multicolumn{2}{c}{$V=30$} & \multicolumn{2}{c}{$V=40$} \\
\cmidrule(lr){2-3} \cmidrule(lr){4-5}
& \multicolumn{2}{c}{$G=(5,5,5,5,5,5)$} & \multicolumn{2}{c}{$G=(5,5,5,5,5,5,5,5)$} \\
& \multicolumn{2}{c}{\makecell{$\exp(-\boldsymbol{\rho})=[0.001,0.2,0.4,$\\$0.5,0.6,0.8]$}}
& \multicolumn{2}{c}{\makecell{$\exp(-\boldsymbol{\rho})=[0.001,0.1,0.2,0.3,$\\$0.4,0.5,0.6,0.7]$}} \\
\cmidrule(lr){2-3} \cmidrule(lr){4-5}
& S1 & S2 & S1 & S2 \\
\midrule
normalized error & 0.0001 & 0.0001 & 0.00004 & 0.00004 \\
ARI & 1.00 & 1.00 & 1.00 & 1.00 \\
sensitivity & 100\% & 100\% & 100\% & 100\% \\
specificity & 100\% & 100\% & 100\% & 100\% \\
Frobenius norm & 0.00253 & 0.00247 & 0.00164 & 0.00167 \\
\bottomrule
\end{tabular*}
\end{table*}

%===========================================================
%===========================================================
\begin{figure*}[tp!]  
    \centering  
    \resizebox{\textwidth}{!}{%
        \setlength{\tabcolsep}{1pt}  
        \newcommand{\subfiglabel}[1]{\textbf{(#1)}}
        \newcommand{\rowtitle}[1]{\multicolumn{1}{m{4em}}{\raggedleft\bfseries #1}} 
        
        \begin{tabular}{@{}m{4em}cccc@{}}
            & \subfiglabel{a} & \subfiglabel{b} & \subfiglabel{c} & \subfiglabel{d} \\[0.5ex]

            \rowtitle{Trace plots} 
            & \adjustimage{width=4cm,valign=c}{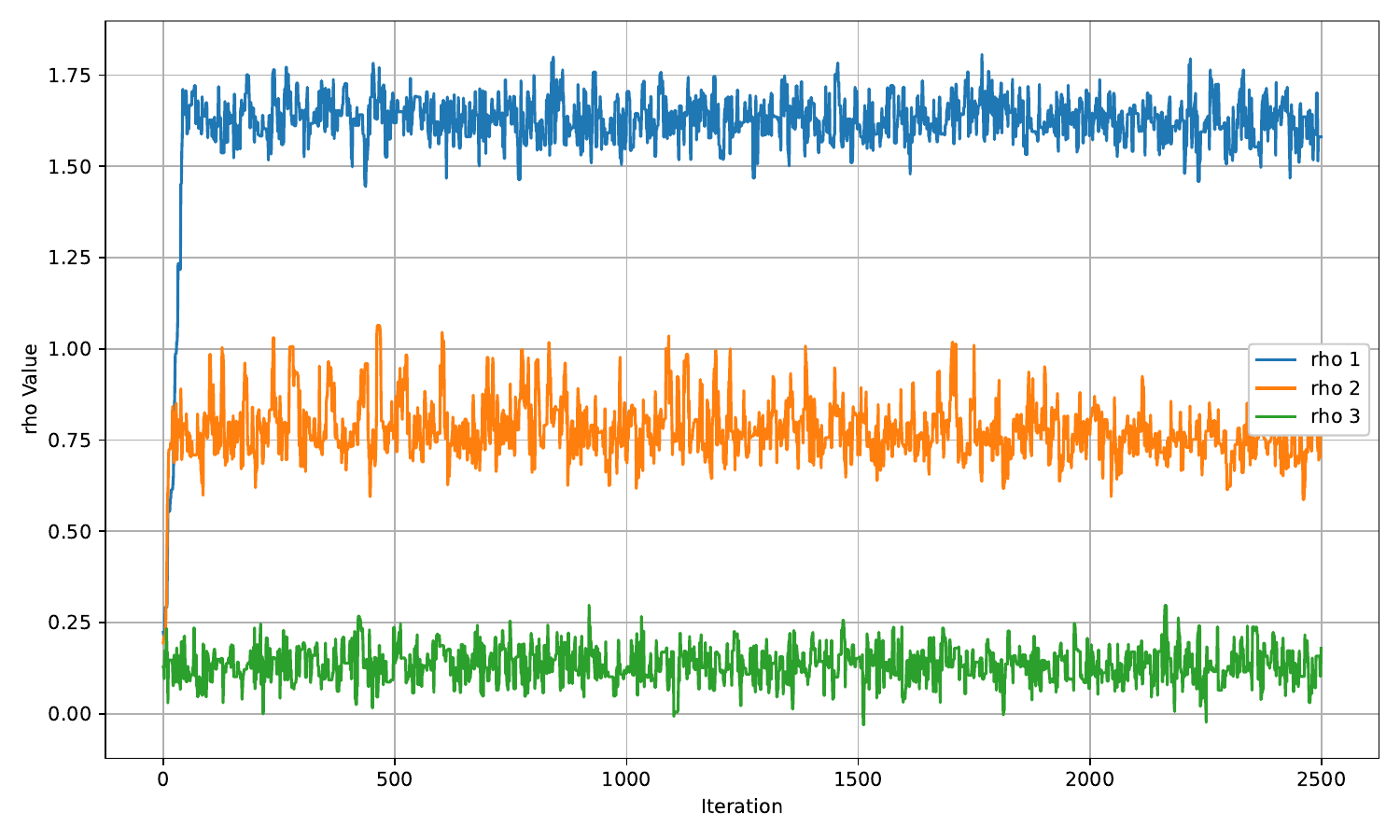}
            & \adjustimage{width=4cm,valign=c}{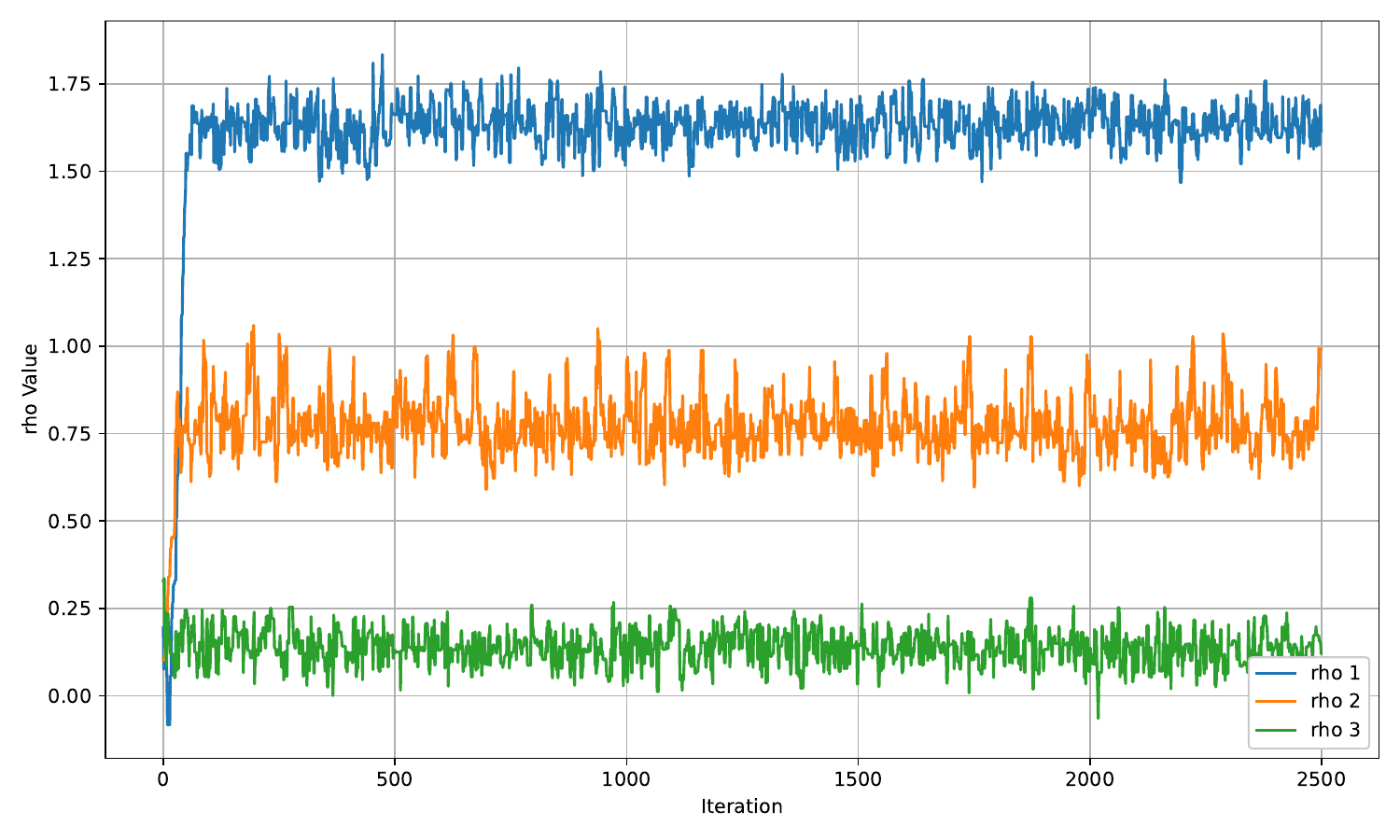}
            & \adjustimage{width=4cm,valign=c}{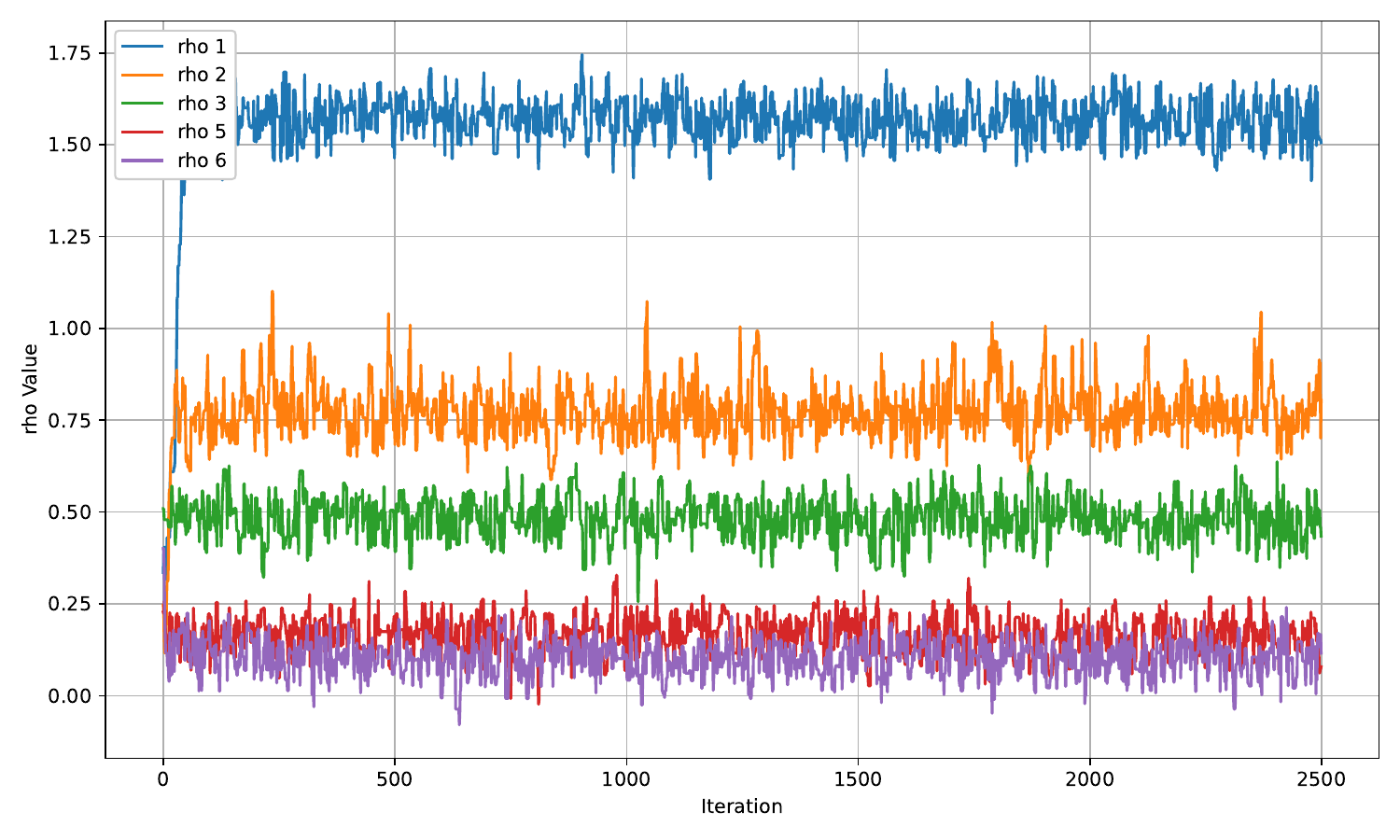}
            & \adjustimage{width=4cm,valign=c}{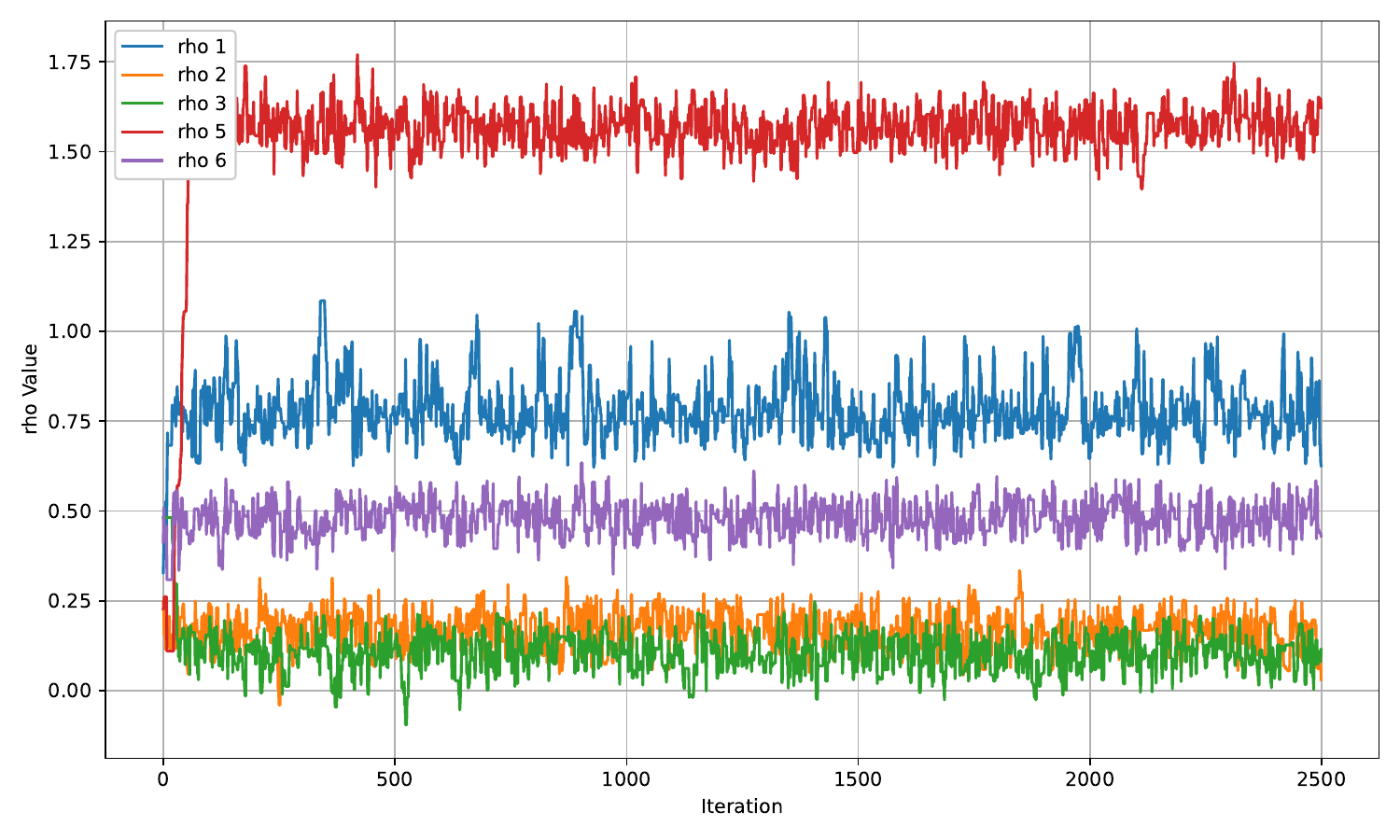} \\[1ex]
            
            \rowtitle{$\mathbf{\Lambda}$} 
            & \adjustimage{width=4cm,valign=c}{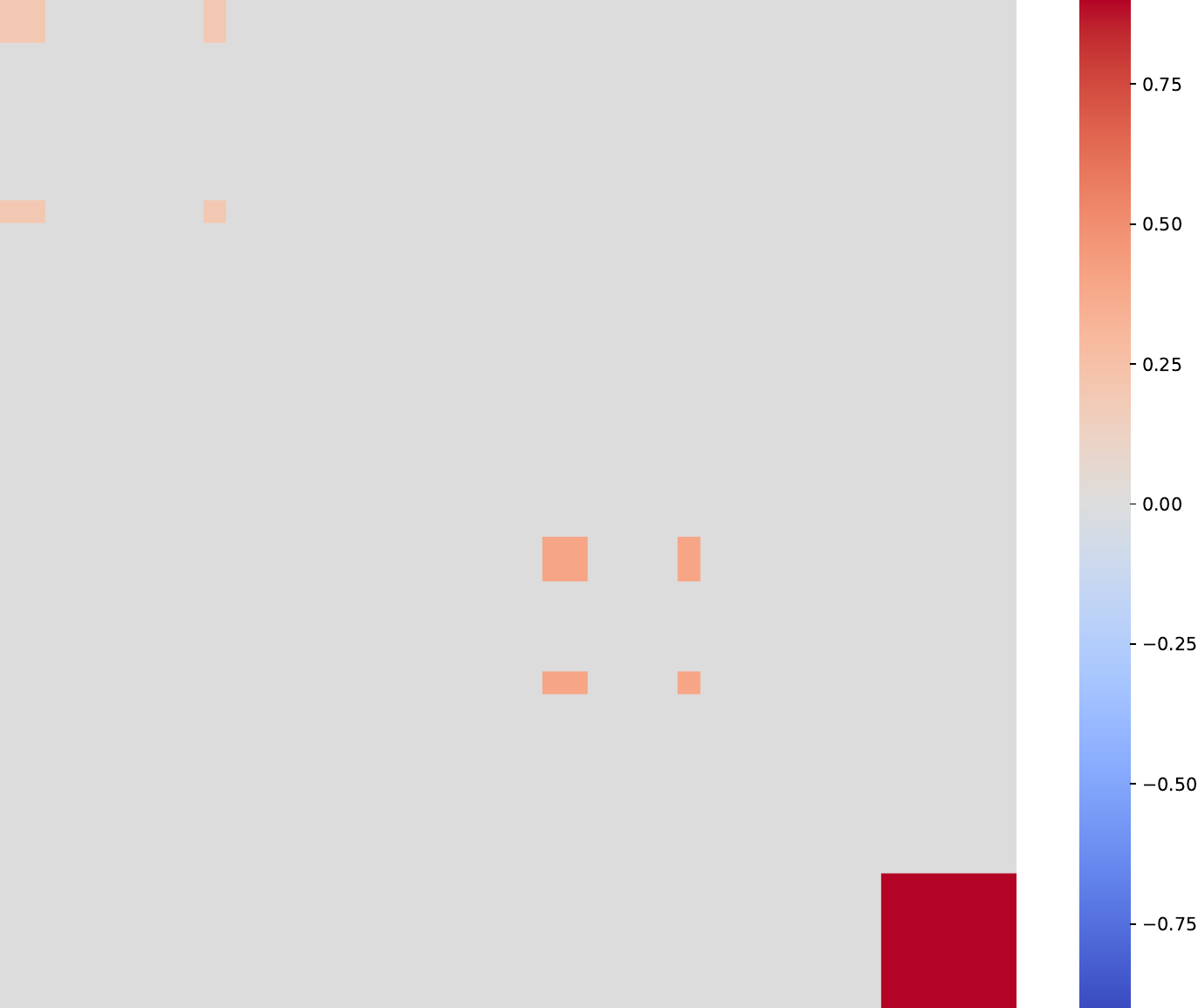}
            & \adjustimage{width=4cm,valign=c}{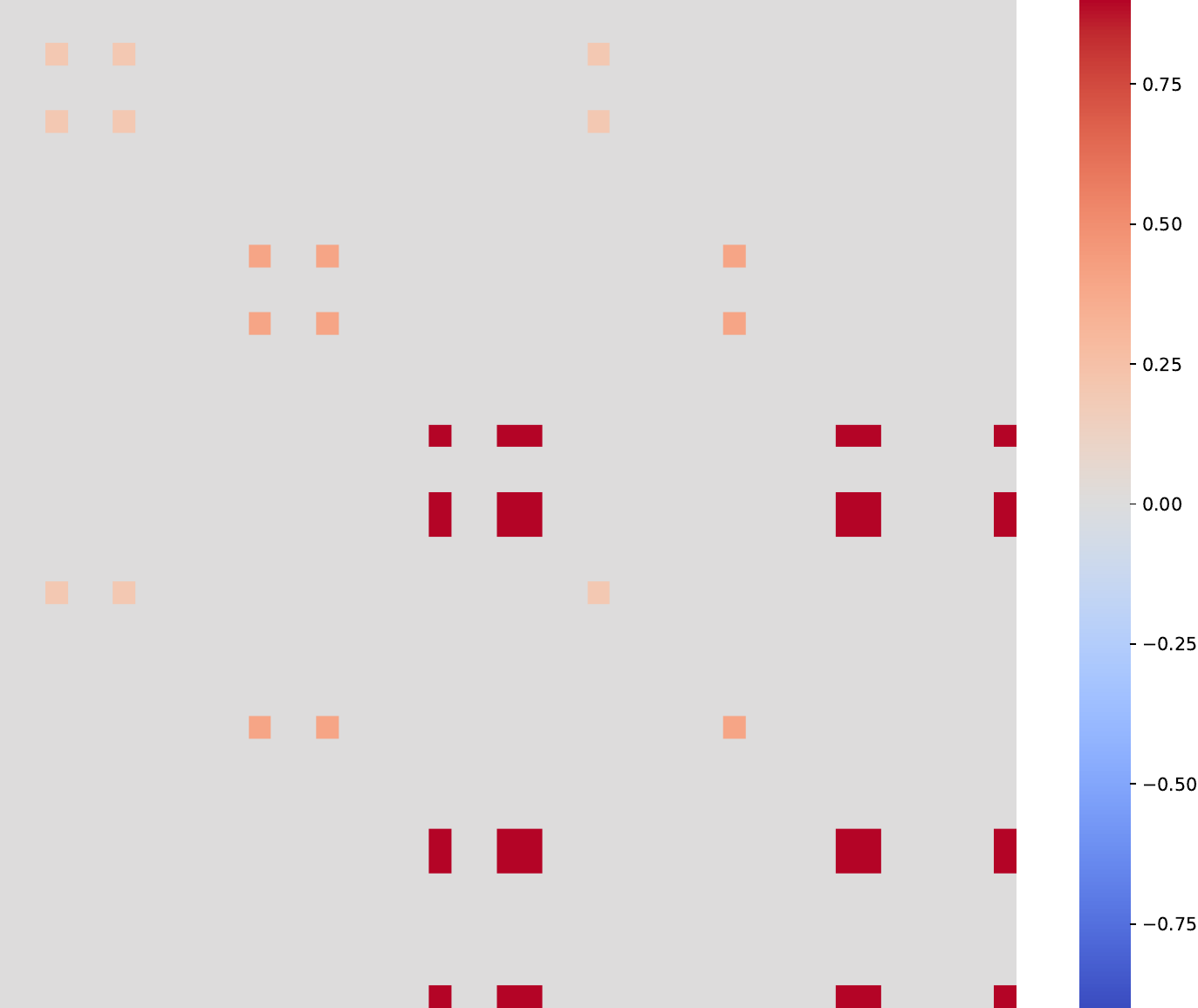}
            & \adjustimage{width=4cm,valign=c}{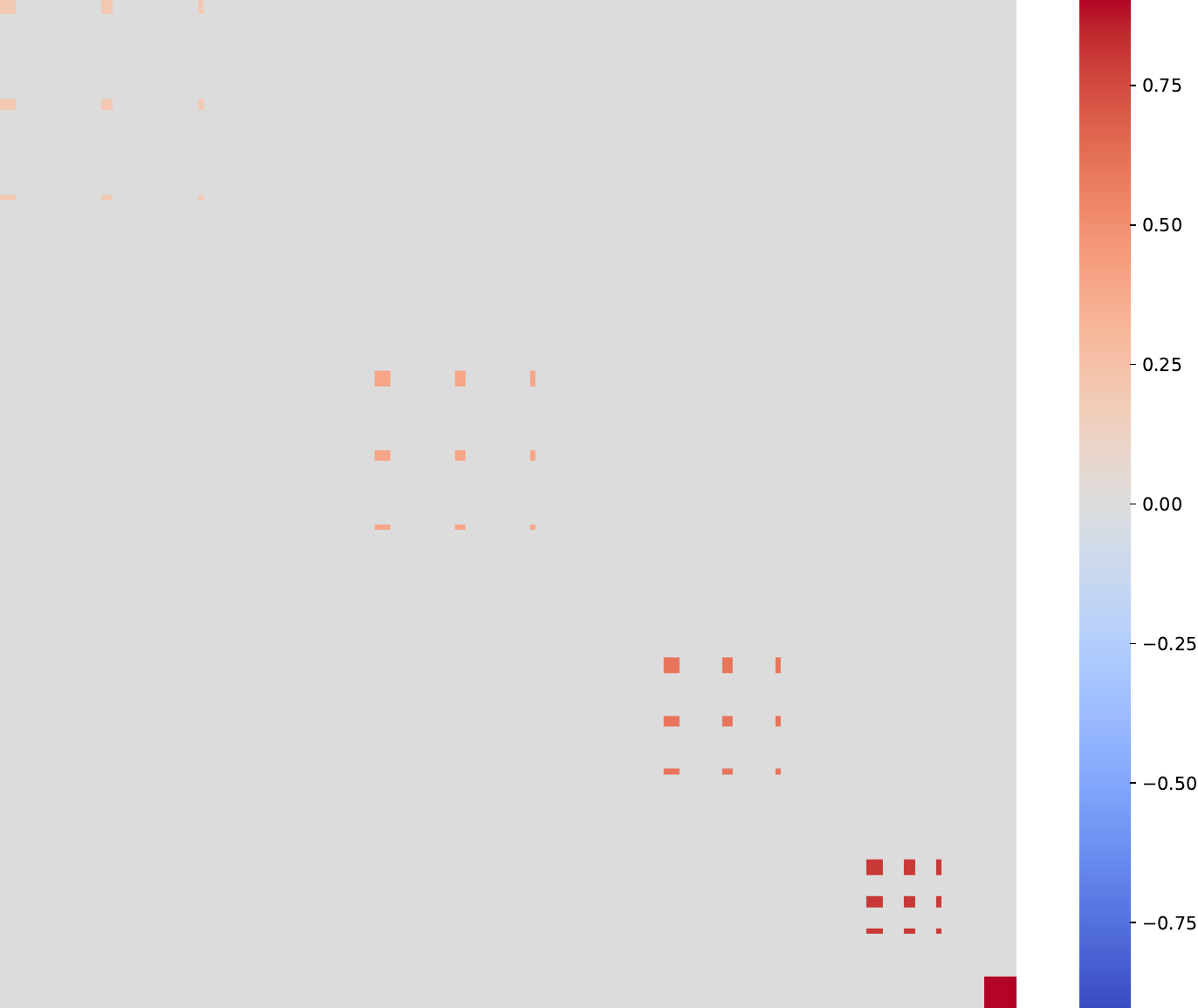}
            & \adjustimage{width=4cm,valign=c}{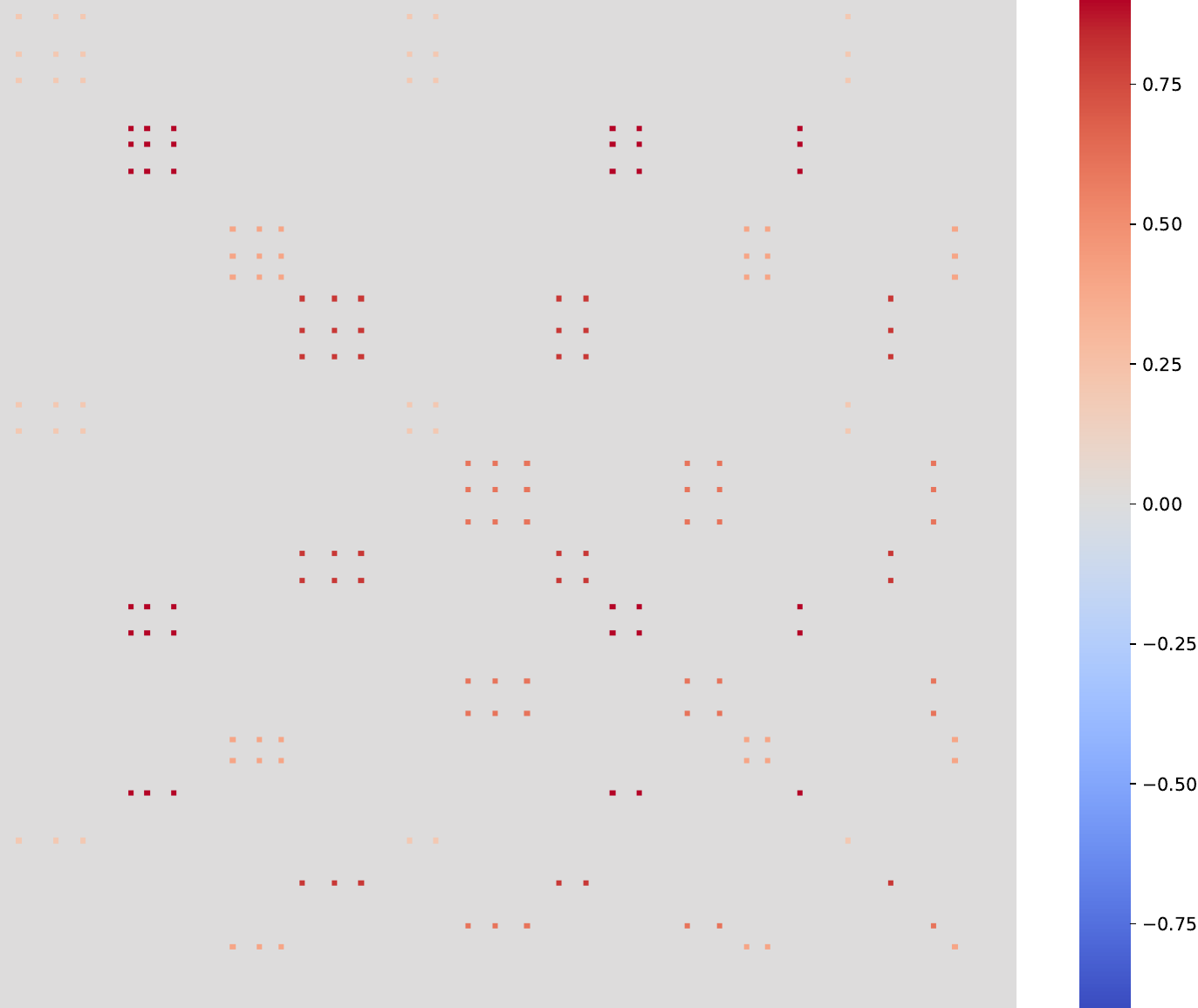} \\[1ex]
            
            \rowtitle{$\hat{\mathbf{\Lambda}}$} 
            & \adjustimage{width=4cm,valign=c}{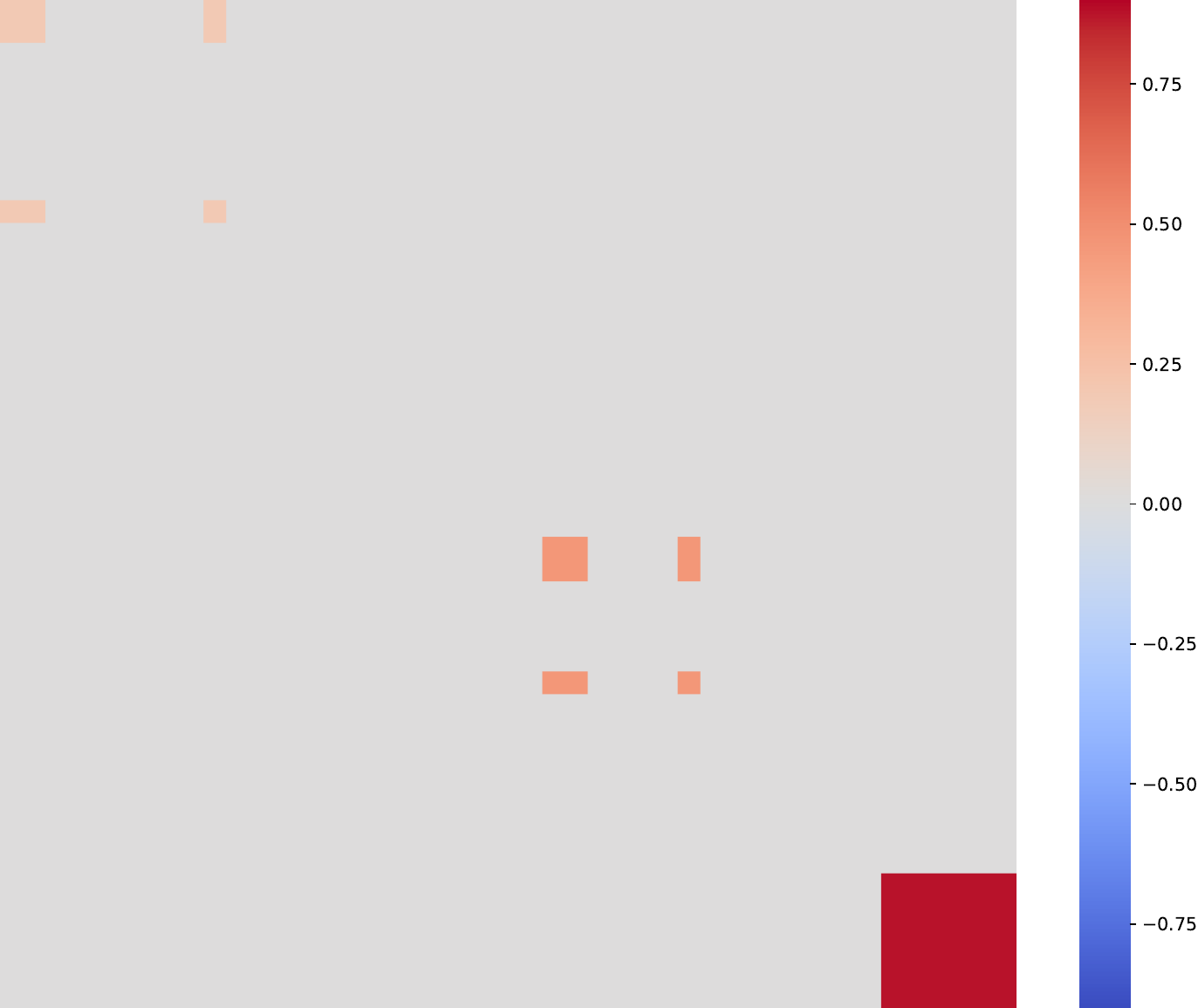}
            & \adjustimage{width=4cm,valign=c}{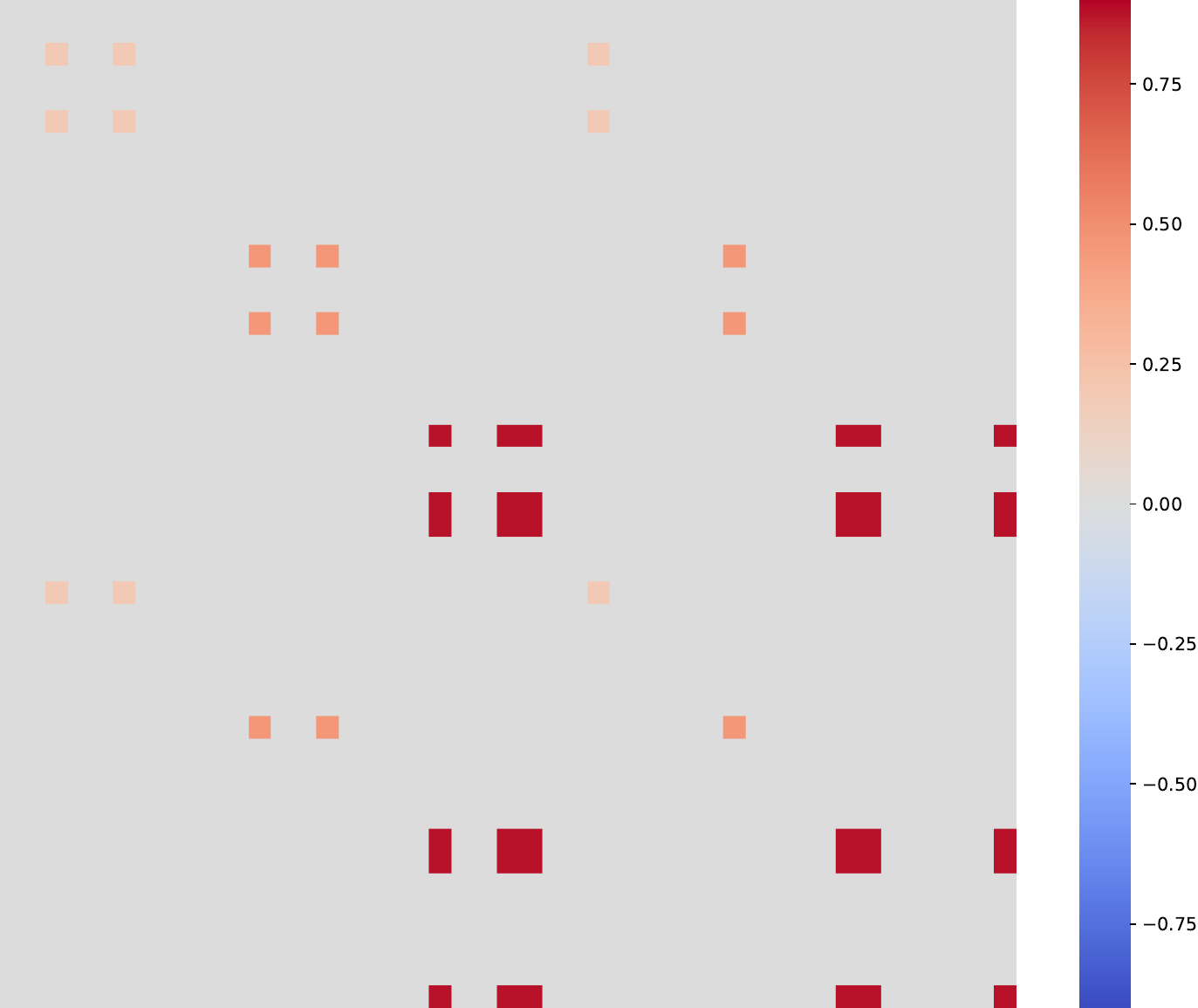}
            & \adjustimage{width=4cm,valign=c}{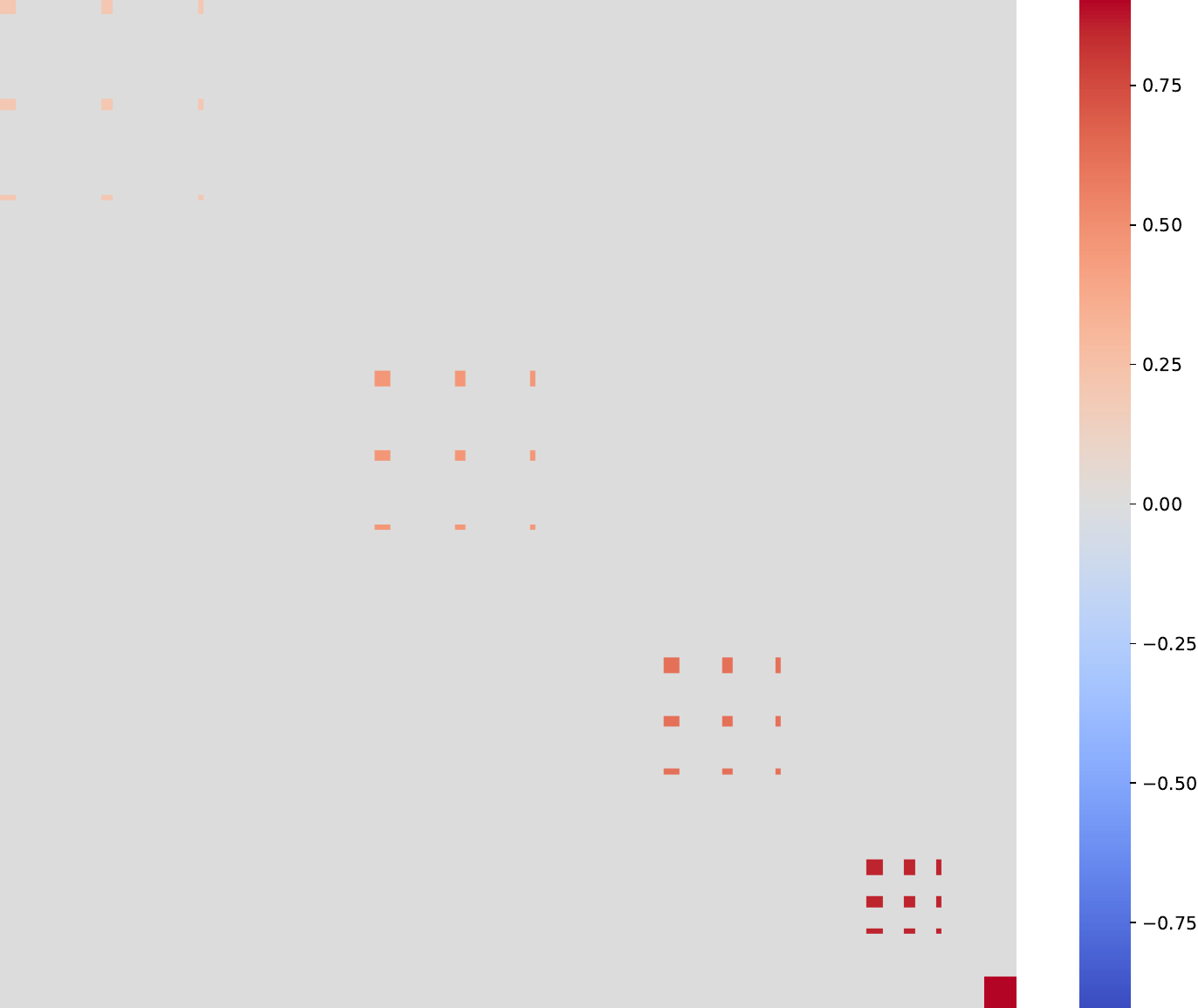}
            & \adjustimage{width=4cm,valign=c}{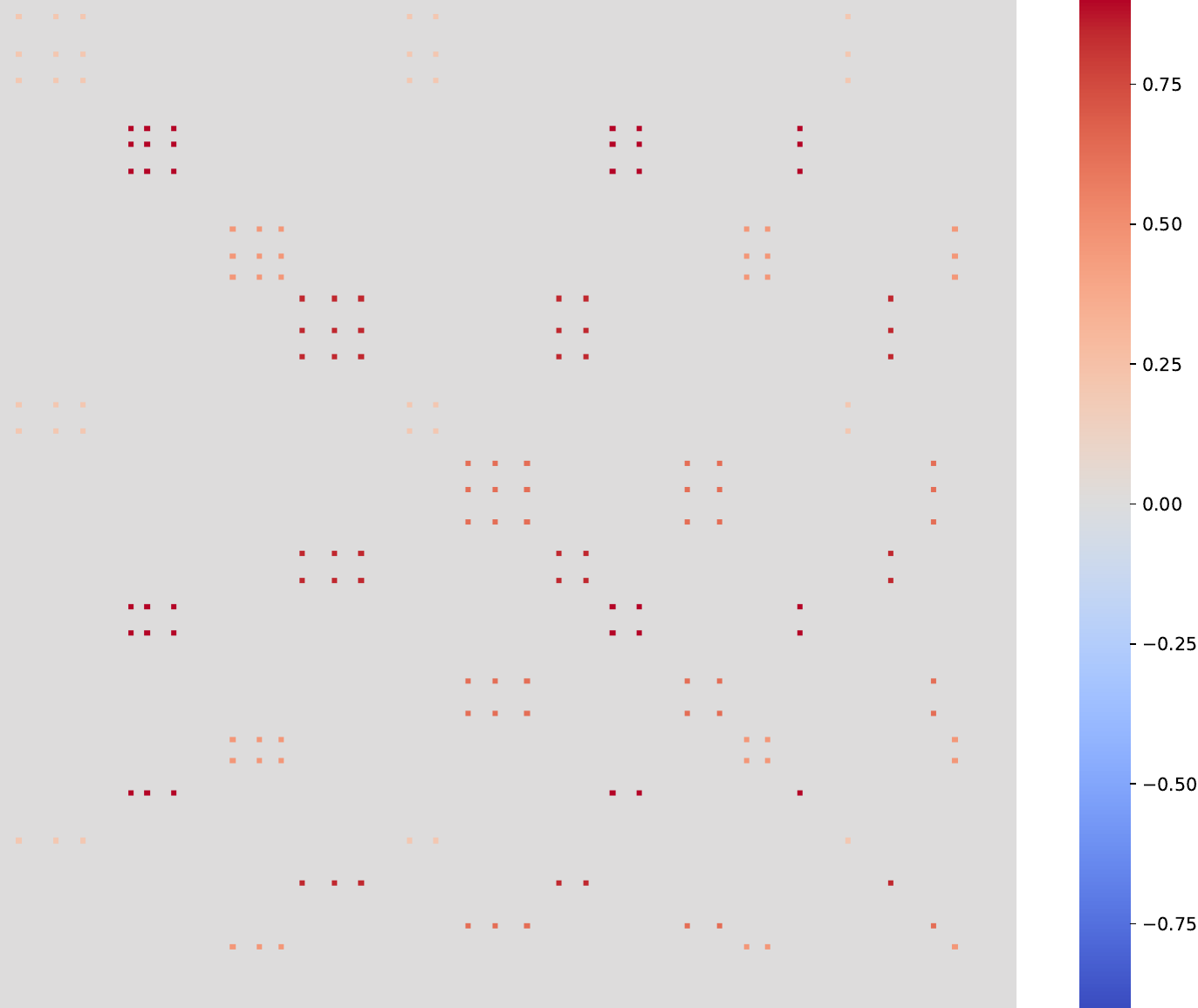} \\[1ex]
            
            \rowtitle{$\mathbf{\Lambda}-\hat{\mathbf{\Lambda}}$} 
            & \adjustimage{width=4cm,valign=c}{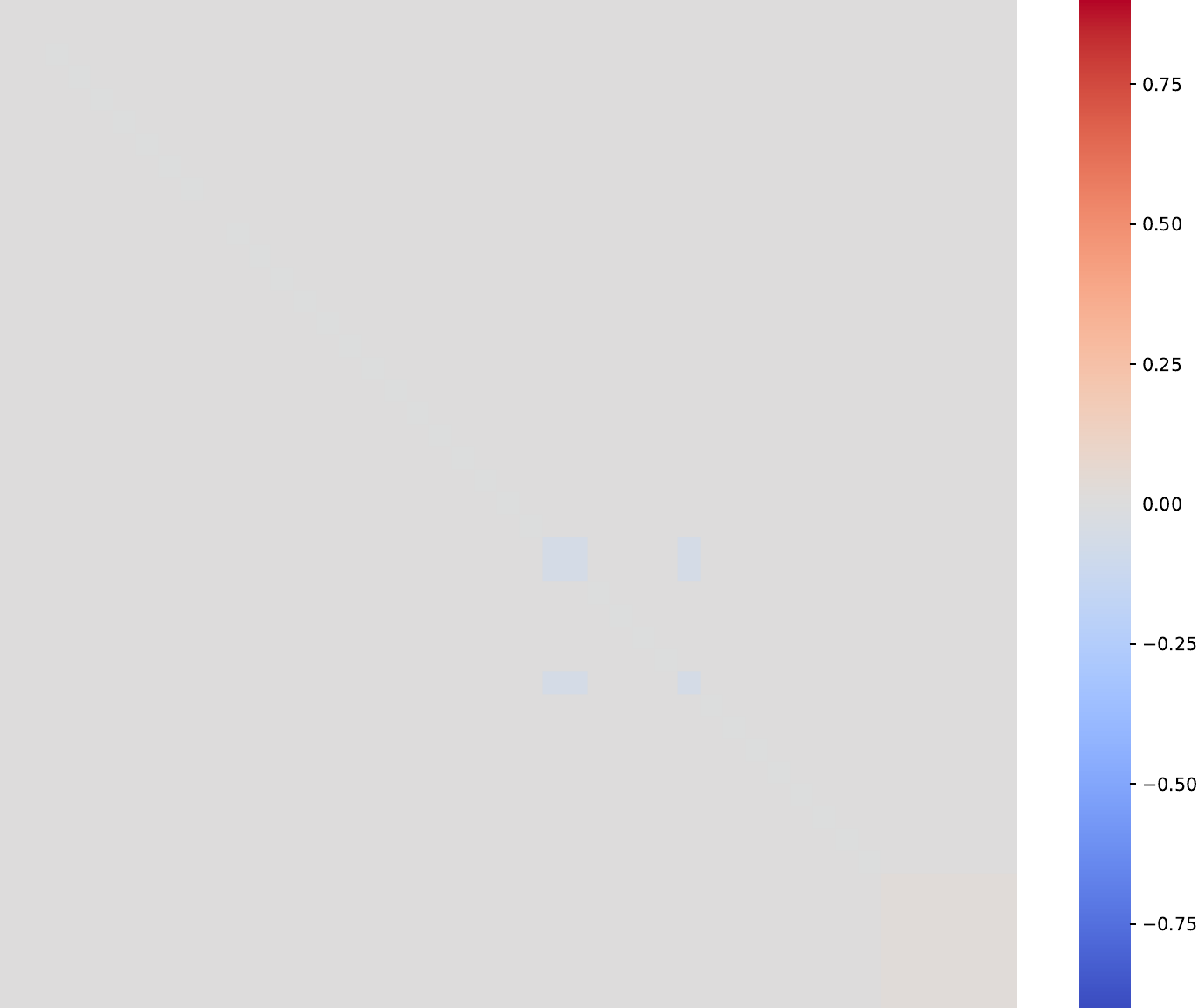}
            & \adjustimage{width=4cm,valign=c}{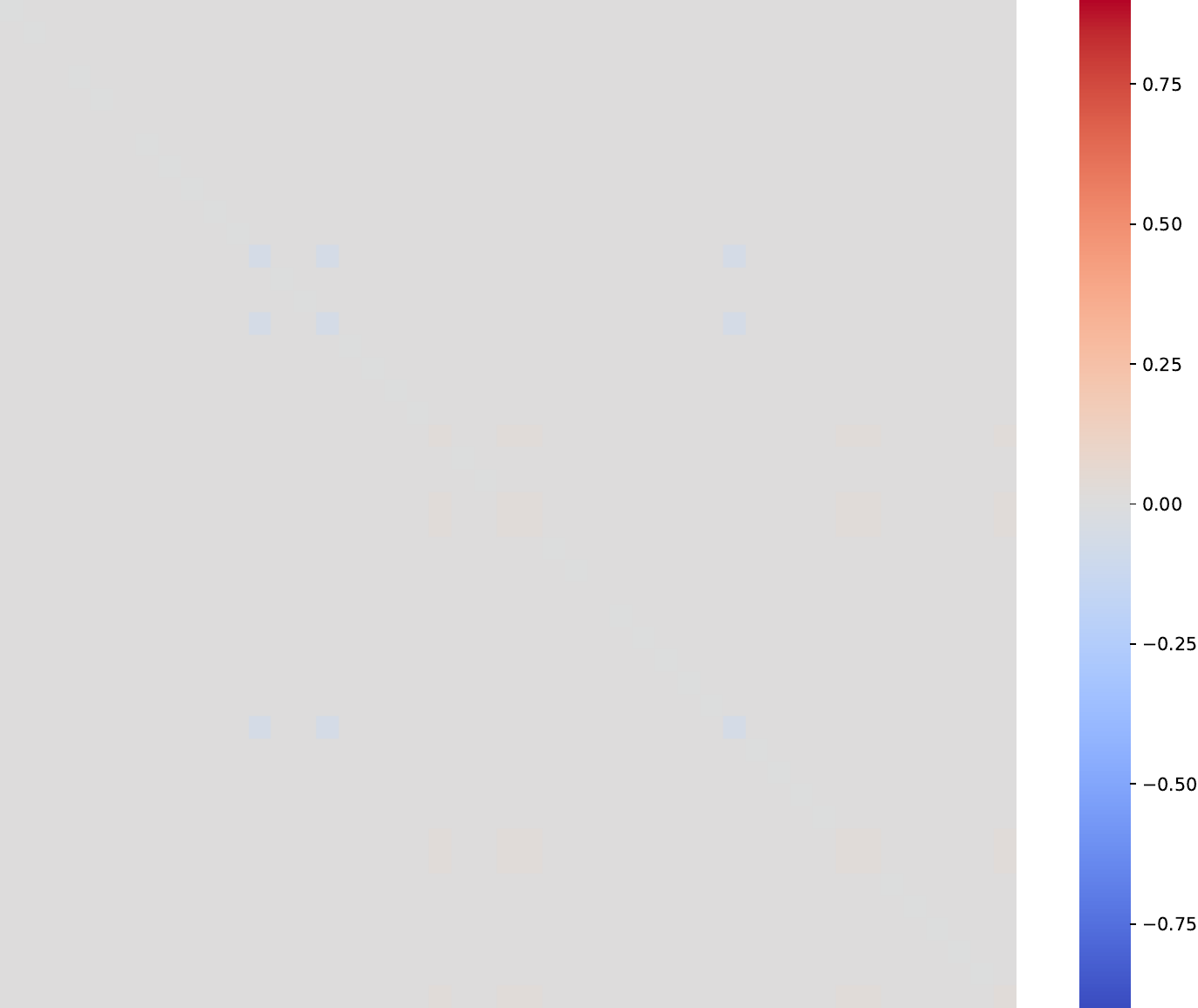}
            & \adjustimage{width=4cm,valign=c}{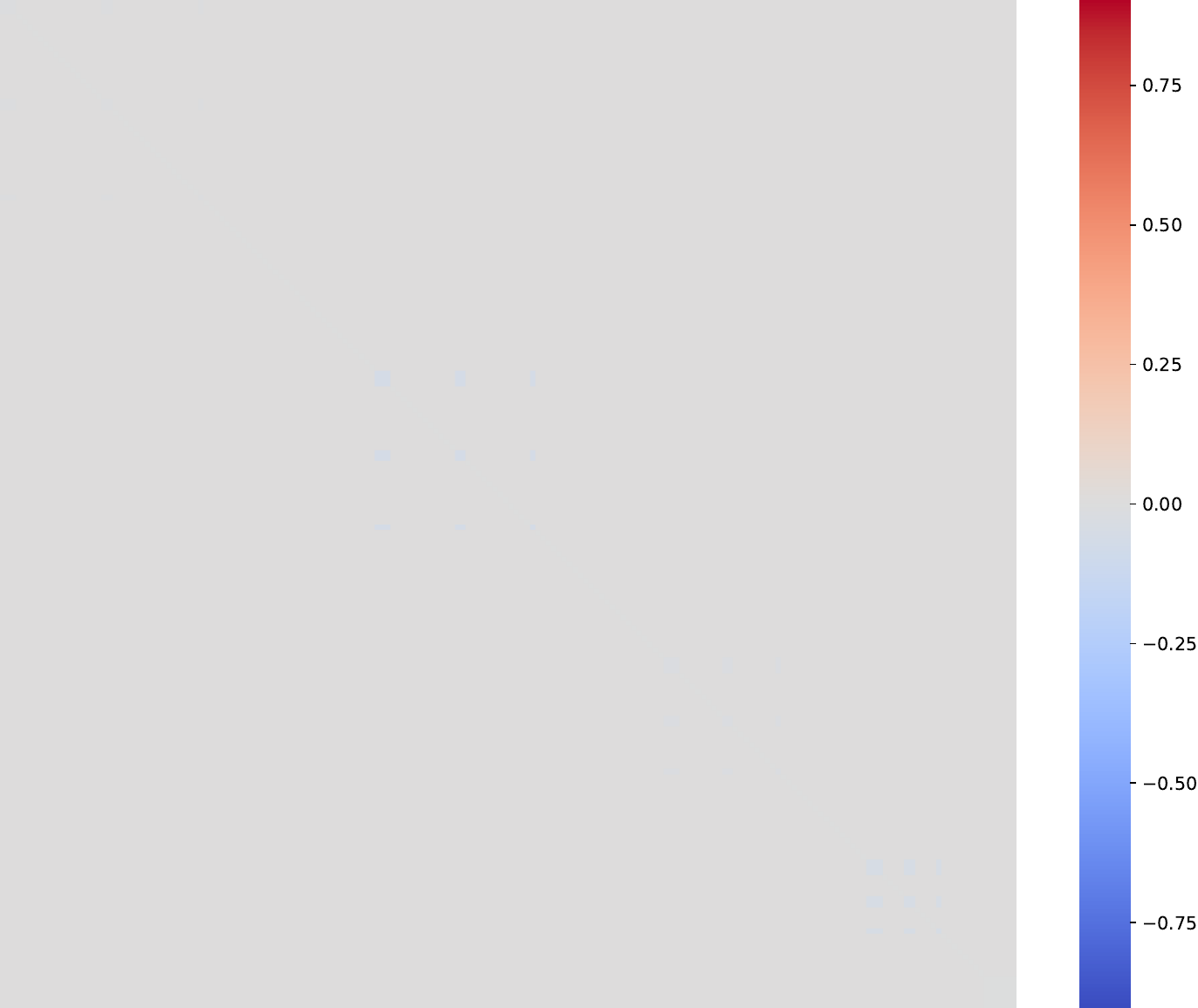}
            & \adjustimage{width=4cm,valign=c}{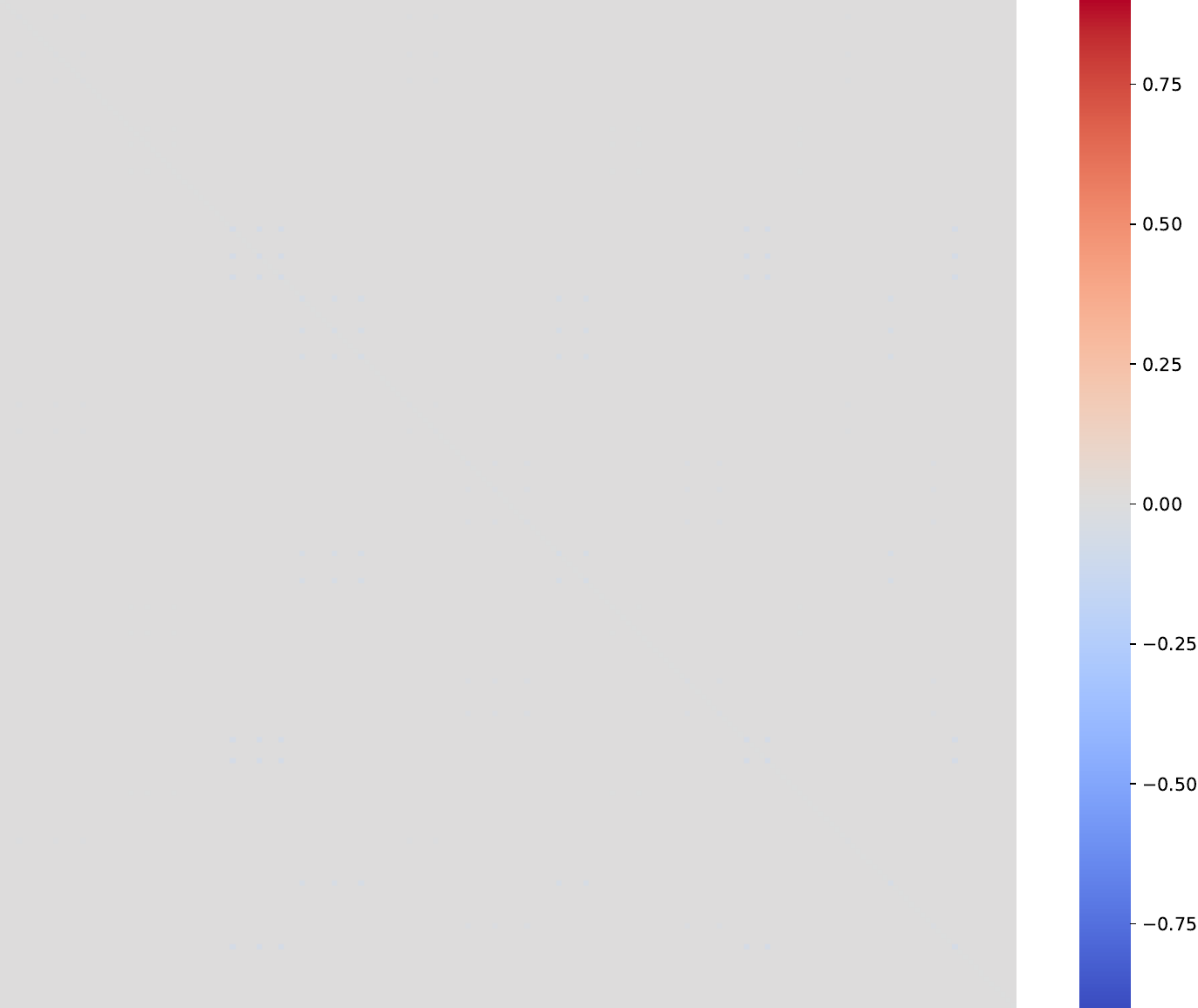} \\
        \end{tabular}%
    }
    \caption{Visualization of model estimation and convergence for different network topologies. Rows show: (1) trace plots for convergence diagnostics; (2) true connectivity matrices ($\mathbf{\Lambda}$); (3) estimated covariance matrices ($\hat{\mathbf{\Lambda}}$); (4) estimation error matrices ($\mathbf{\Lambda} - \hat{\mathbf{\Lambda}}$). Columns (a-d) correspond to different network configurations ($V=10,20$) and topologies (S1, S2).}
    \label{fig:brain_analysis}
\end{figure*}

%===========================================================
% 
%===========================================================
\begin{figure*}[tp!]
    \centering
    \resizebox{\textwidth}{!}{%
        \setlength{\tabcolsep}{1pt}
        \newcommand{\subfiglabel}[1]{\textbf{(#1)}}
        \newcommand{\rowtitle}[1]{\multicolumn{1}{m{4em}}{\raggedleft\bfseries #1}}
        
        \begin{tabular}{@{}m{4em}cccc@{}}
            
            & \subfiglabel{a} & \subfiglabel{b} & \subfiglabel{c} & \subfiglabel{d} \\[0.5ex]

            \rowtitle{Trace plots} 
            & \adjustimage{width=4cm,valign=c}{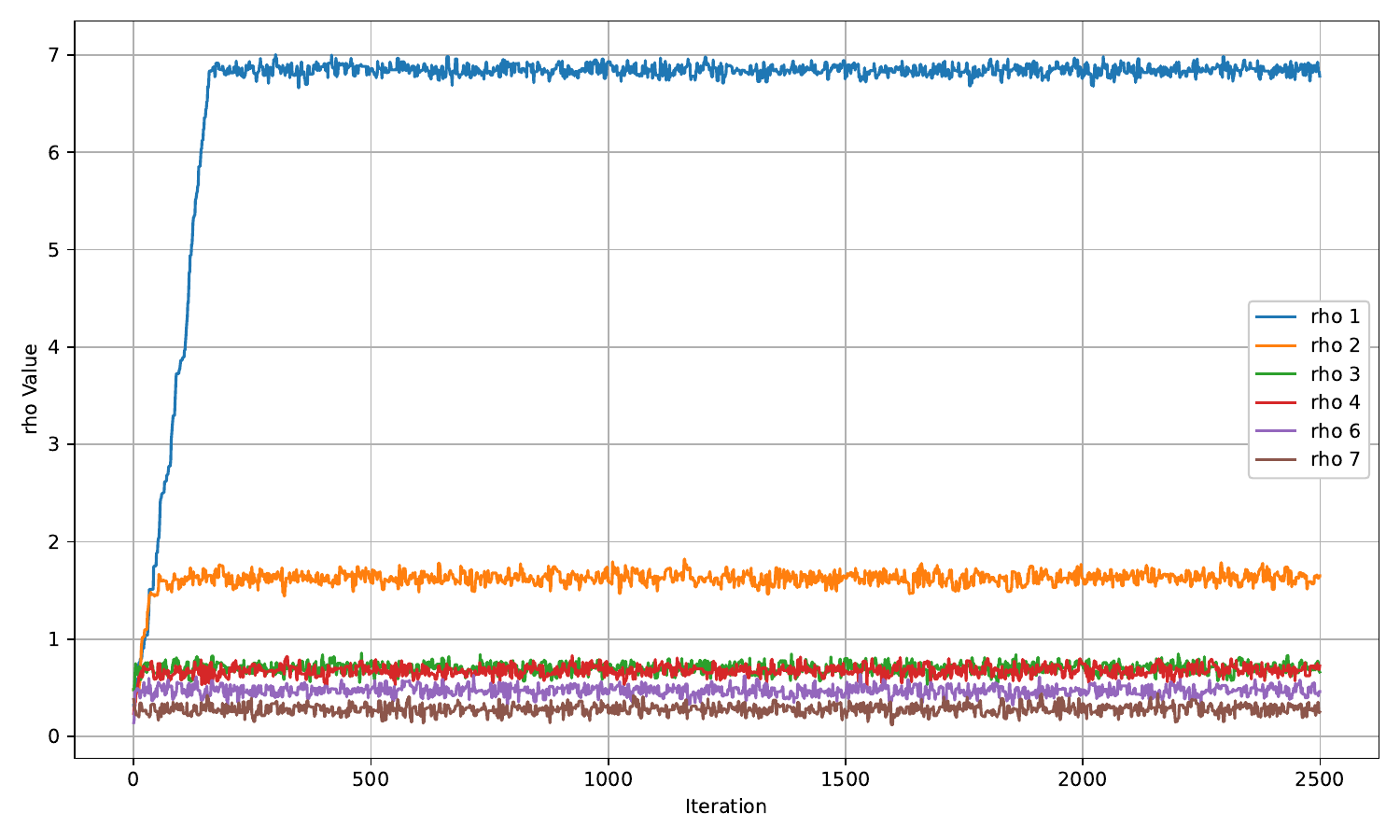}
            & \adjustimage{width=4cm,valign=c}{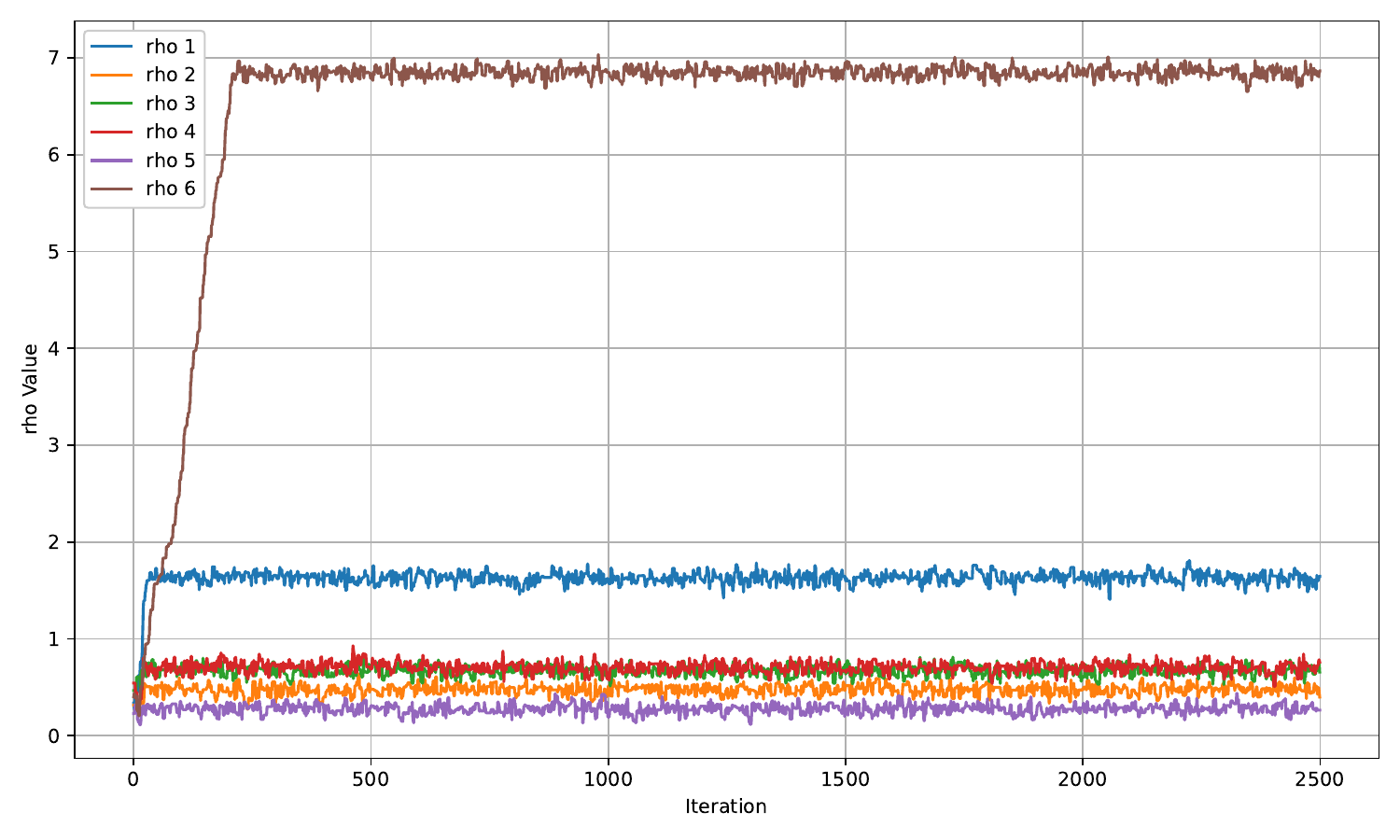}
            & \adjustimage{width=4cm,valign=c}{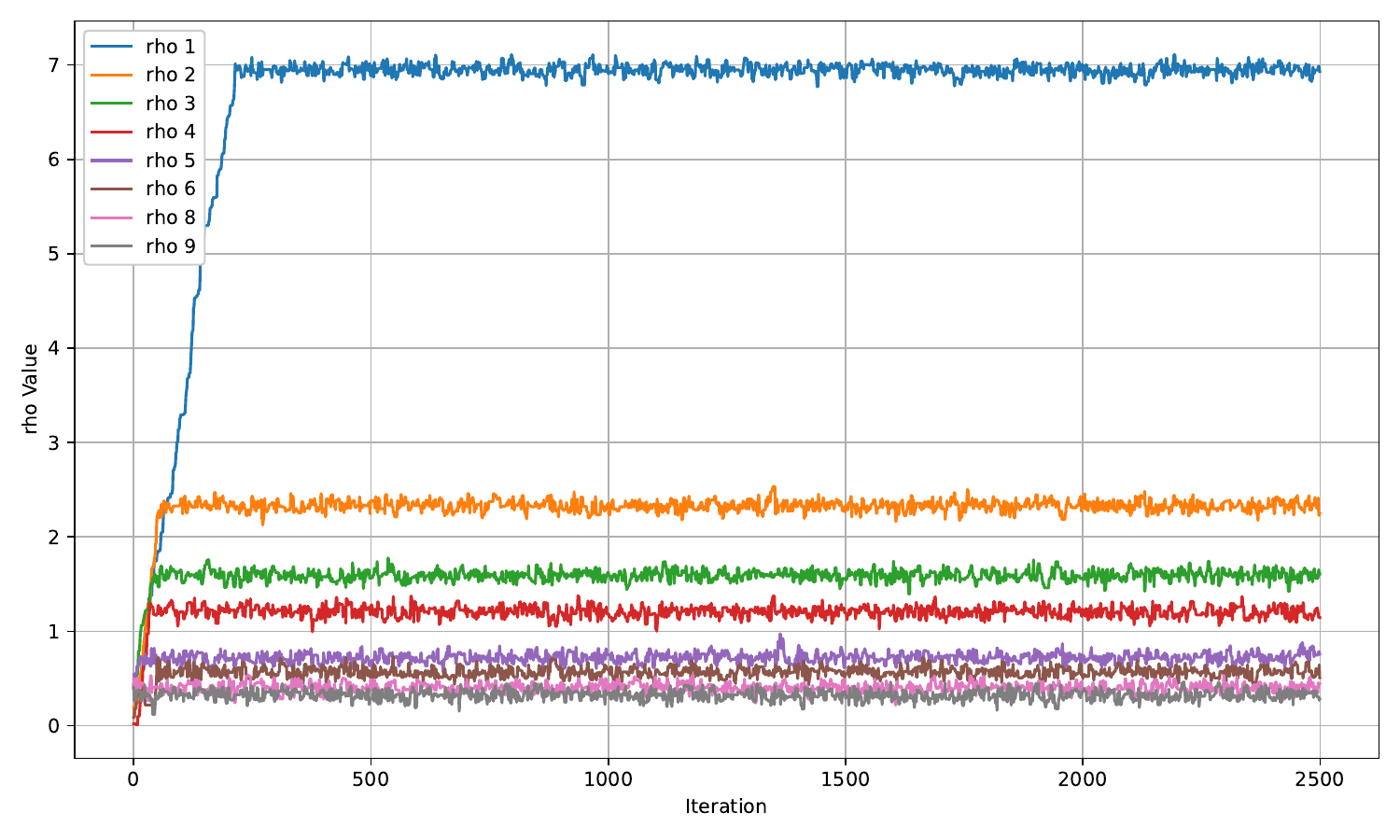}
            & \adjustimage{width=4cm,valign=c}{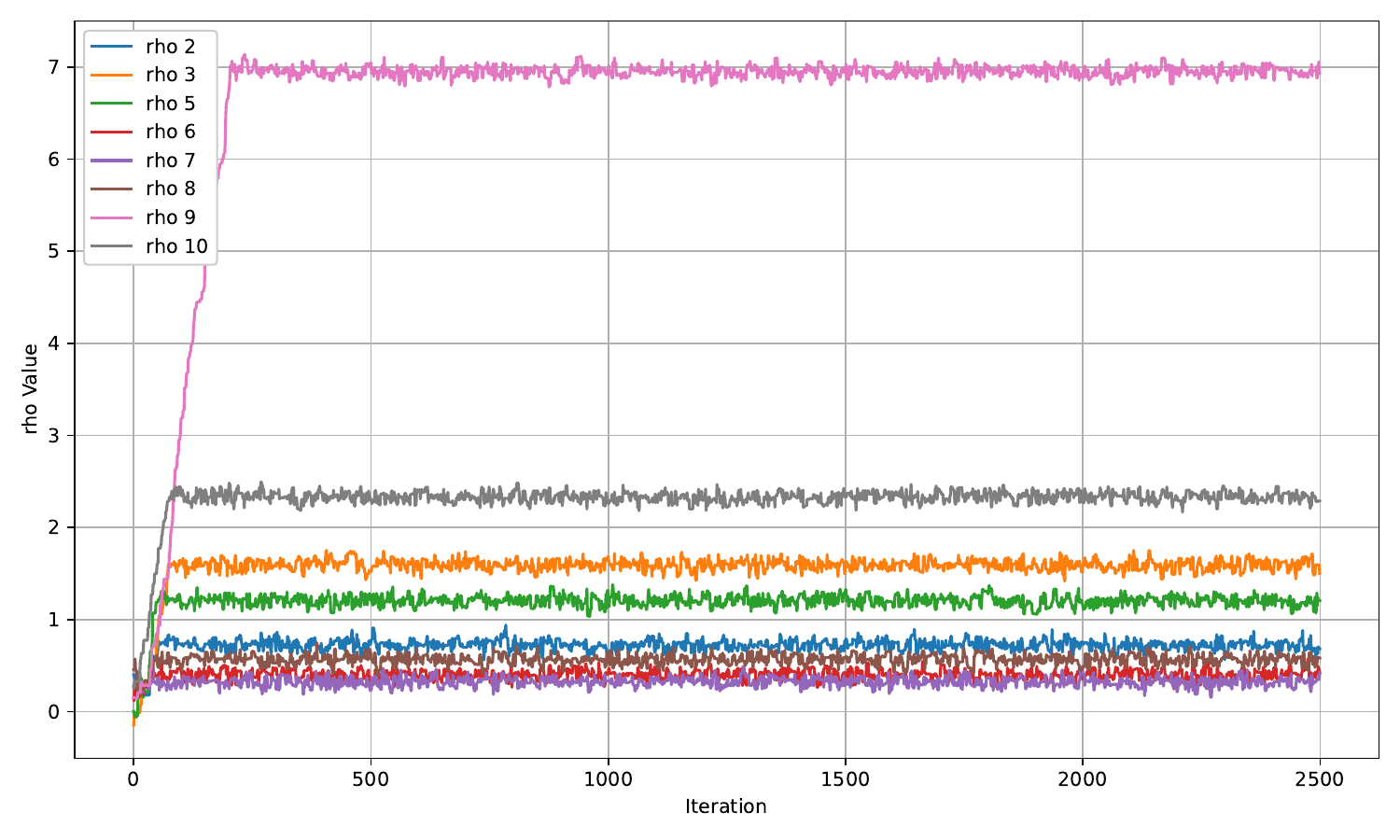} \\[1ex]
            
            \rowtitle{$\mathbf{\Lambda}$} 
            & \adjustimage{width=4cm,valign=c}{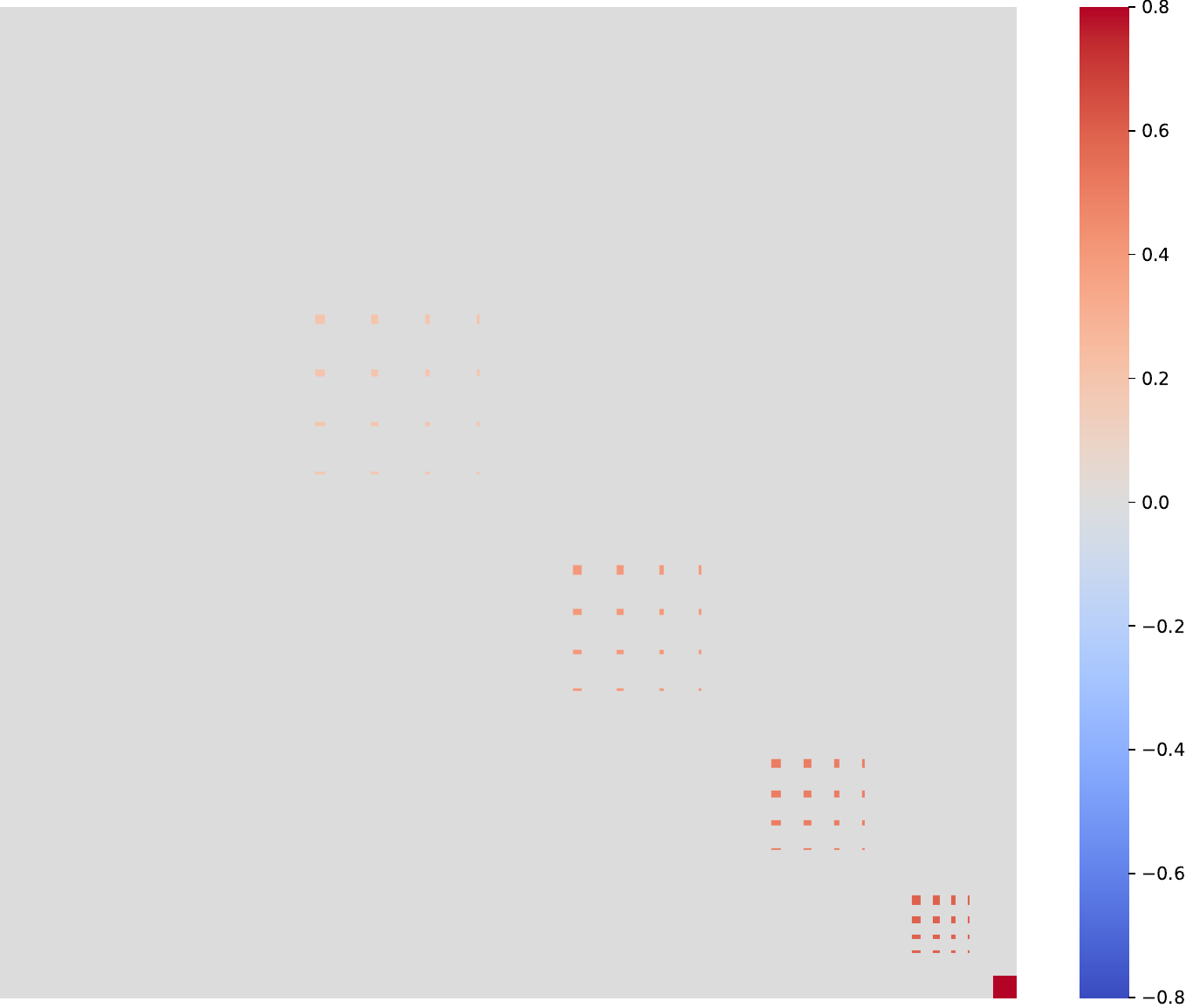}
            & \adjustimage{width=4cm,valign=c}{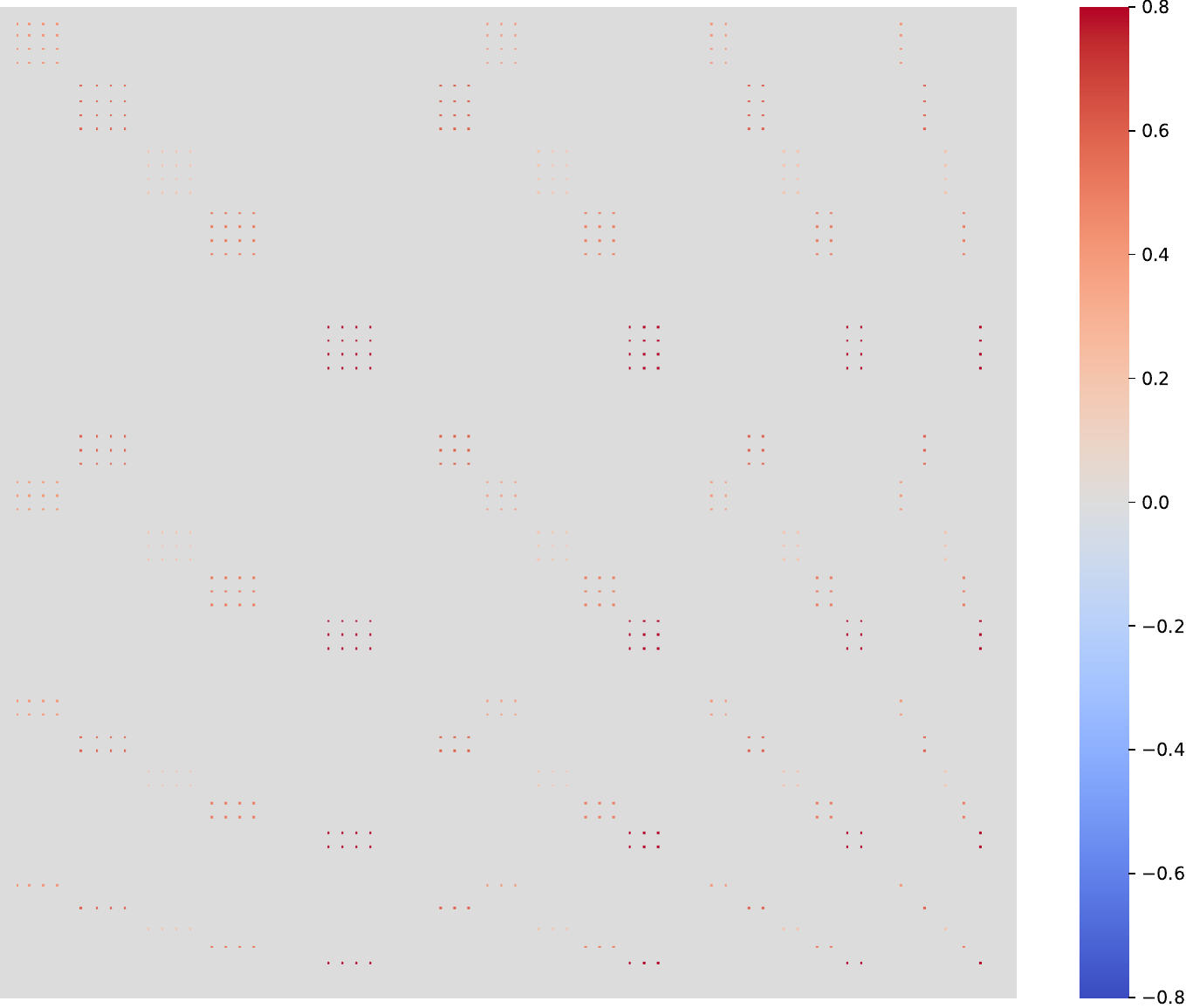}
            & \adjustimage{width=4cm,valign=c}{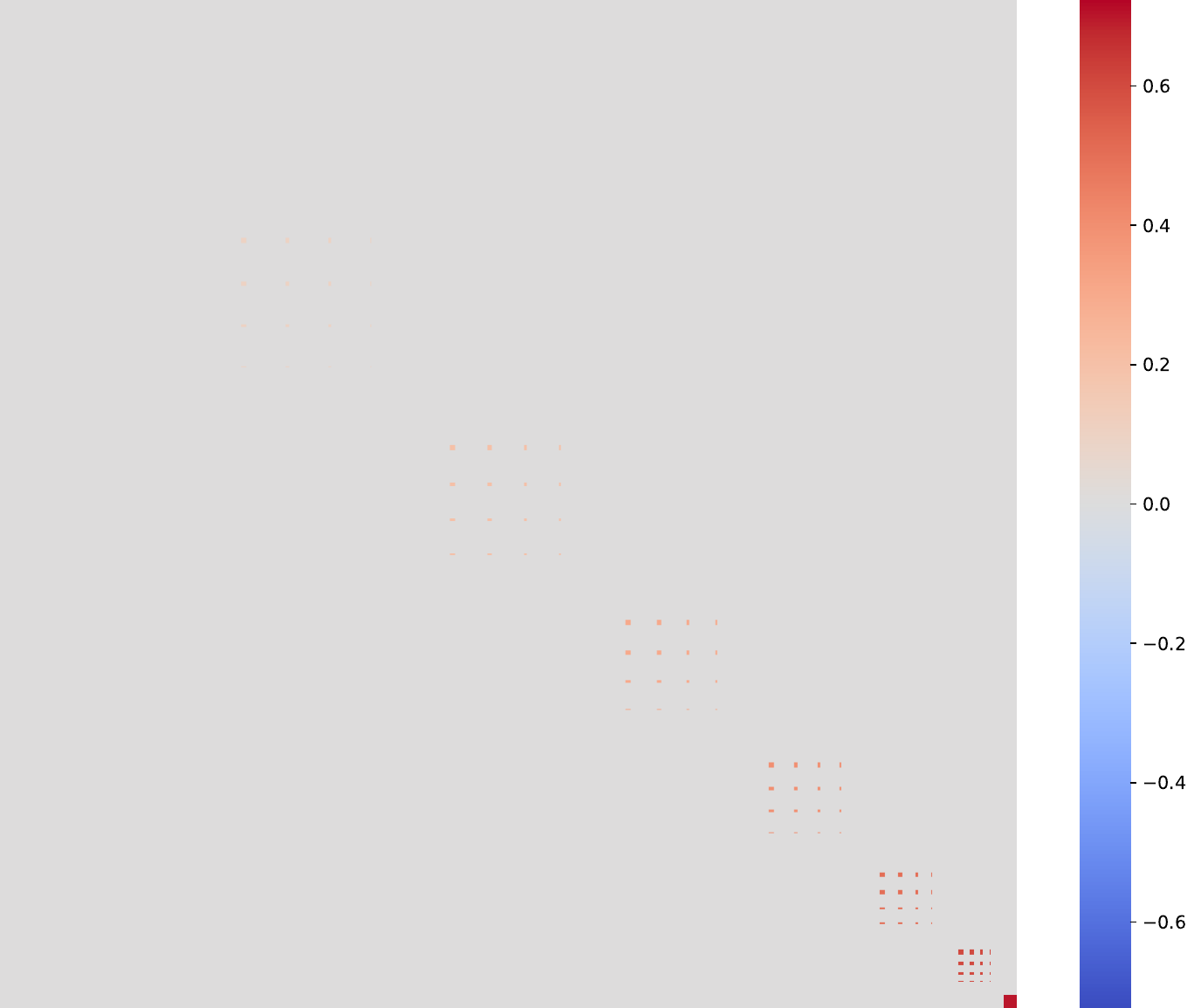}
            & \adjustimage{width=4cm,valign=c}{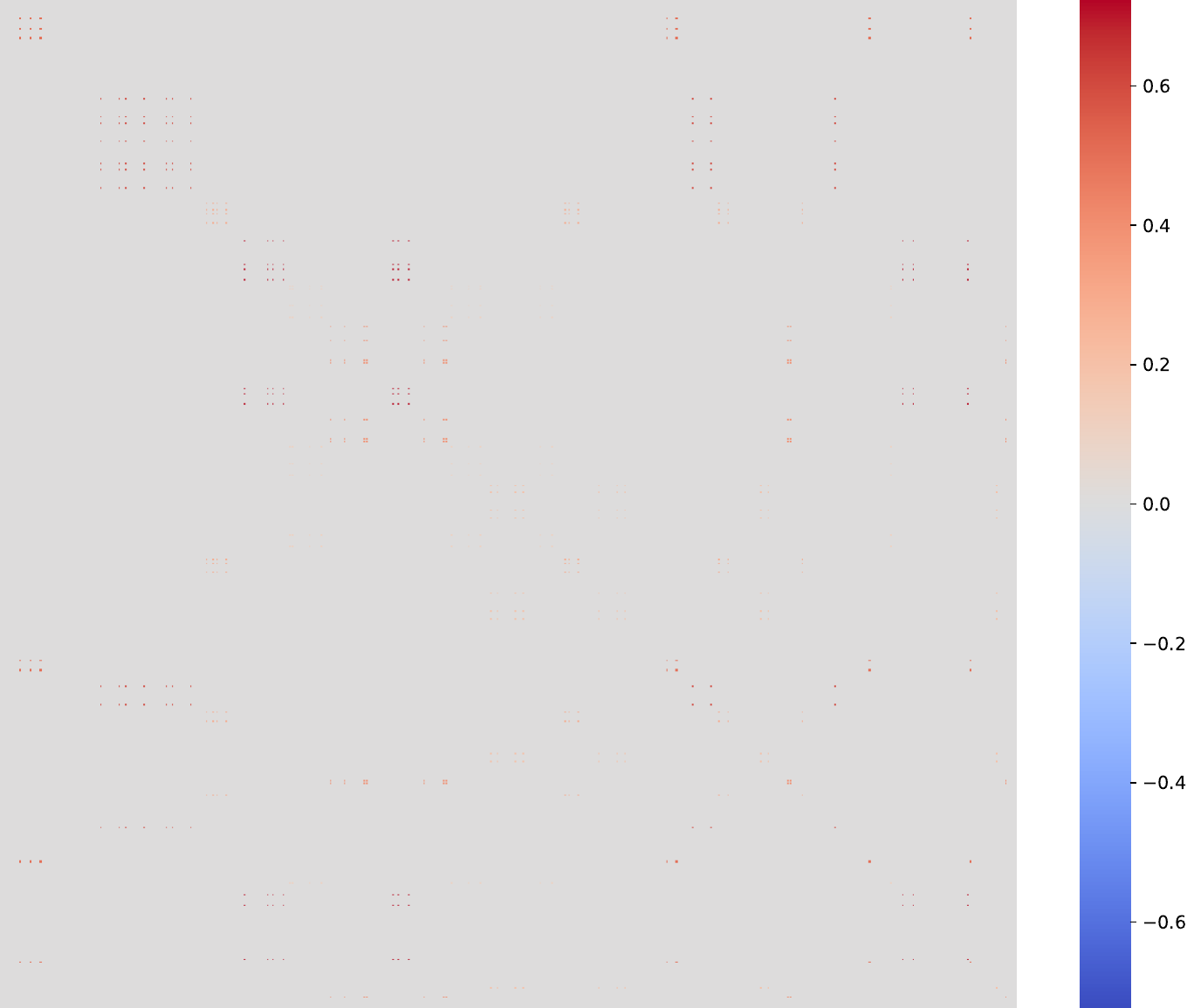} \\[1ex]
            
            \rowtitle{$\hat{\mathbf{\Lambda}}$} 
            & \adjustimage{width=4cm,valign=c}{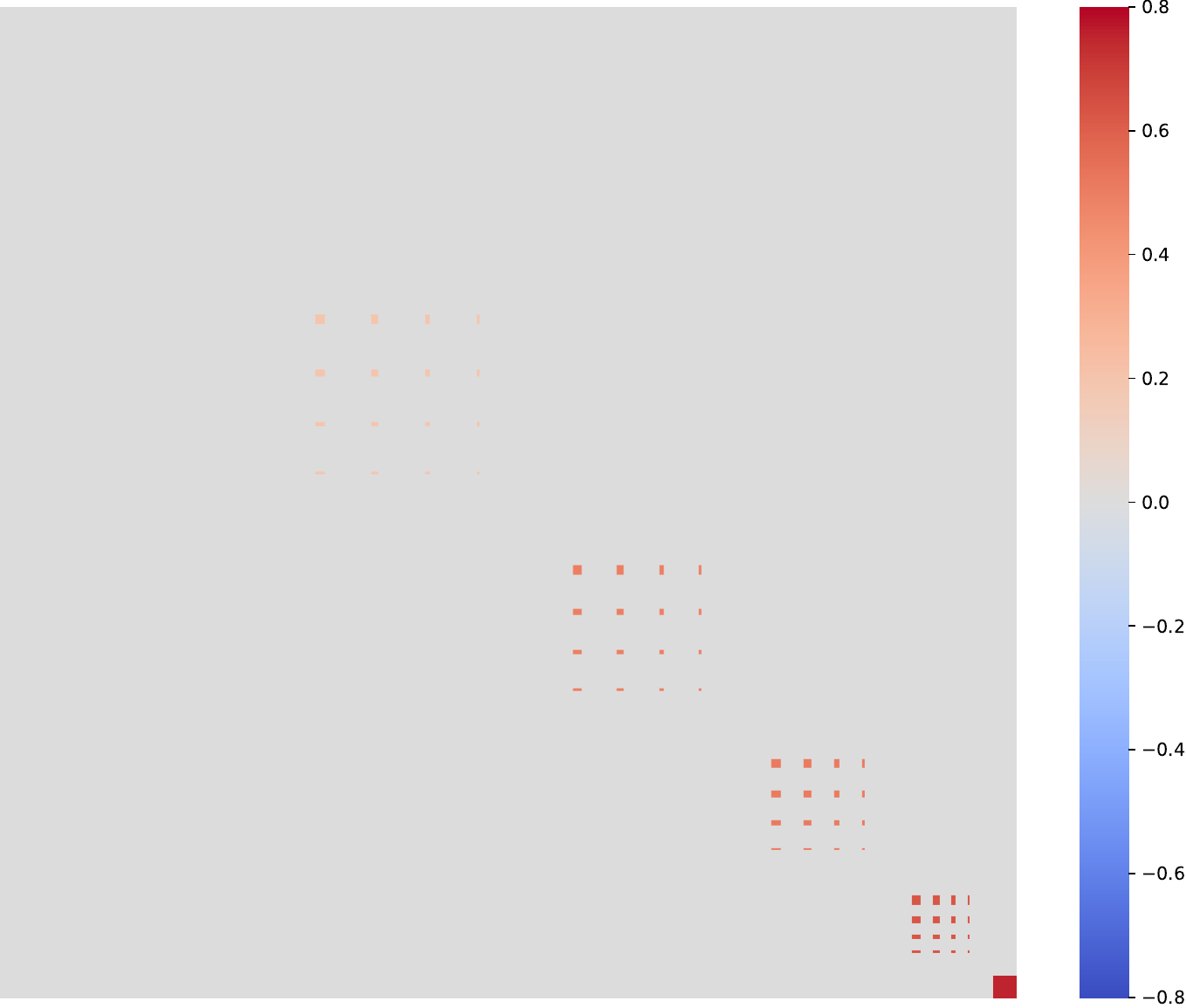}
            & \adjustimage{width=4cm,valign=c}{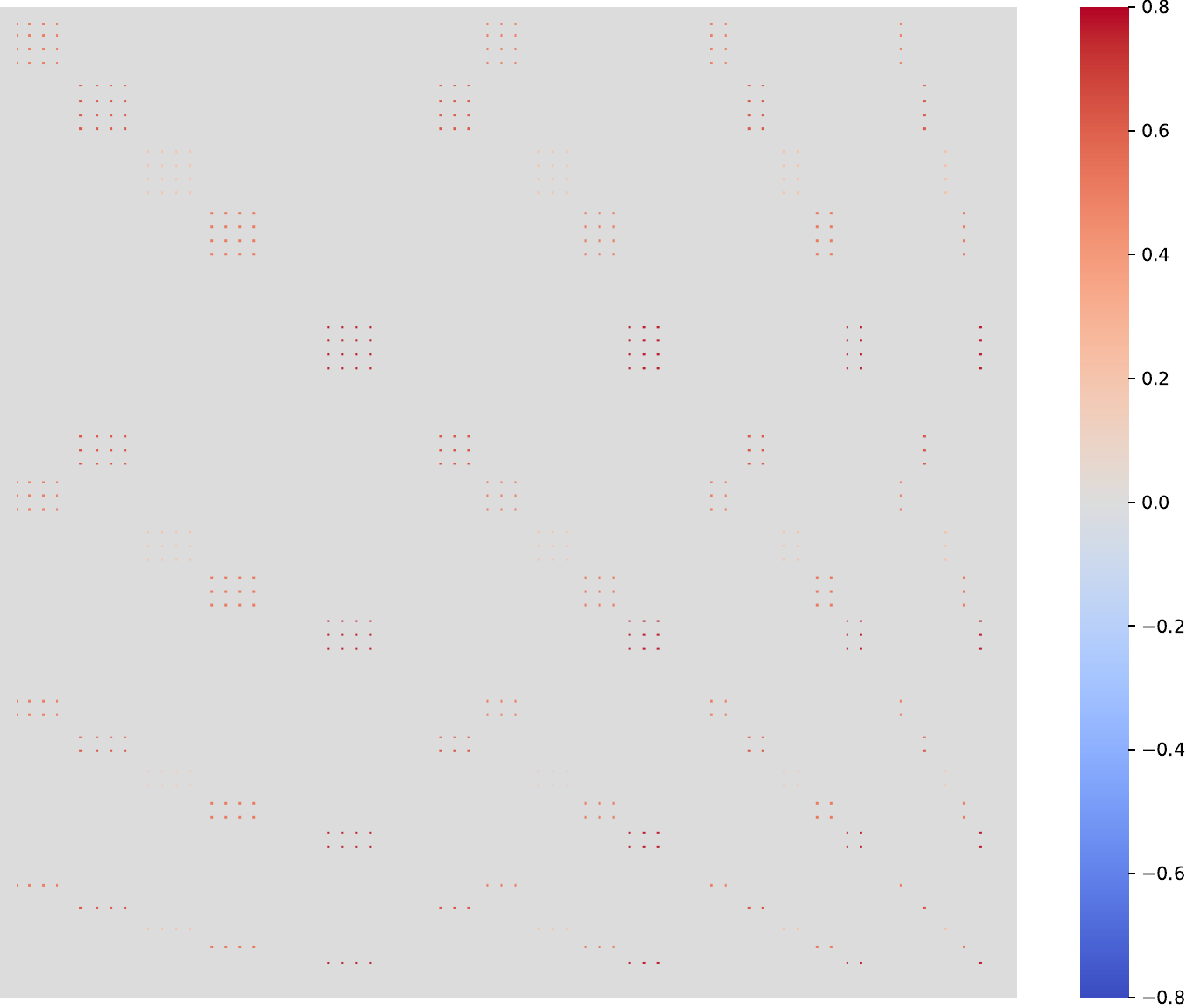}
            & \adjustimage{width=4cm,valign=c}{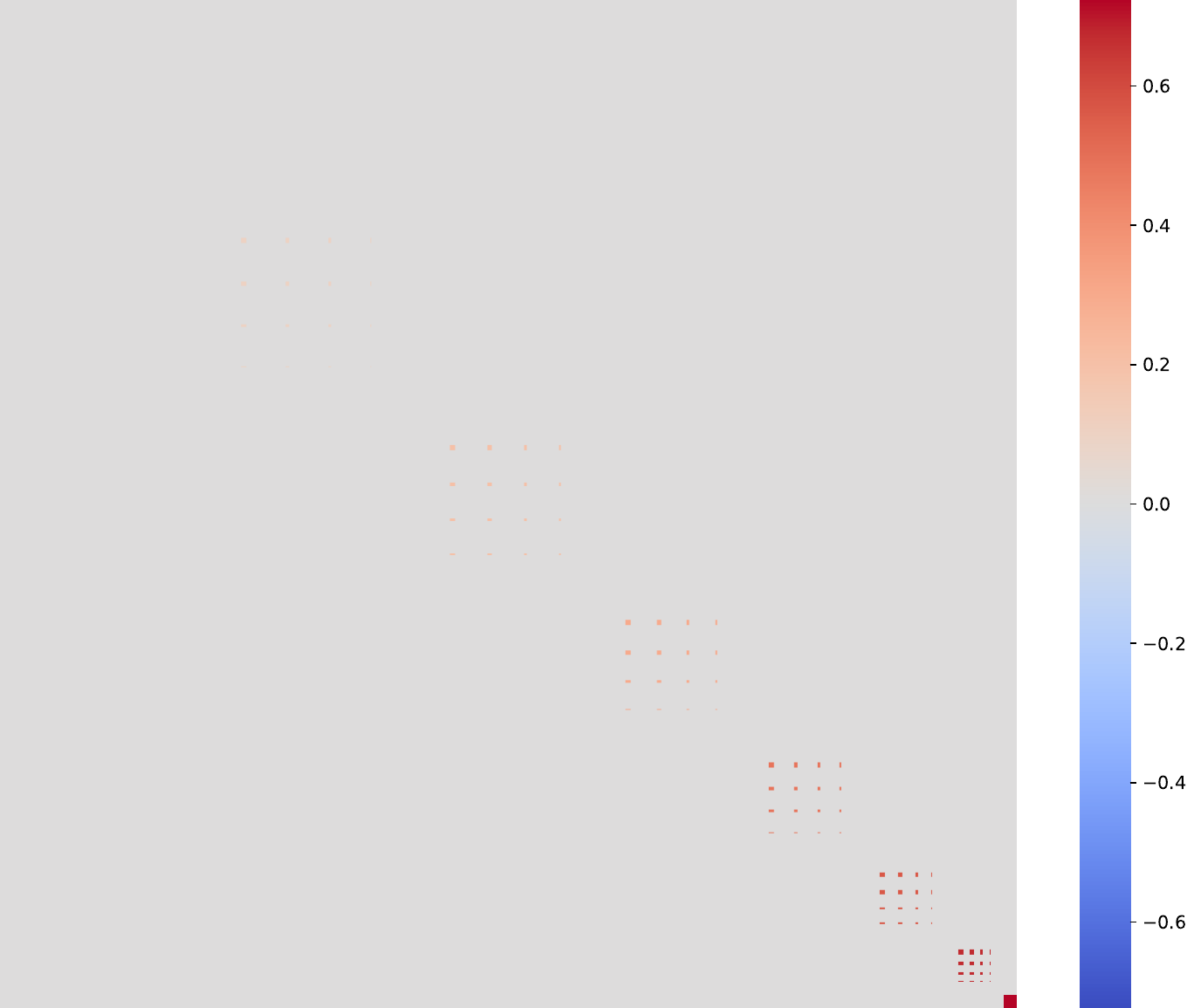}
            & \adjustimage{width=4cm,valign=c}{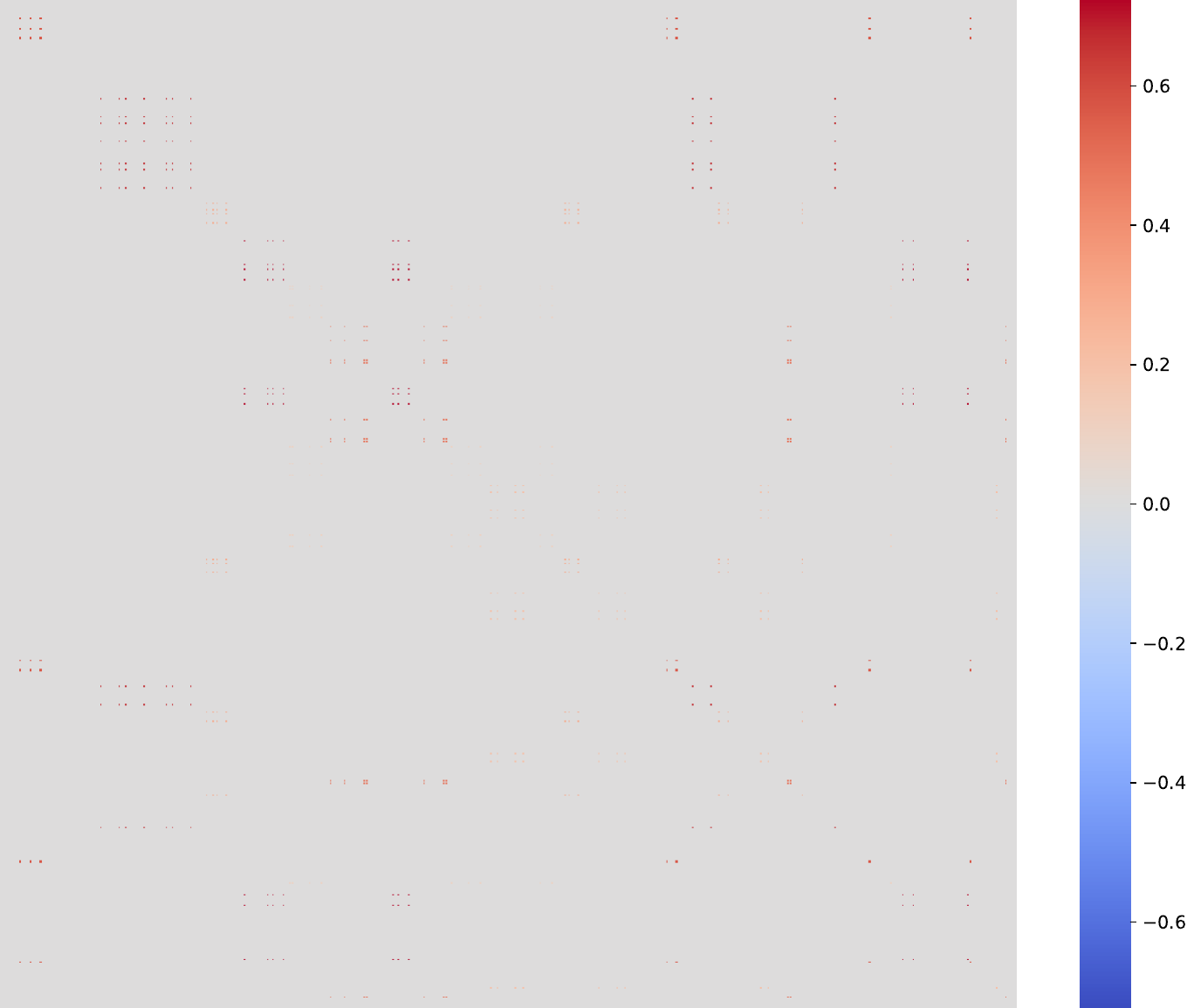} \\[1ex]
            
            \rowtitle{$\mathbf{\Lambda}-\hat{\mathbf{\Lambda}}$} 
            & \adjustimage{width=4cm,valign=c}{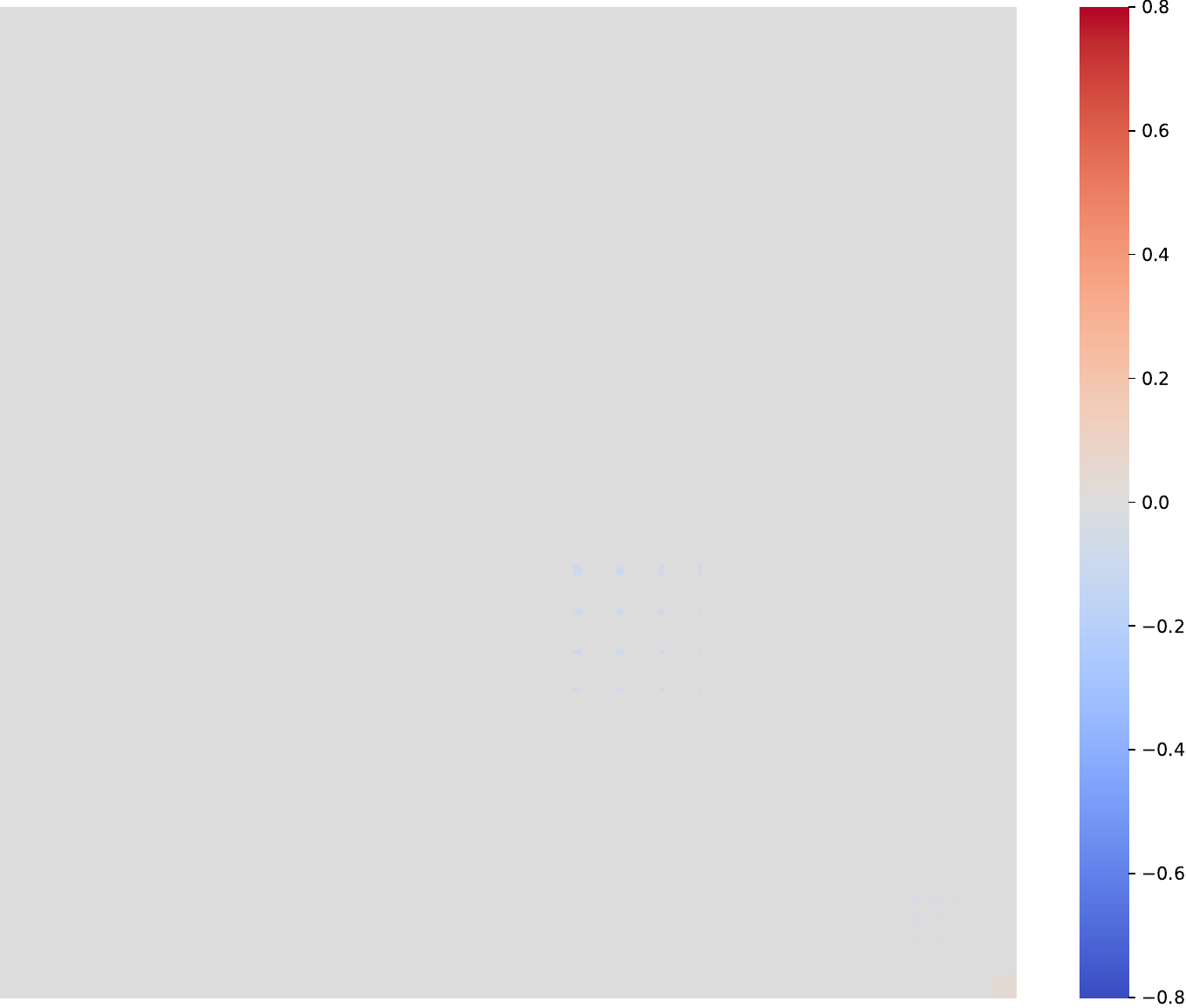}
            & \adjustimage{width=4cm,valign=c}{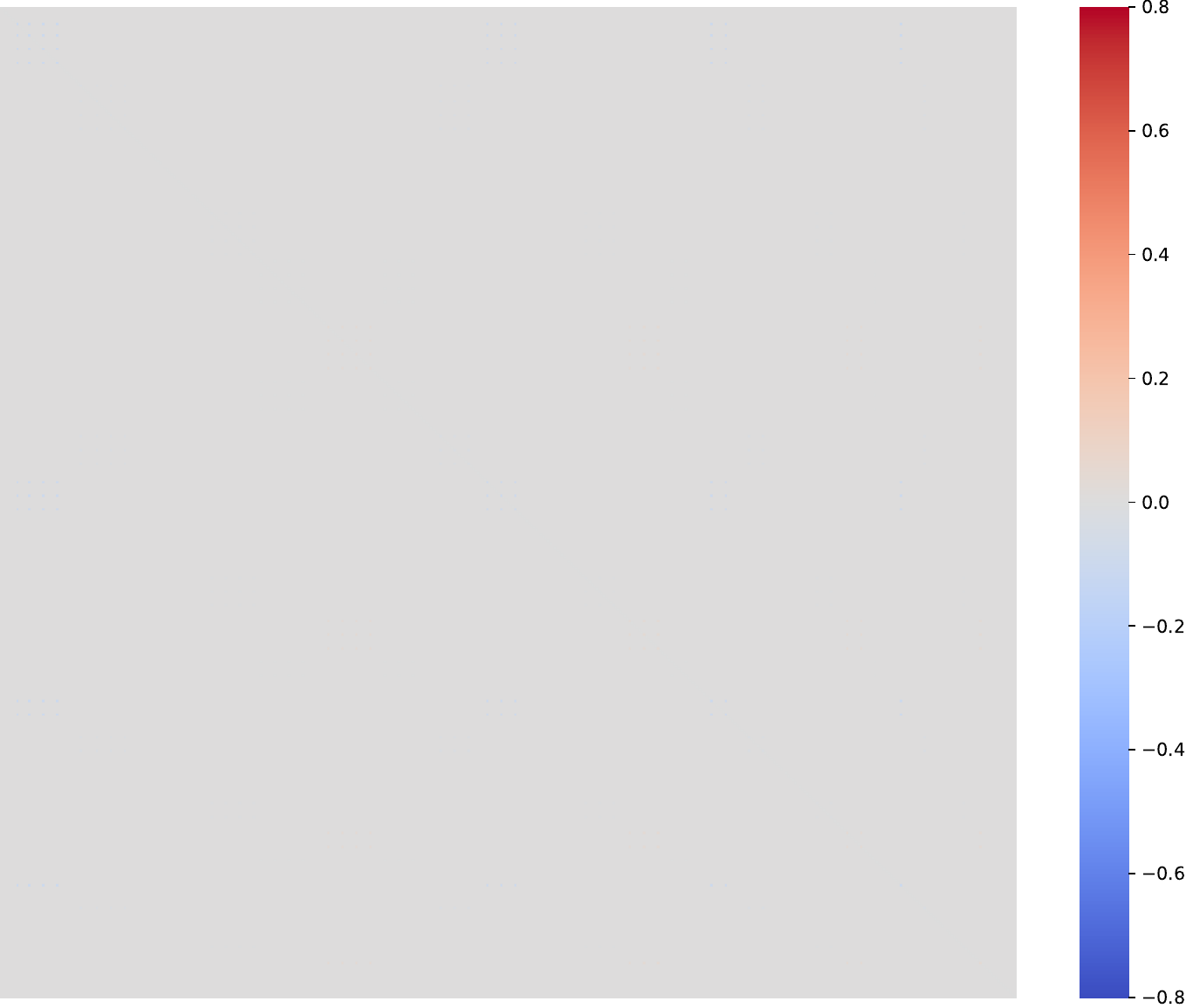}
            & \adjustimage{width=4cm,valign=c}{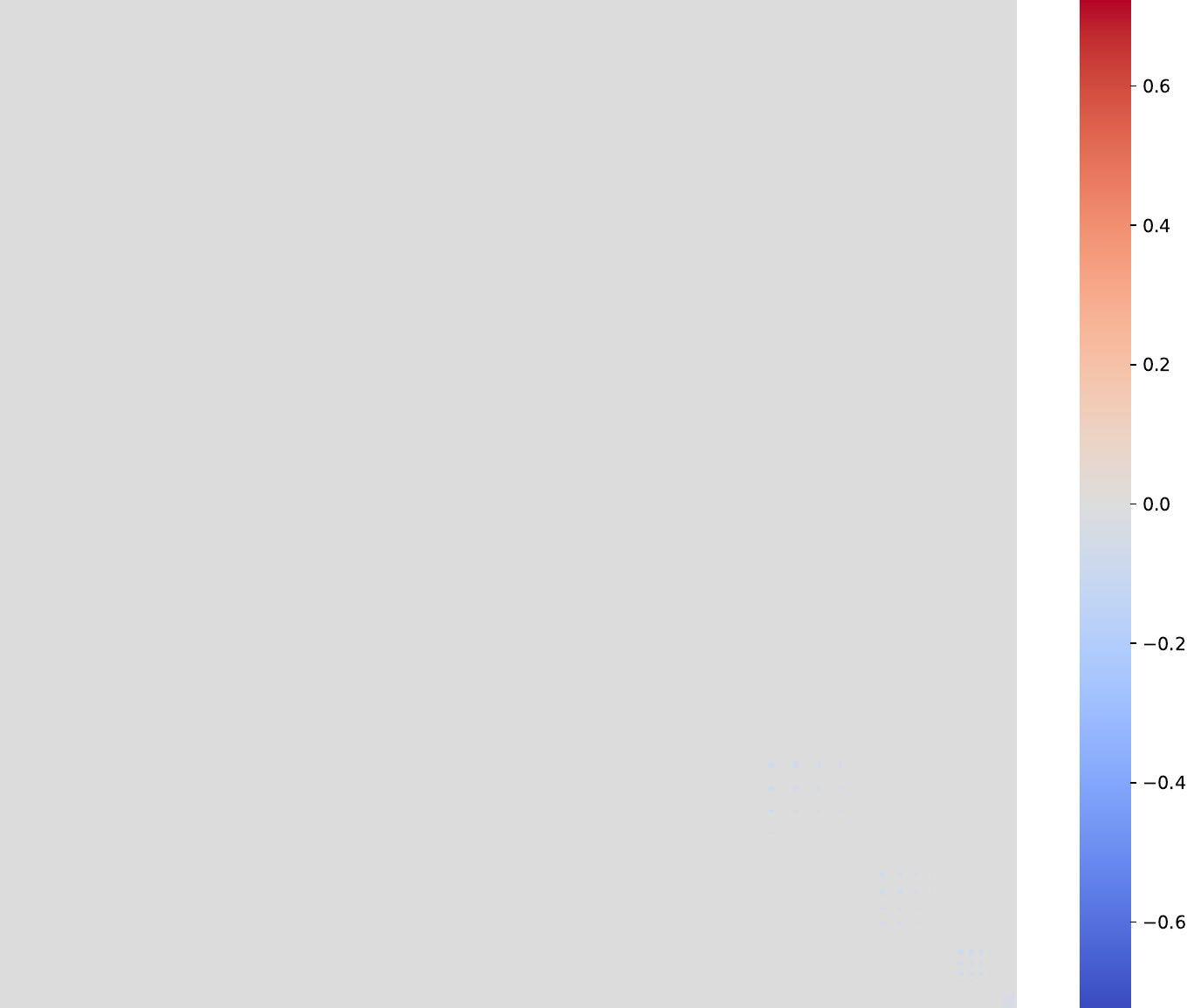}
            & \adjustimage{width=4cm,valign=c}{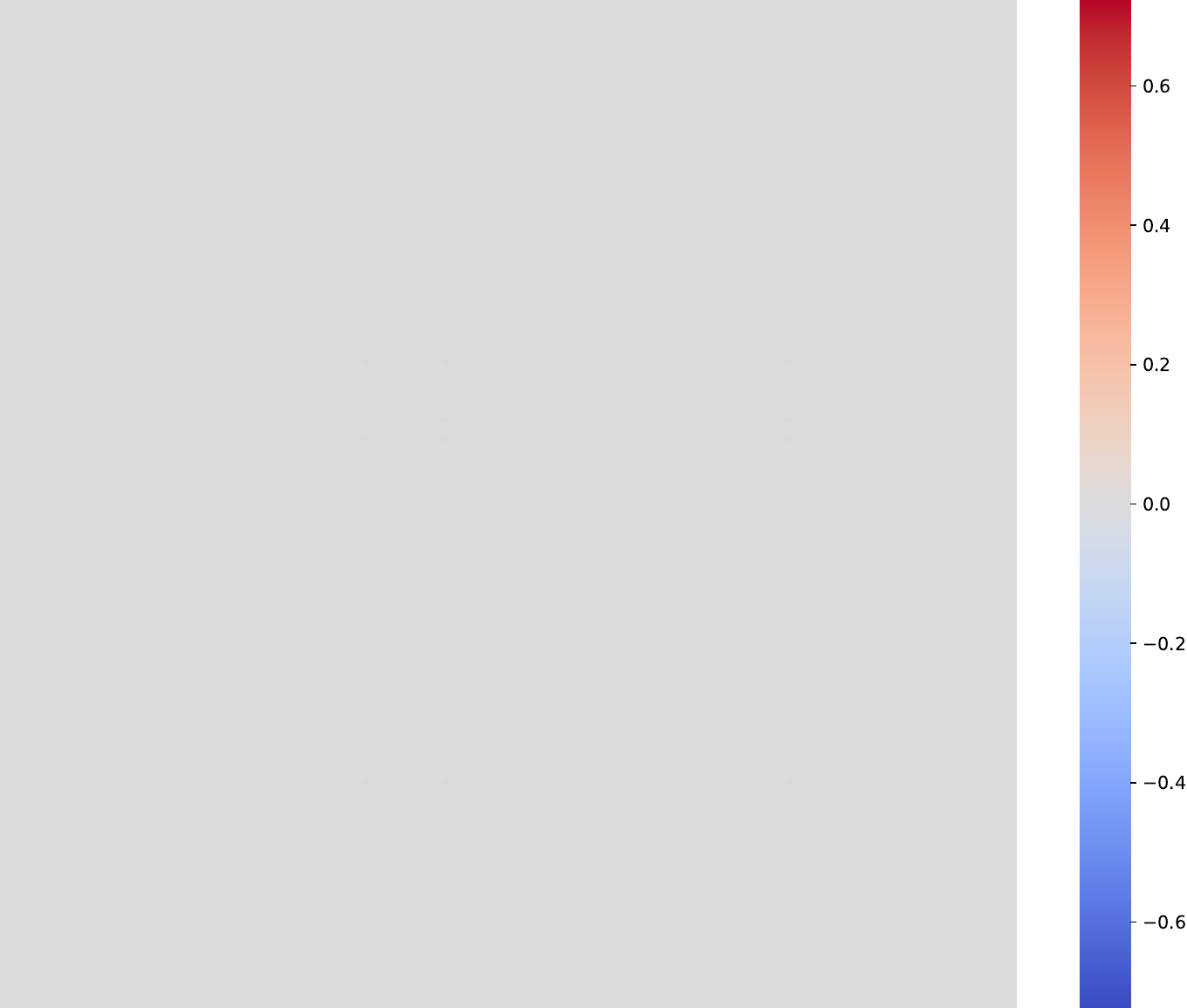} \\
        \end{tabular}%
    }
    \caption{Visualization of model estimation and convergence for large-scale network topologies. Rows show: (1) trace plots for convergence diagnostics; (2) true connectivity matrices ($\mathbf{\Lambda}$); (3) estimated covariance matrices ($\hat{\mathbf{\Lambda}}$); (4) estimation error matrices ($\mathbf{\Lambda} - \hat{\mathbf{\Lambda}}$). Columns (a-d) represent different large-scale network configurations ($V=30,40$) and topologies (S1, S2).}
    \label{fig:large_scale_networks}
\end{figure*}

Furthermore, we also conducted related simulation experiments on large-scale networks with $V=100$, please refer to \nameref{append4} for detailed results. The results show that the model can still infer the topological structure and correlation parameters in large-scale networks, and the parameter estimation can be completed within an acceptable timeframe. Additionally, we evaluated the model's performance under model misspecification by introducing multivariate t-distributed noise and skew-normal distribution noise. Detailed simulation settings and results are provided in \nameref{append6}. The result indicated that the model maintains stability in model inference and parameter estimation, even when heavy-tailed noise were introduced or when samples perturbed by a certain proportion of asymmetric abnormal samples.
%This indicates that the model maintains stability in model inference and parameter estimation, even when perturbed by a certain proportion of asymmetric abnormal samples.

Following the analysis of the model's performance under various parameter settings, we further conducted quantitative comparisons with existing mainstream methods under identical simulation conditions to evaluate the performance of the proposed model in network inference and parameter estimation. Specifically, we compared the proposed BBeC model, based on a finite-dimensional Dirichlet distribution, with two baseline methods: (1) the Graphical Lasso (GLasso) algorithm \cite{friedman2008sparse}; and (2) a Bayesian model based on the Dirichlet Process (CDP) \cite{chen2020bayesian}.

First, we compared the BBeC model with the GLasso method. While BBeC simultaneously infers network topology and estimates edge-related covariance matrices, GLasso is primarily designed for estimating sparse precision matrices and lacks an inherent node clustering mechanism. Therefore, we implemented a two-step strategy for this baseline: covariance matrices were first estimated using GLasso, followed by K-Means clustering to infer the topological structure.

Specifically, the input data were first standardized to prevent unbalanced penalty weights in the $L_1$ regularization of GLasso due to differences in feature scales. Subsequently, GLasso with 3-fold cross-validation was employed to estimate the marginal covariance matrix, with the optimal sparsity regularization parameter determined via grid search.

To infer node topology from the estimated covariance matrix, node-wise features were constructed. For each node, the rows corresponding to its connected edges were extracted from the covariance matrix, and their mean vector was calculated as the connectivity feature for that node. Finally, these features were standardized, and K-Means clustering was applied to obtain the final class labels.

In these experiments, to assess the recovery capability of the baseline method under ideal conditions, the value of $K$ in K-Means was set to the true number of clusters (i.e., $K=3, 5, 6, 8$ for $V=10, 20, 30, 40$, respectively). By providing the ground truth $K$, we aimed to test whether this two-step approach could accurately recover the latent topology when the number of classes is known. The quantitative results are summarized in Tables \ref{tab:comparison_Lasso_V10}, \ref{tab:comparison_Lasso_V20}, \ref{tab:comparison_Lasso_V30}, and \ref{tab:comparison_Lasso_V40}.

As shown in the Tables \ref{tab:comparison_Lasso_V10}, \ref{tab:comparison_Lasso_V20}, \ref{tab:comparison_Lasso_V30}, and \ref{tab:comparison_Lasso_V40}, the GLasso + K-Means pipeline successfully recovers the correct topology with acceptable estimation errors when provided with the true number of clusters. However, a practical limitation of this baseline is its strict dependence on a known $K$, which is typically unavailable in real-world scenarios. In contrast, the BBeC model offers a practical advantage by not requiring a prior specification of the exact number of classes; the latent topology can be automatically inferred as long as the initial $K$ is set slightly larger than the expected value. Furthermore, BBeC demonstrates higher precision in covariance matrix estimation, with errors typically one order of magnitude lower than those of GLasso. Finally, as a Bayesian inference framework, BBeC provides uncertainty quantification for parameter estimates, a feature generally not available in point-estimation methods like GLasso.

\begin{table}[htbp]
\centering
\caption{Comparison of network estimation accuracy between the GLasso + K-Means baseline and the BBeC model for $V=10$.}
\label{tab:comparison_Lasso_V10}
\begin{tabular*}{\linewidth}{@{\extracolsep{\fill}} l c c c c}
\toprule
\multicolumn{5}{c}{$V=10$,~ $G=(3,3,4), ~\exp(-\boldsymbol{\rho})=[0.2,0.4,0.9]$} \\
%\multicolumn{5}{c}{} \\
\midrule
& \multicolumn{2}{c}{S1} & \multicolumn{2}{c}{S2} \\
\cmidrule(lr){2-3} \cmidrule(lr){4-5}
& GLasso & BBeC & GLasso & BBeC \\
\midrule
normalized error & 0.0234  & 0.0013  & 0.0234  & 0.0003  \\
ARI              & 1.00    & 1.00    & 1.00    & 1.00    \\
sensitivity      & 100\%   & 100\%   & 100\%   & 100\%   \\
specificity      & 100\%   & 100\%   & 100\%   & 100\%   \\
Frobenius norm   & 0.13946 & 0.00532 & 0.13946 & 0.00547 \\
\bottomrule
\end{tabular*}
\end{table}

\begin{table}[htbp]
\centering
\caption{Comparison of network estimation accuracy between the GLasso + K-Means baseline and the BBeC model for $V=20$.}
\label{tab:comparison_Lasso_V20}
\begin{tabular*}{\linewidth}{@{\extracolsep{\fill}} l c c c c}
\toprule
\multicolumn{5}{c}{$V=20,~ G=(4,4,4,4,4),~\exp(-\boldsymbol{\rho})=[0.2,0.4,0.6,0.8,0.9]$} \\
%\multicolumn{5}{c}{$$} \\
\midrule
& \multicolumn{2}{c}{S1} & \multicolumn{2}{c}{S2} \\
\cmidrule(lr){2-3} \cmidrule(lr){4-5}
& GLasso & BBeC & GLasso & BBeC \\
\midrule
normalized error & 0.0058  & 0.0002  & 0.0059  & 0.0003  \\
ARI              & 1.00    & 1.00    & 1.00    & 1.00    \\
sensitivity      & 100\%   & 100\%   & 100\%   & 100\%   \\
specificity      & 100\%   & 100\%   & 100\%   & 100\%   \\
Frobenius norm   & 0.07123 & 0.00254 & 0.07131 & 0.00237 \\
\bottomrule
\end{tabular*}
\end{table}

\begin{table}[htbp]
\centering
\caption{Comparison of network estimation accuracy between the GLasso + K-Means baseline and the BBeC model for $V=30$.}
\label{tab:comparison_Lasso_V30}
\begin{tabular*}{\linewidth}{@{\extracolsep{\fill}} l c c c c}
\toprule
%\multicolumn{5}{c}{$V=30$} \\
\multicolumn{5}{c}{\small $V=30,~ G=(5,5,5,5,5,5),~\exp(-\boldsymbol{\rho})=[0.001,0.2,0.4,0.5,0.6,0.8]$} \\
\midrule
& \multicolumn{2}{c}{S1} & \multicolumn{2}{c}{S2} \\
\cmidrule(lr){2-3} \cmidrule(lr){4-5}
& GLasso & BBeC & GLasso & BBeC \\
\midrule
normalized error & 0.0034  & 0.0001  & 0.0033  & 0.0001  \\
ARI              & 1.00    & 1.00    & 1.00    & 1.00    \\
sensitivity      & 100\%   & 100\%   & 100\%   & 100\%   \\
specificity      & 100\%   & 100\%   & 100\%   & 100\%   \\
Frobenius norm   & 0.05185 & 0.00253 & 0.05168 & 0.00247 \\
\bottomrule
\end{tabular*}
\end{table}

\begin{table}[htbp]
\centering
\caption{Comparison of network estimation accuracy between the GLasso + K-Means baseline and the BBeC model for $V=40$.}
\label{tab:comparison_Lasso_V40}
\begin{tabular*}{\linewidth}{@{\extracolsep{\fill}} l c c c c}
\toprule
%\multicolumn{5}{c}{$V=40$} \\
\multicolumn{5}{c}{ $V=40,~G=(5,5,5,5,5,5,5,5),~ \exp(-\boldsymbol{\rho})=[0.001,0.1,0.2,0.3,0.4,0.5,0.6,0.7]$} \\
%\multicolumn{5}{c}{ $$} \\
\midrule
& \multicolumn{2}{c}{S1} & \multicolumn{2}{c}{S2} \\
\cmidrule(lr){2-3} \cmidrule(lr){4-5}
& GLasso & BBeC & GLasso & BBeC \\
\midrule
normalized error & 0.0018  & 0.00004 & 0.0018  & 0.00004 \\
ARI              & 1.00    & 1.00    & 1.00    & 1.00    \\
sensitivity      & 100\%   & 100\%   & 100\%   & 100\%   \\
specificity      & 100\%   & 100\%   & 100\%   & 100\%   \\
Frobenius norm   & 0.03859 & 0.00164 & 0.03858 & 0.00167 \\
\bottomrule
\end{tabular*}
\end{table}

Next, we compared the BBeC model with the CDP model proposed by Chen et al. \cite{chen2020bayesian}. Unlike the finite-dimensional Dirichlet distribution approach adopted in this paper, CDP utilizes a non-parametric Bayesian framework, which theoretically allows the model to infer the node clustering structure from any initialization state.

During the iterative inference process of the CDP model, a node can either be assigned to an existing class or independently form a new class. Specifically, the full conditional posterior probability of node $i$ being assigned to a newly generated class is as follows:
\begin{equation*}
\begin{aligned}
p(\omega_i \neq \omega_j \text{ for all } j \neq i \mid \boldsymbol{\omega}_{-i}, \boldsymbol{\rho}, \mathbf{R})
&\propto\; \exp\Bigg\{ -\frac{S}{2}\log\left(\det\left(\mathbf{\Lambda} + \lambda \mathbf{I}\right)\right)\\
&\hspace{-3em} -\frac{S}{2}\,\mathrm{tr}\left(\mathbf{H}\left(\mathbf{\Lambda} + \lambda \mathbf{I}\right)^{-1}\right) \Bigg\}
\times \frac{\alpha}{V-1+\alpha}, \quad i=1,\ldots, V,
\end{aligned}
\end{equation*}
where $\alpha$ is the concentration parameter used to adjust the probability of generating a new class. In this experiment, $\alpha$ is set to $10$.

Given that both BBeC and CDP are built upon the Bayesian framework, it is necessary to evaluate the convergence of the MCMC algorithms. We used the split-$\hat{R}$ metric for convergence diagnostics. Since both the topological parameter $\boldsymbol{\omega}$ and the correlation parameter $\boldsymbol{\rho}$ are multi-dimensional vectors, reporting only the average split-$\hat{R}$ metric might obscure convergence issues in specific parameter components. Therefore, the \textbf{maximum} split-$\hat{R}$ values observed among all components of $\boldsymbol{\omega}$ and $\boldsymbol{\rho}$ are reported. This approach adopts a stricter criterion: the model is considered to have converged as long as the parameter component with the worst convergence meets the convergence threshold.

In the comparative analysis, all simulated datasets were generated based on the S1 topology. In addition to the previously introduced parameter configurations, the following settings were included: 
For $V=10$, the group structure is set to $G=(5,5)$ with intra-group correlation strengths $\exp(-\boldsymbol{\rho})=[0.2, 0.4]$; 
For $V=20$, $G=(10, 5, 5)$ and $\exp(-\boldsymbol{\rho})=[0.2, 0.4, 0.6]$; 
For $V=30$, $G=(10, 10, 10)$ and $\exp(-\boldsymbol{\rho})=[0.001, 0.2, 0.4]$; 
Finally, for $V=40$, $G=(20, 10, 10)$ and $\exp(-\boldsymbol{\rho})=[0.001, 0.1, 0.2]$.

The quantitative results of the comparative experiments are summarized in Tables~\ref{tab:comparison_CB_V10}, 
\ref{tab:comparison_CB_V20}, \ref{tab:comparison_CB_V30}, and 
\ref{tab:comparison_CB_V40}. The results show that for network sizes of 
$V=10$ and $20$, when the underlying true clustering structure is 
relatively simple (e.g., containing only $3$ classes), the CDP model can
 achieve node clustering and recover the network topology. However, as 
the scale increases to $V=30$ and $40$, even if the true number of 
classes remains at $3$, the estimation accuracy of the CDP model 
decreases. Furthermore, when the true number of classes increases and 
the network topology becomes more complex, the clustering accuracy of 
the CDP model reduced substantially. Specifically, as shown in Tables~\ref{tab:comparison_CB_V20}, \ref{tab:comparison_CB_V30}, and \ref{tab:comparison_CB_V40}, when $V$ is $20$, $30$, and $40$, and the number of groups is greater than $3$, the ARI values of the CDP model are $0.42$, $0.29$, and $0.42$, respectively. In contrast, all the ARI values of the proposed BBeC model remain at $1.00$.

\begin{table}[htbp]
\centering
\caption{Comparison of network estimation accuracy and convergence diagnostics between the CDP and BBeC models for $V=10$. For the BBeC model, the parameter $K$ was set to $4$ for both examples.}
\label{tab:comparison_CB_V10}
\begin{tabular*}{\linewidth}{@{\extracolsep{\fill}} l c c c c}
\toprule
\multicolumn{5}{c}{$V=10$} \\
\midrule
& \multicolumn{2}{c}{$G=(3,3,4)$} & \multicolumn{2}{c}{$G=(5,5)$} \\
& \multicolumn{2}{c}{$\exp(-\boldsymbol{\rho})=[0.2, 0.4, 0.9]$} & \multicolumn{2}{c}{$\exp(-\boldsymbol{\rho})=[0.2, 0.4]$} \\
\cmidrule(lr){2-3} \cmidrule(lr){4-5}
& CDP & BBeC & CDP & BBeC \\
\midrule
normalized error & 0.0007 & 0.0005 & 0.7098 & 0.0049 \\
ARI              & 1.00   & 1.00   & 0.00   & 1.00   \\
sensitivity      & 100\%  & 100\%  & 100\%  & 100\%  \\
specificity      & 100\%  & 100\%  & 0\%    & 100\%  \\
Frobenius norm   & 0.00536 & 0.00552 & 0.74907 & 0.01604 \\
Max $\hat{R}$ for $\boldsymbol{\omega}$ & 1.0000 & 1.0000 & 1.0000 & 1.0000 \\
Max $\hat{R}$ for $\boldsymbol{\rho}$   & 1.0035 & 1.0035 & 1.2274 & 1.0027 \\
\bottomrule
\end{tabular*}
\end{table}

\begin{table}[htbp]
\centering
\caption{Comparison of network estimation accuracy and convergence diagnostics between the CDP and BBeC models for $V=20$. For the BBeC model, the parameter $K$ was set to $6$ and $4$, respectively.}
\label{tab:comparison_CB_V20}
\begin{tabular*}{\linewidth}{@{\extracolsep{\fill}} l c c c c}
\toprule
\multicolumn{5}{c}{$V=20$} \\
\midrule
& \multicolumn{2}{c}{$G=(4,4,4,4,4)$} & \multicolumn{2}{c}{$G=(10,5,5)$} \\
& \multicolumn{2}{c}{$\exp(-\boldsymbol{\rho})=[0.2, 0.4, 0.6, 0.8, 0.9]$} & \multicolumn{2}{c}{$\exp(-\boldsymbol{\rho})=[0.2, 0.4, 0.6]$} \\
\cmidrule(lr){2-3} \cmidrule(lr){4-5}
& CDP & BBeC & CDP & BBeC \\
\midrule
normalized error & 0.0638 & 0.0002 & 0.0005 & 0.0015 \\
ARI              & 0.42   & 1.00   & 1.00   & 1.00   \\
sensitivity      & 100\%  & 100\%  & 100\%  & 100\%  \\
specificity      & 88.17\%  & 100\%  & 100\%  & 100\%  \\
Frobenius norm   & 0.18742 & 0.00254 & 0.00480 & 0.00502 \\
Max $\hat{R}$ for $\boldsymbol{\omega}$ & 1.0000 & 1.0000 & 1.0000 & 1.0000 \\
Max $\hat{R}$ for $\boldsymbol{\rho}$   & 2.1426 & 1.0107 & 1.0056 & 1.0144 \\
\bottomrule
\end{tabular*}
\end{table}

\begin{table}[htbp]
\centering
\caption{Comparison of network estimation accuracy and convergence diagnostics between the CDP and BBeC models for $V=30$. For the BBeC model, the parameter $K$ was set to $7$ and $4$, respectively.}
\label{tab:comparison_CB_V30}
\begin{tabular*}{\linewidth}{@{\extracolsep{\fill}} l c c c c}
\toprule
\multicolumn{5}{c}{$V=30$} \\
\midrule
& \multicolumn{2}{c}{$G=(5,5,5,5,5,5)$} & \multicolumn{2}{c}{$G=(10,10,10)$} \\
& \multicolumn{2}{c}{\begin{tabular}{@{}c@{}}$\exp(-\boldsymbol{\rho})=[0.001, 0.2, 0.4,$ \\ $0.5, 0.6, 0.8]$\end{tabular}} 
& \multicolumn{2}{c}{$\exp(-\boldsymbol{\rho})=[0.001, 0.2, 0.4]$} \\[2ex]
\cmidrule(lr){2-3} \cmidrule(lr){4-5}
& CDP & BBeC & CDP & BBeC \\
\midrule
normalized error & 0.2122 & 0.0001 & 0.1454 & 0.0005 \\
ARI              & 0.29   & 1.00   & 0.55   & 1.00   \\
sensitivity      & 100\%  & 100\%  & 100\%  & 100\%  \\
specificity      & 81.07\%  & 100\%  & 82.5\%   & 100\%  \\
Frobenius norm   & 0.48594 & 0.00253 & 0.42219 & 0.00121 \\
Max $\hat{R}$ for $\boldsymbol{\omega}$ & 1.0000 & 1.0000 & 1.0000 & 1.0000 \\
Max $\hat{R}$ for $\boldsymbol{\rho}$   & 1.0929 & 1.0064 & 1.3068 & 1.0123 \\
\bottomrule
\end{tabular*}
\end{table}

\begin{table}[htbp]
\centering
\caption{Comparison of network estimation accuracy and convergence diagnostics between the CDP and BBeC models for $V=40$. For the BBeC model, the parameter $K$ was set to $9$ and $4$, respectively.}
\label{tab:comparison_CB_V40}
\begin{tabular*}{\linewidth}{@{\extracolsep{\fill}} l c c c c}
\toprule
\multicolumn{5}{c}{$V=40$} \\
\midrule
& \multicolumn{2}{c}{$G=(5,5,5,5,5,5,5,5)$} & \multicolumn{2}{c}{$G=(20,10,10)$} \\
& \multicolumn{2}{c}{\begin{tabular}{@{}c@{}}$\exp(-\boldsymbol{\rho})=[0.001, 0.1, 0.2, 0.3,$ \\ $0.4, 0.5, 0.6, 0.7]$\end{tabular}} 
& \multicolumn{2}{c}{$\exp(-\boldsymbol{\rho})=[0.001, 0.1, 0.2]$} \\[2ex]
\cmidrule(lr){2-3} \cmidrule(lr){4-5}
& CDP & BBeC & CDP & BBeC \\
\midrule
normalized error & 0.1210 & 0.00004 & 0.0287 & 0.00008 \\
ARI              & 0.42   & 1.00    & 0.74   & 1.00   \\
sensitivity      & 100\%  & 100\%   & 100\%  & 100\%  \\
specificity      & 94.12\%  & 100\%   & 94.36\%  & 100\%  \\
Frobenius norm   & 0.32648 & 0.00164  & 0.11949 & 0.00030\\
Max $\hat{R}$ for $\boldsymbol{\omega}$ & 1.0000 & 1.0000 & 1.0000 & 1.0000 \\
Max $\hat{R}$ for $\boldsymbol{\rho}$   & 2.0434 & 1.0200 & 1.7895 & 1.0093 \\
\bottomrule
\end{tabular*}
\end{table}

Regarding the convergence diagnostics, when clustering deviations occur, the maximum split-$\hat{R}$ value for the correlation parameter $\boldsymbol{\rho}$ of the CDP model is typically greater than $1.1$, indicating that the parameter estimation does not fully converge. Meanwhile, in all experiments, the maximum split-$\hat{R}$ value for the topological allocation parameter $\boldsymbol{\omega}$ is shown to be $1.00$. The split-$\hat{R}$ values
stay near $1$ for $\boldsymbol{\omega}$ yet exceed $1.1$ for $\boldsymbol{\rho}$, indicating sampling stagnation in the model: after the burn-in period, node allocation states barely change, resulting in nearly negligible inter-chain variance. In such cases, the ARI serves as a reliable metric for objective clustering evaluation.

In contrast, the BBeC model maintains accuracy in parameter estimation. By setting the initial $K$ value to a larger value, BBeC is able to recover the topology and correlation parameters. In addition, the maximum split-$\hat{R}$ values for its correlation parameter $\boldsymbol{\rho}$ are consistently below $1.03$, reflecting better convergence characteristics. These results also verify the applicability of the BBeC model across different scenarios.

\begin{figure}[htbp]
    \centering
    \includegraphics[width=0.8\textwidth]{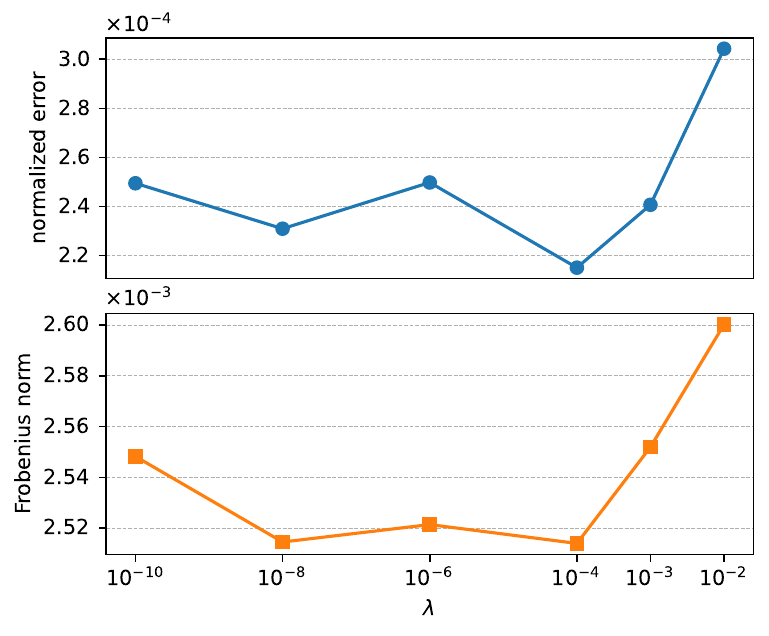}
    \caption{Sensitivity analysis of the regularization parameter $\lambda$ for $V=20$. The plot shows the variations in normalized error and Frobenius norm across different $\lambda$ values.}
    \label{fig:lambda_sens_20}
\end{figure}

\begin{figure}[htbp]
    \centering
    \includegraphics[width=0.8\textwidth]{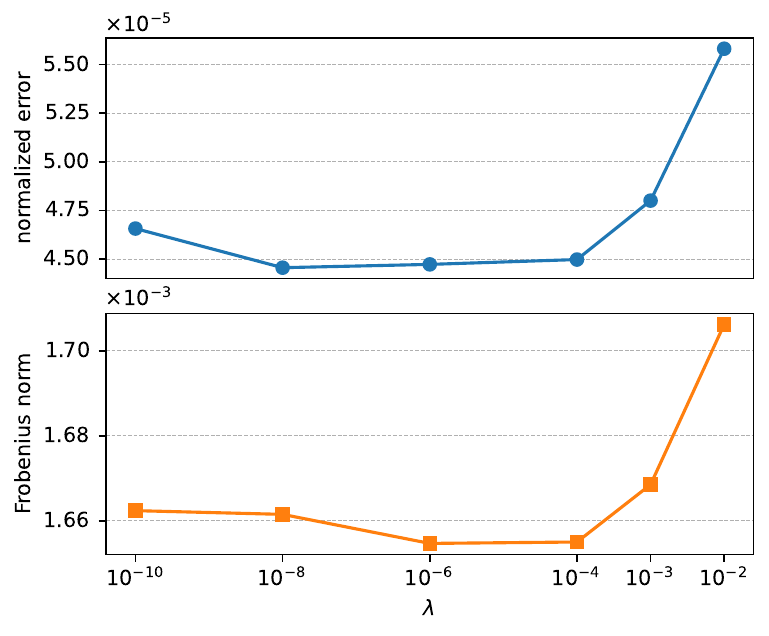}
    \caption{Sensitivity analysis of the regularization parameter $\lambda$ for $V=40$. The plot shows the variations in normalized error and Frobenius norm across different $\lambda$ values.}
    \label{fig:lambda_sens_40}
\end{figure}

Furthermore, we investigated the impact of the regularization parameter $\lambda$ on the estimation accuracy of the model. In the experimental setup, the initial number of clusters $K$ was set to $6$ for the case with $V=20$ nodes, and $K=9$ for $V=40$ nodes. The detailed quantitative results are summarized in Table~\ref{tab:result_V20_lambda} and Table~\ref{tab:result_V40_lambda}. The corresponding trends for the Normalized Error and Frobenius Norm are illustrated in Figure~\ref{fig:lambda_sens_20} and Figure~\ref{fig:lambda_sens_40}, respectively.

The results indicate that the BBeC model is generally robust to the choice of the regularization parameter $\lambda$. The estimation metrics remain stable across several orders of magnitude. However, a slight increase in error is observed when $\lambda$ reaches $10^{-2}$. Notably, in the $V=40$ scenario, setting $\lambda = 10^{-2}$ leads to convergence issues for the correlation parameter $\rho$, with the maximum split-$\hat{R}$ value exceeding $1.1$. This instability likely arises because an excessively large $\lambda$ may exceed the actual minimum connection strength within the network, thereby introducing bias into the parameter inference. Despite these localized convergence challenges, the model maintains high accuracy in node clustering and topology recovery, as evidenced by the ARI values. In conclusion, while the structural identification is relatively insensitive to $\lambda$, we recommend selecting a smaller $\lambda$ value in practical applications to ensure stable and accurate parameter estimation.

\begin{table}[htbp]
\centering
\caption{Estimation accuracy of the BBeC model under different regularization parameters $\lambda$ ($V=20$).}
\label{tab:result_V20_lambda}
\small
\begin{tabular*}{\textwidth}{@{\extracolsep{\fill}}lcccccc}
\toprule
\multicolumn{7}{c}{\textbf{$V=20,~G=(4,4,4,4,4),~\exp(-\boldsymbol{\rho})=[0.2, 0.4, 0.6, 0.8, 0.9]$}} \\
%\multicolumn{7}{c}{\small $G=(4,4,4,4,4)$, $\exp(-\boldsymbol{\rho})=[0.2, 0.4, 0.6, 0.8, 0.9]$} \\
\cmidrule(lr){1-7}
Value of $\lambda$ & $10^{-10}$ & $10^{-8}$ & $10^{-6}$ & $10^{-4}$ & $10^{-3}$ & $10^{-2}$ \\
\midrule
normalized error                  & 0.00024 & 0.00023 & 0.00024 & 0.00021 & 0.00024 & 0.0003 \\
ARI                               & 1.00    & 1.00    & 1.00    & 1.00    & 1.00    & 1.00   \\
sensitivity                       & 100\%   & 100\%   & 100\%   & 100\%   & 100\%   & 100\%  \\
specificity                       & 100\%   & 100\%   & 100\%   & 100\%   & 100\%   & 100\%  \\
Frobenius norm                    & 0.00254 & 0.00251 & 0.00252 & 0.00251 & 0.00255 & 0.0026 \\
Max $\hat{R}$ for $\boldsymbol{\omega}$ & 1.0000  & 1.0000  & 1.0000  & 1.0000  & 1.0000  & 1.0000 \\
Max $\hat{R}$ for $\boldsymbol{\rho}$   & 1.0107  & 1.012   & 1.0107  & 1.0124  & 1.0117  & 1.0101 \\
\bottomrule
\end{tabular*}
\end{table}

\begin{table}[htbp]
\centering
\caption{Estimation accuracy of the BBeC model under different regularization parameters $\lambda$ ($V=40$).}
\label{tab:result_V40_lambda}
\small
\begin{tabular*}{\textwidth}{@{\extracolsep{\fill}}lcccccc}
\toprule
\multicolumn{7}{c}{\textbf{$V=40,~G=(5,5,5,5,5,5,5,5),~\exp(-\boldsymbol{\rho})=[0.001, 0.1, 0.2, 0.3, 0.4, 0.5, 0.6, 0.7]$}} \\
%\multicolumn{7}{c}{\small $G=(5,5,5,5,5,5,5,5)$} \\
%\multicolumn{7}{c}{\small $\exp(-\boldsymbol{\rho})=[0.001, 0.1, 0.2, 0.3, 0.4, 0.5, 0.6, 0.7]$} \\
\cmidrule(lr){1-7}
Value of $\lambda$ & $10^{-10}$ & $10^{-8}$ & $10^{-6}$ & $10^{-4}$ & $10^{-3}$ & $10^{-2}$ \\
\midrule
normalized error                  & 0.00004 & 0.00004 & 0.00004 & 0.00004 & 0.00004 & 0.00005 \\
ARI                               & 1.00    & 1.00    & 1.00    & 1.00    & 1.00    & 1.00   \\
sensitivity                       & 100\%   & 100\%   & 100\%   & 100\%   & 100\%   & 100\%  \\
specificity                       & 100\%   & 100\%   & 100\%   & 100\%   & 100\%   & 100\%  \\
Frobenius norm                    & 0.00166 & 0.00166 & 0.00165 & 0.00165 & 0.00166 & 0.0017 \\
Max $\hat{R}$ for $\boldsymbol{\omega}$ & 1.0000  & 1.0000  & 1.0000  & 1.0000  & 1.0058  & 1.0000 \\
Max $\hat{R}$ for $\boldsymbol{\rho}$   & 1.0102  & 1.0102  & 1.0059  & 1.0059  & 1.0058  & 1.1570 \\
\bottomrule
\end{tabular*}
\end{table}

\section{Application}\label{sec5}

The proposed methodology was applied to a sMRI dataset collected from the Alzheimer's Disease Neuroimaging Initiative (ADNI) database\footnote{Data used in the preparation of this article were obtained from the Alzheimer’s Disease Neuroimaging Initiative (ADNI) database (adni.loni.usc.edu). The ADNI was launched in 2003 as a public-private partnership, led by Principal Investigator Michael W. Weiner, MD. The primary goal of ADNI has been to test whether serial magnetic resonance imaging (MRI), positron emission tomography (PET), other biological markers, and clinical and neuropsychological assessment can be combined to measure the progression of mild cognitive impairment (MCI) and early Alzheimer’s disease (AD).}. The data  contained $n=632$ subjects and $10$ paired ROIs considered in Szefer et al. (2017), Greenlaw et al. (2017) and Song et al. (2022) on the left and right hemispheres of the brain~\cite{szefer2017multivariate, greenlaw2017bayesian, song2022bayesian}.  As described in the literature \cite{song2022bayesian,szefer2017multivariate,szefer2014joint}, the response variables in the ADNI sMRI dataset used in this study were already adjusted during the preprocessing stage. Specifically, the data were adjusted not only for confounding factors such as age, gender, education, handedness, baseline total intracranial volume, and potential population stratification, but also for clinical disease status, including control, mild cognitive impairment, and Alzheimer's disease.

In this study, we focused on modeling the edge-based connectivity among $20$ specific ROIs  and the justification for selecting the 20 ROIs is provided in \nameref{append7}. These regions include anatomical landmarks in the cerebral cortex, including the primary sensorimotor cortex, parietal association areas, and frontal regions associated with higher-order cognition and research indicated that these regions constitute structural covariance networks and exhibit certain connectivity patterns~\cite{seeley2009neurodegenerative,alexander2013imaging}. These $20$ ROIs included $6$ composite metrics, such as the mean cortical thickness of the precentral and postcentral gyri.  These composite metrics allowed us to verify whether the model can spontaneously assign anatomical regions to the same category as their corresponding composite means, thereby assessing the accuracy of the clustering. Furthermore, preliminary K-Means clustering performed on the covariance matrix of these $20$ ROIs revealed that these regions constitute a subset with a rich internal covariance structure (see \nameref{append3}). A detailed description  of all $20$ ROIs is given in Table~\ref{tab:roi_description}. The cortical thickness measures from these ROIs constituted the feature vector for subsequent analysis.

\begin{table*}[tp!]
\centering
\caption{The imaging phenotypes were defined as cortical thickness measurements of $20$ ROIs, derived from the $10$ bilateral parcellations generated by automated FreeSurfer processing.}
\label{tab:roi_description}
\begin{tabularx}{\textwidth}{l l >{\raggedright\arraybackslash}X >{\raggedright\arraybackslash}X}
\toprule 
\textbf{ID} & \textbf{Measure} & \textbf{ROI Name} & \textbf{Core Function} \\
\midrule
Precentral & Thickness & Precentral Gyrus & Primary motor cortex, controls voluntary movement~\cite{banker2019neuroanatomy} \\
Postcentral & Thickness & Postcentral Gyrus & Primary somatosensory cortex, processes bodily sensations~\cite{diguiseppi2023neuroanatomy} \\
Precuneus & Thickness & Precuneus & Core hub of the default mode network, linked to self-awareness~\cite{cavanna2006precuneus} \\
Supramarg & Thickness & Supramarginal Gyrus & Involved in language and spatial cognition~\cite{MANES2025342,struiksma2017tell} \\  
InfParietal & Thickness & Inferior Parietal Gyrus & Integrates multimodal sensory information~\cite{li2020left} \\
SupParietal & Thickness & Superior Parietal Gyrus & Related to visuomotor coordination~\cite{de2022superior} \\
SupFrontal & Thickness & Superior Frontal Gyrus & Associated with working memory and executive functions~\cite{nissim2017frontal} \\
\addlinespace 
MeanSensMotor & Mean thickness & Precentral and postcentral gyri & Composite measure of the sensorimotor network \\
MeanPar & Mean thickness & Inferior parietal, superior parietal, supramarginal gyri, and precuneus & Composite measure of key parietal regions \\
MeanFront & Mean thickness & Caudal middle frontal, rostral middle frontal, and superior frontal gyri & Composite measure of key frontal regions \\
\bottomrule
\end{tabularx}
\end{table*}

We standardized the $20$-dimensional ROI feature vectors for all $632$ subjects by subtracting the mean and dividing by the standard deviation for each feature to eliminate scale differences. Then, for each subject, we generated a symmetric matrix by computing the outer product of their $20$-dimensional feature vector, then used the flattened upper triangular part ($190$ elements) as the final feature input for the model. These feature vectors from all $632$ subjects were stacked to form the complete input matrix for the proposed model.

\begin{figure*}[tp!]
    \centering
    \includegraphics[width=0.8\textwidth]{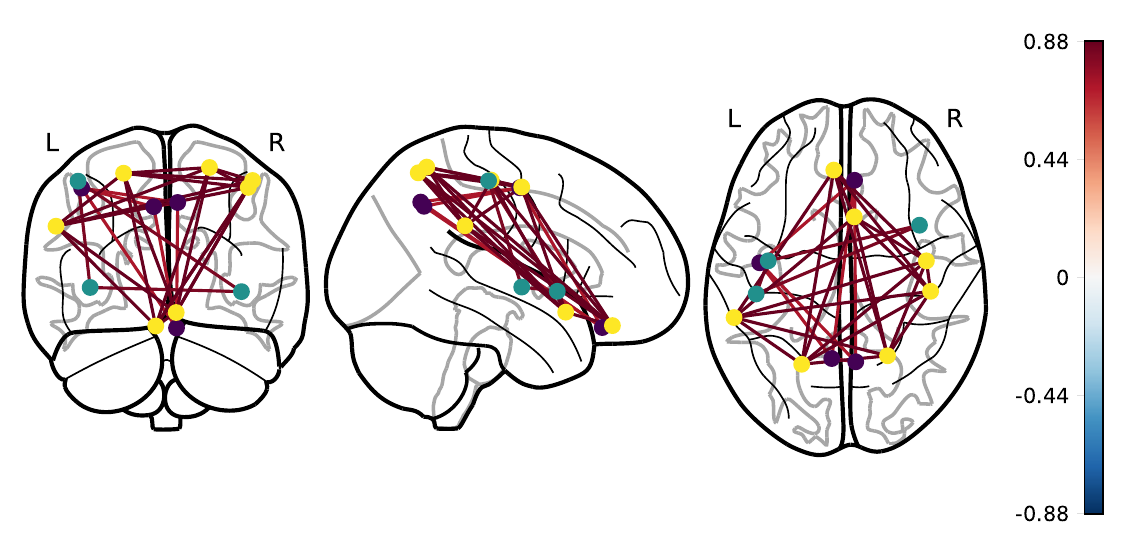}
    \caption{A brain connectivity demonstration - anatomical views of three structural brain modules (i.e. ROI groups) identified by the proposed model: coronal (left), sagittal (middle), and axial (right). Node color indicates group assignment. The deep purple nodes, encompassing the precentral gyri, postcentral gyri, and their composite regions across both cerebral hemispheres—collectively represent the sensorimotor module. The green nodes comprise the left superior parietal gyrus, bilateral superior frontal gyri and composite of key frontal regions, representing the Fronto-Parietal anatomical module. And the yellow nodes contain the bilateral precuneus gyri, supramarginal gyri, inferior parietal gyri, right superior parietal gyrus, and bilateral composite regions derived from key parietal areas, corresponding to the Medial-Lateral Parietal Module. Edge color represents connection strength, scaled according to the color bar on the right. The composite metrics were excluded from this visualization as they lack specific anatomical coordinates.}
    \label{fig:brain_clusters}
\end{figure*}

For the model specification, we tuned the number of clusters for the brain region topology, i.e., $K=5$. For the prior of the correlation parameters $\boldsymbol{\rho}$, we specified a normal prior distribution with a mean of $0.25$ and a standard deviation of $0.01$. The parameters $\boldsymbol{\rho}$ were updated using normal proposal distribution with random-walk, where the step size was set to $0.01$. To determine the initial values for $\rho_k$ and $\rho_0$, we performed a preliminary K-Means clustering on the sample covariance matrix (shown in Fig.~\ref{fig:kmeans_matrix} in Appendix 3); the average correlations within the identified groups and the average correlations between distinct groups were used as the starting points, respectively. The MH algorithm was run for $2500$ iterations.  The results revealed a topological structure among the $20$ selected ROIs. Although the model was tuned with $K=5$ clusters, its inference converged to $\hat{K}=3$ clusters, as shown in Table ~\ref{tab:cluster_rois}. Fig.   ~\ref{fig:brain_clusters} provides  anatomical views of three structural brain modules (i.e. ROI groups) identified by the proposed model.

The identified clusters showed that all composite metrics were grouped within the same clusters as their constituent brain regions. Furthermore, these clustering patterns were consistent with established findings in the literature~\cite{zielinski2010network,chen2008revealing,spreng2013structural}. Together, these results demonstrate the model's reliability.

As illustrated in Fig.~\ref{fig:brain_clusters} and Table~\ref{tab:cluster_rois}, the identified clusters conform to well-established neuroanatomical networks. For instance, Cluster 1 (represented by deep purple nodes), which consists of the bilateral precentral gyri, postcentral gyri, and a composite sensorimotor region, demonstrates the tight structural linkage characteristic of the primary sensorimotor cortex~\cite{zielinski2010network}. Cluster 2 (denoted by green nodes), encompassing the bilateral superior frontal gyri, a composite frontal region, and the left superior parietal gyrus, aligns with the anatomical architecture of fronto-parietal circuitry~\cite{chen2008revealing}. Cluster 3 (indicated by bright yellow nodes), incorporating the bilateral precuneus, supramarginal gyri, inferior parietal gyri, a composite parietal region, and the right superior parietal gyrus, represents key hubs of the default mode network, which are known to exhibit pronounced structural covariance~\cite{spreng2013structural}.

\begin{table}[htbp]
\centering
\caption{List of the three identified clusters.}
\label{tab:cluster_rois}
\begin{tabular}{l p{8cm}}
\toprule
\textbf{Cluster} & \textbf{ROI IDs} \\
\midrule
Cluster 1 & Precentral (Left/Right), Postcentral (Left/Right), MeanSensMotor (Left/Right) \\
\addlinespace
Cluster 2 & SupFrontal (Left/Right), MeanFront (Left/Right), SupParietal (Left) \\
\addlinespace
Cluster 3 & Precuneus (Left/Right), Supramarg (Left/Right), InfParietal (Left/Right), MeanPar (Left/Right), SupParietal (Right) \\
\bottomrule
\end{tabular}
\end{table}

\begin{figure*}[tp!]
    \centering
    \includegraphics[width=0.8\textwidth]{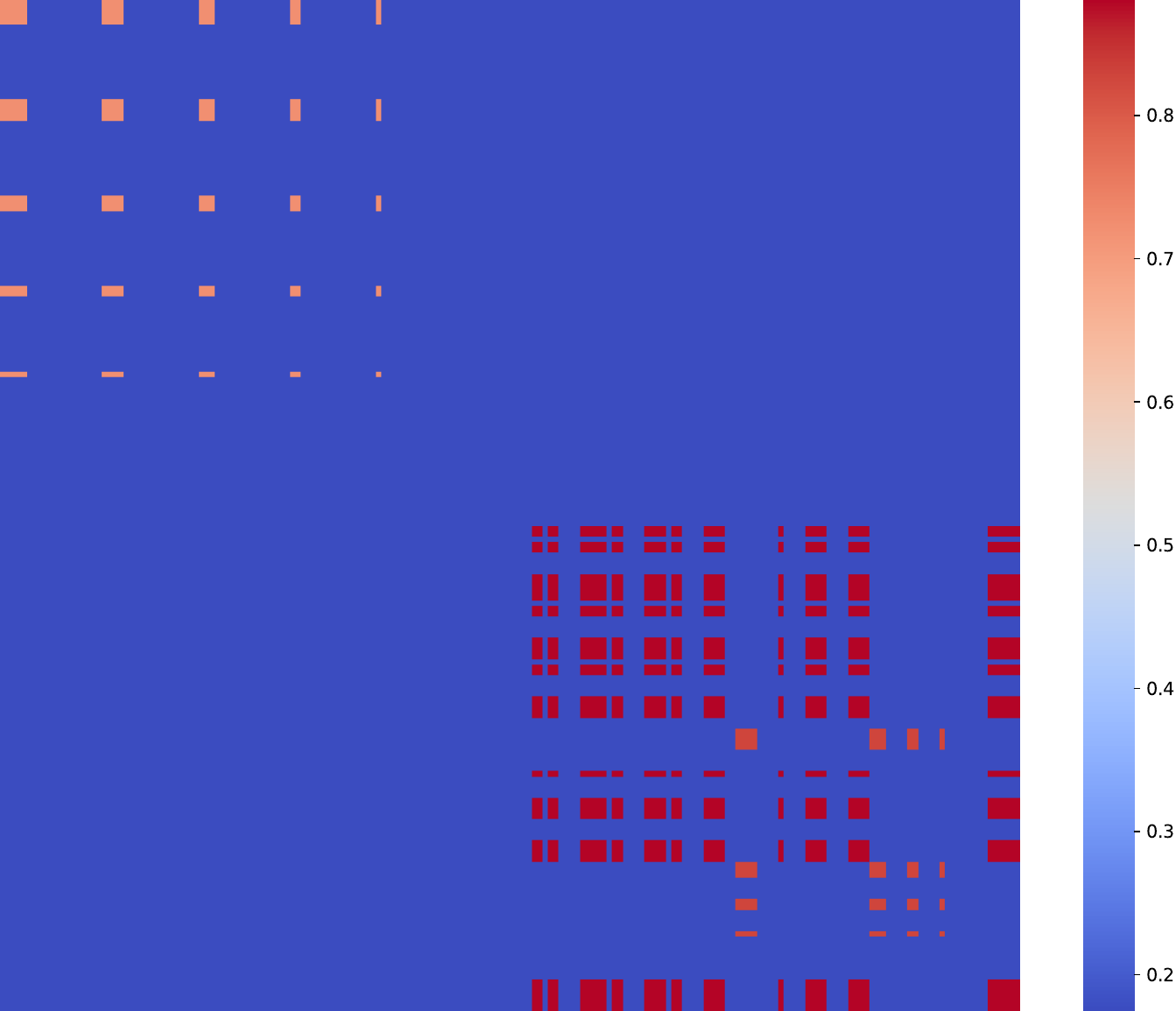}
    \caption{The estimated edge based correlation matrix ($\boldsymbol{\hat{\Lambda}}$). The top-left orange points indicate edge correlation within Cluster 1. In the bottom-right, the orange-red points correspond to Cluster 2, while the dark red points correspond to Cluster 3.}
    \label{fig:brain_correlation}
\end{figure*}

Beyond identifying these anatomical partitions, our model also quantified the average strength of correlations within and between these groups. The results indicate significant internal structural correlation: the estimated covariance strength was $0.8806$ for Cluster 3, $0.8275$ for the Cluster 2, and $0.7224$ for the Cluster 1. In contrast, the correlation strength between distinct groups was estimated to be a lower value of $0.1747$. It is evident that the connection strength within groups far exceeds that between groups, revealing a clear pattern that brain regions within the same group maintain tight structural links while remaining relatively distinct from others. The distinct block-diagonal structure evident in the correlation matrix (Fig.~\ref{fig:brain_correlation}) visually confirms this result, marked by strong intra-group correlations alongside weak inter-group associations.

Furthermore, to evaluate the edge-based correlations across different clinical conditions, we conducted a stratified analysis on the entire cohort, which consists of cognitively normal (CN) individuals, late mild cognitive impairment (LMCI) individuals, and Alzheimer's disease (AD) patients. The sample sizes are $179$, $309$, and $144$, respectively. For each clinical group, we computed the outer product of the region-of-interest vectors and extracted the upper triangular elements to form the edge data. We then applied our model to each group independently with $K$ set to $5$.

For the LMCI and AD groups, the identified clustering structures are identical to those presented in Table \ref{tab:cluster_rois}. The CN group showed a distinct clustering pattern for the left supramarginal gyrus: this region was assigned to Cluster 1, corresponding to the sensorimotor network, rather than to Cluster 3, as observed in the LMCI and AD groups. From a neurobiological perspective, the supramarginal gyrus is an associative area involved in sensory-motor integration. In a healthy aging brain, it maintains structural covariance with primary sensorimotor areas to support intact sensory processing and movement coordination. As the disease pathology progresses, the inferior parietal lobule becomes vulnerable \cite{pini2016brain}. This pathological burden likely causes the supramarginal gyrus to disconnect from the core sensorimotor network and instead co-atrophy with the broader default mode and parietal networks \cite{seeley2009neurodegenerative}, explaining its shift to Cluster 3 in the impaired groups.

Furthermore, we examined the intra-group and inter-group correlation parameters across the three clinical stages. The estimated correlation parameters are summarized in Table \ref{tab:response_correlation_params}, and the corresponding estimated correlation matrices are presented in \nameref{append7}.

\begin{table}[htbp]
\centering
\caption{Estimated correlation parameters across different clinical groups.}
\label{tab:response_correlation_params}
\begin{tabular}{lcccc}
\toprule
\textbf{Clinical Group} & $\exp(-\rho_0)$ & $\exp(-\rho_1)$ & $\exp(-\rho_2)$ & $\exp(-\rho_3)$ \\
\midrule
CN & 0.2398 & 0.6779 & 0.5720 & 0.7244 \\
LMCI & 0.1623 & 0.7351 & 0.6816 & 0.6633 \\
AD & 0.1607 & 0.9660 & 0.7353 & 0.7791 \\
\bottomrule
\end{tabular}
\end{table}

As shown in Table \ref{tab:response_correlation_params}, the inter-cluster relationship, represented by $\exp{(-\rho_0)}$, shows a gradual decline  across disease progression. Conversely, the intra-cluster relationships, specifically within  Clusters 1, 2 (represented by $\exp{(-\rho_1)}$ and $\exp{(-\rho_2)}$, respectively), exhibit a progressive strengthening trend.  However, the intra-cluster within  Cluster 3 (represented by $\exp{(-\rho_3)}$) shows a non-monotonic pattern: it decreases from the CN group to the LMCI group, but increases again from the LMCI group to the AD group.   These progressive strengthening trends are consistent with recent pathophysiological findings. Alzheimer's disease is recognized as a disconnection syndrome characterized by the gradual deterioration of long-range inter-network integration \cite{delbeuck2003alzheimer,brier2014network}. Meanwhile, the observed increase in intra-network connectivity during the early and intermediate stages is commonly interpreted in neuroimaging literature as a compensatory hyperconnectivity mechanism, where local networks increase their synchronization to maintain cognitive output despite ongoing structural degeneration \cite{hillary2015hyperconnectivity}. 

\section{Discussion and conclusion}\label{sec6}

In this study, we proposed a Bayesian model designed to infer latent topological structures and estimate high-dimensional covariance matrices from MRI data. By applying this model to both simulated datasets and the ADNI cohort, we demonstrated its efficacy in identifying structural subnetworks. Simulations indicated its accuracy in recovering predefined node topology and correlation parameters. Notably, in the ADNI data analysis within selected ROIs, the model identified three major assemblies: the sensorimotor module, a fronto-parietal circuit, and the medial-lateral parietal module. Crucially, the biological validity of these clusters was supported by composite anatomical metrics.

In our clustering results, while Cluster $1$ successfully captured the structural link between left and right sensorimotor regions described by Zielinski et al. (2010)~\cite{zielinski2010network}, while  Clusters 2 and 3 exhibited a divergence regarding the superior parietal gyrus (SupParietal) , where the model assigned the left and right regions to different groups. This finding is consistent with the significant lateralization difference of the superior parietal gyrus reported in ~\cite{gotts2013two}. Specifically, the left SupParietal region was grouped in Cluster 2. This aligns with Chen et al (2008), who defined the superior frontal gyrus as a distinct structural core~\cite{chen2008revealing}, while the inclusion of the parietal node in our model likely reflected the expansion and integration of frontal and parietal structural connections observed in the mature brain~\cite{zielinski2010network}. In contrast, the right SupParietal region was assigned to Cluster $3$, clustering with the precuneus and inferior parietal gyrus. Spreng and Turner (2013) have identified the precuneus and inferior parietal regions as core areas of structural covariance~\cite{spreng2013structural}. In the ADNI cohort, this likely reflects that the right superior parietal gyrus shows a similar pattern of atrophy to these core regions due to Alzheimer's pathology.

One of the main contributions of this work is its effective resolution of the curse of dimensionality inherent in edge-based connectivity analyses by explicitly modeling the latent topological structure. Traditional pairwise approaches often suffer from reduced statistical power due to the massive number of multiple comparisons required for high-resolution atlases ~\cite{zalesky2010network}. By contrast, our model operates by first grouping brain regions into latent clusters and subsequently estimating the covariance parameters based on these cluster assignments. This efficiency in unsupervised structure learning was rigorously validated in the simulation studies, where the model successfully recovered the ground-truth node topology and connection strengths. This simplified representation facilitates the interpretation of complex connectomes and leads to robust parameter estimation, particularly in datasets with limited sample sizes. Specifically, in high-dimensional settings where the number of potential connections often exceeds the number of subjects, standard sample covariance may be singular. By reducing the effective degrees of freedom to a small set of cluster-specific parameters, our model aggregates statistical information across topologically similar edges to stabilize parameter estimation.  

A distinct feature of our model is that, unlike the non-parametric Bayesian approach based on the DP employed by Chen et al.~\cite{chen2020bayesian}, this study utilizes a finite-dimensional Dirichlet distribution to model network node assignments. Although the DP allows for an infinite number of clusters in theory, it introduces variable dimensionality during MCMC iterations. This variation may hinder algorithm convergence and cause label-switching problems. In our simulation experiments, the CDP model only achieved accurate topological inference in a few simple structural scenarios and struggled to cluster effectively in more complex cases. In contrast, our approach stabilizes the parameter space, thereby reducing the risk of the Markov chain being trapped in local optima and improving sampling efficiency. This stability enables the reasonably accurate estimation of both topological structures and correlation parameters.

Furthermore, the proposed model provides direct clustering inference. Comparisons with the GLasso model indicate that our approach holds an advantage in controlling parameter estimation errors. Moreover, the GLasso model does not inherently perform clustering and relies on external algorithms, such as K-Means, to infer the internal topology of the data. However, in practical applications, it is difficult to accurately specify the true number of clusters in advance for K-Means. Our model, by contrast, only requires setting the initial cluster number $K$ to a value slightly larger than the expected true number. Through iterations, it can automatically identify the topological structure, making it more applicable in practice.

The sensitivity analysis of the regularization parameter $\lambda$ shows that the model is generally insensitive to its selection, and changes the order of magnitude of this parameter generally do not obviously affect the overall estimation accuracy. Although larger $\lambda$ values introduce a slight increase in estimation error, the errors remain within an acceptable range. But, if the selected $\lambda$ exceeds the actual minimum connection strength present in the covariance matrix, the Markov chain corresponding to that connection strength may fail to converge. This non-convergence manifests as the parameter estimation diverging after an initial period of fluctuation and stability, leading to local estimation bias. Nevertheless, even in the presence of such local parameter divergence, the model continues to correctly identify the node clustering within the data, demonstrating its robustness in topological inference. In a summary,  while structural identification is relatively insensitive to $\lambda$, we recommend selecting a smaller value in practical applications to ensure stable and accurate parameter estimation.

In addition, we investigated the impact of the sampling order (i.e., whether to prioritize the sampling of clustering labels $\omega$ or correlation parameters $\rho$) on the inference. The results indicate that, given an appropriate setting of the initial $K$, the model yields consistent conclusions regardless the sequence of the sampling series.  However, in certain complex scenarios, a slightly larger initial $K$ can help the model achieve better convergence and estimation performance for the reverse sampling order strategy. Detailed quantitative results are provided in \nameref{append5}.

Regarding the implementation, determining the number of latent brain networks ($K$) presents a practical consideration. The simulation results indicate that the algorithm's convergence is highly sensitive to this tuning parameter. Failure to specify a sufficient number of networks (i.e., fewer than the ground truth) can prevent convergence of the MCMC algorithm. In contrast, setting an overly large $K$ is still likely to converge to the true number of clusters, as the superfluous groups will be effectively pruned away during the progression of the Markov chains.  This finding suggests a practical strategy for real-world applications: setting $K$ to a moderately more overspecified value than anticipated. This practice strengthens algorithmic robustness without substantially compromising estimation performance. To further validate this strategy, we additionally evaluated the model's performance using different criteria. In addition to WAIC, we explored the Bayesian Information Criterion (BIC). Although these two criteria had different theoretical foundations and calculation methods, the experimental results showed that the differences between the BIC values across different $K$ settings, as well as the differences between the WAIC values, were negligible. This occurred because, in the BBeC model, superfluous clusters were automatically emptied during the MCMC iterations as long as the initial $K$ was set larger than the true number of clusters. Consequently, the final estimated correlation parameters and topological assignments were nearly identical across all adequate $K$ settings, leading to very little difference in the resulting likelihood values and evaluation metrics. These findings further confirmed that an adequate initial $K$ allowed the model to automatically converge to the correct topological structure, eliminating the need to meticulously adjust $K$ based on minor variations in the evaluation criteria.

Despite these strengths, the current framework relies on specific modeling assumptions that warrant further refinement. First, the model currently assumes that the correlation between groups is positive. However, neuroimaging studies suggest that brain connectivity is complex and may involve negative correlations between competing networks. In scenarios where negative associations exist, the current model specification might fail to accurately capture the true topological structure. Future extensions could incorporate signed stochastic block models ~\cite{aicher2015learning}, which explicitly account for both positive and negative interactions between functional modules.In this context, a limitation of the present study is that we primarily considered positive correlation structures, focusing on the major positive connections for methodological verification and model evaluation. We acknowledge that negative correlations, including weak negative associations, may also be informative in certain downstream clinical or biological analyses. Therefore, future work extending the proposed framework will systematically evaluate the contribution of these negative correlation structures to model performance and interpretability. Second, regarding highly sophisticated hub nodes, from a neurobiological perspective, connection strengths between different nodes within the same functional network typically vary. The current modeling approach may average out these subtle connectivity differences, potentially obscuring local disease-related features. To address this limitation, future research could consider adopting degree-corrected stochastic block models ~\cite{passino2022spectral} to identify and quantify the distribution of connection strength variations among nodes within the same cluster. Third,  
regarding isolated nodes within biological networks, the current hard assignment framework still exhibits certain limitations. Real brain networks contain hub nodes that connect to multiple other clusters. Under the current model specification, if a hub node is assigned to a single-node cluster, it forms no internal edges; thus, the covariance matrix lacks corresponding correlation information, making it difficult for the model to identify such singleton clusters. Therefore, future work could explore methods to enable the model to identify these isolated nodes, thereby improving the inference of brain hubs. Fourth, the current model still requires a manual initial setting for the number of clusters $K$ in practical applications, and cannot yet fully rely on the data itself to automatically infer the topological structure. Although the preceding comparative experiments indicate that traditional Dirichlet process-based models face convergence difficulties in complex scenarios, there is still potential to overcome this bottleneck by introducing more flexible non-parametric Bayesian frameworks in the future. For example, more advanced merge-split MCMC algorithms ~\cite{peixoto2020merge} could be considered. Finally, regarding computational inference, while the MH algorithm ensures asymptotic exactness, it can be computationally intensive for large-scale datasets. To address this, future work could adopt variational inference (VI) techniques ~\cite{blei2017variational}.

In summary, this study introduces a Bayesian framework for inferring brain network topology and high-dimensional covariance structures. The model configuration ensures the positive definiteness of the covariance matrix. Furthermore, by introducing a finite-dimensional Dirichlet distribution to model the latent topological structure of brain regions, the proposed approach effectively reduces the parameter dimensionality, thereby alleviating challenges associated with high-dimensional estimation.   The framework thus establishes an efficient and reliable tool for investigating the intrinsic connectivity of the brain in large-scale neuroimaging studies.

\backmatter

\section*{Acknowledgments}

Data collection and sharing for this project was funded by the Alzheimer's Disease Neuroimaging Initiative
(ADNI) (National Institutes of Health Grant U01 AG024904) and DOD ADNI (Department of Defense award
number W81XWH-12-2-0012). ADNI is funded by the National Institute on Aging, the National Institute of
Biomedical Imaging and Bioengineering, and through generous contributions from the following: AbbVie,
Alzheimer’s Association; Alzheimer’s Drug Discovery Foundation; Araclon Biotech; BioClinica, Inc.; Biogen;
Bristol-Myers Squibb Company; CereSpir, Inc.; Cogstate; Eisai Inc.; Elan Pharmaceuticals, Inc.; Eli Lilly and
Company; EuroImmun; F. Hoffmann-La Roche Ltd and its affiliated company Genentech, Inc.; Fujirebio; GE
Healthcare; IXICO Ltd.; Janssen Alzheimer Immunotherapy Research \& Development, LLC.; Johnson \&
Johnson Pharmaceutical Research \& Development LLC.; Lumosity; Lundbeck; Merck \& Co., Inc.; Meso
Scale Diagnostics, LLC.; NeuroRx Research; Neurotrack Technologies; Novartis Pharmaceuticals
Corporation; Pfizer Inc.; Piramal Imaging; Servier; Takeda Pharmaceutical Company; and Transition
Therapeutics. The Canadian Institutes of Health Research is providing funds to support ADNI clinical sites
in Canada. Private sector contributions are facilitated by the Foundation for the National Institutes of Health
(\url{www.fnih.org}). The grantee organization is the Northern California Institute for Research and Education,
and the study is coordinated by the Alzheimer’s Therapeutic Research Institute at the University of Southern
California. ADNI data are disseminated by the Laboratory for Neuro Imaging at the University of Southern
California.

The authors thank ShanghaiTech University for supporting this work through the startup fund and the HPC platform.

\section*{Declarations}
\subsection*{Funding}
This project was supported by National Natural Science Foundation of China (12401383), and the startup fund of ShanghaiTech University.

\subsection*{Consent for publication}
Not applicable.

\subsection*{Ethical approval and consent to participate}

Not applicable.

\subsection*{Availability of data and materials}

The datasets analyzed during the current study are available from the Alzheimer’s Disease Neuroimaging Initiative (ADNI) repository at \url{https://adni.loni.usc.edu}. In accordance with the ADNI data use agreement, the authors cannot publicly share the data. Access for authorized investigators is granted upon application to the ADNI Data Sharing and Publications Committee.

The synthetic data and results, and  code are available at: \url{https://github.com/mimi6501/BBeC}.

\subsection*{Conflict of Interests}
The authors declare no conflict of interest.

\subsection*{Author contribution}
Z. L. - Conceptualization, Data Generation, Formal Analysis, Methodology, Coding, Visualization, Original draft, Review and Editing.  C. Z. - Data Generation, Formal Analysis,  Original draft.    S. G. - Conceptualization, Data Curation, Formal Analysis, Methodology, Coding Evaluation, Project Administration, Supervision, Review and Editing.  All authors read and approved the final manuscript.

\subsection*{Corresponding author}
Correspondence to Shufei Ge, \url{geshf@shanghaitech.edu.cn}.

\bibliography{sn-bibliography}

\newpage
\section*{Appendix 1}\label{append1}

\noindent{\large{\bf{Derivation of the joint posterior of $G, \boldsymbol{\rho}$ given data  $\boldsymbol{R}$}}}

\noindent Firstly,  the joint distribution of parameters $G, \boldsymbol{\rho}$ and data $\boldsymbol{R}$ is

\[
p(\mathbf{R}, G, \boldsymbol{\rho}) = p(\mathbf{R} \mid G, \boldsymbol{\rho})p(G)p(\rho).
\]

\noindent By the Bayesian formula,  the joint posterior distribution of the $G, \boldsymbol{\rho}$ given data $\mathbf{R}$ is
\begin{align*}
p(G, \boldsymbol{\rho} \mid \mathbf{R}) &\propto p(\mathbf{R}\mid G, \boldsymbol{\rho})p(G, \boldsymbol{\rho}) \\
&\propto p(\mathbf{R}\mid G, \boldsymbol{\rho})p(G)p(\boldsymbol{\rho}).
\end{align*}

\noindent From the model, $\boldsymbol{\Lambda} = f(G,\boldsymbol{\rho})$, $p(\mathbf{R}\mid G, \boldsymbol{\rho})$ is same as $p(\mathbf{R} \mid \boldsymbol{\Lambda})$,  it follows that 
$$\mathbf{R} \mid \boldsymbol{\Lambda} \sim N(0, (\boldsymbol{\Lambda}+\lambda \boldsymbol{I})).$$ 

\noindent Given $S$ independent individuals, the likelihood function can be written as
\begin{align*}
&p(\mathbf{R} \mid \boldsymbol{\Lambda}) \\
&\quad= \prod_{s=1}^{S}p(\mathbf{R}^s\mid \boldsymbol{\Lambda}) \\
&\quad\propto ( \det(\boldsymbol{\Lambda}+\lambda \boldsymbol{I}))^{-\frac{S}{2}} \\
&\qquad \times \exp\left\{-\frac{1}{2}\sum_{s=1}^{S} \mathbf{R}^s (\boldsymbol{\Lambda}+\lambda \boldsymbol{I})^{-1} (\mathbf{R}^s)^T\right\} \\
&\quad \propto \exp \left\{\log( \det(\boldsymbol{\Lambda}+\lambda \boldsymbol{I}))^{-\frac{S}{2}} \right. \\
&\qquad \left. -\frac{1}{2}\sum_{s=1}^{S}\mathbf{R}^s (\boldsymbol{\Lambda}+\lambda \boldsymbol{I})^{-1} (\mathbf{R}^s)^T\right\}\\
\end{align*}

\noindent Denote
\begin{align*}
\mathbf{R} &= \begin{bmatrix}
r_{1,1} & r_{1,2} & \cdots & r_{1,E} \\
r_{2,1} & r_{2,2} & \cdots & r_{2,E} \\
\vdots & \vdots & \ddots & \vdots \\
r_{S,1} & r_{S,2} & \cdots & r_{S,E}
\end{bmatrix}_{S\times E}, \\
\left(\boldsymbol{\Lambda}+\lambda \boldsymbol{I}\right)^{-1} &= \begin{bmatrix}
a_{1,1} & a_{1,2} & \cdots & a_{1,E} \\
a_{2,1} & a_{2,2} & \cdots & a_{2,E} \\
\vdots & \vdots & \ddots & \vdots \\
a_{E,1} & a_{E,2} & \cdots & a_{E,E}
\end{bmatrix}_{E\times E.}
\end{align*}

\noindent We have,
\begin{align*}
&\mathrm{tr}(\mathbf{R}^\top\mathbf{R}\left(\boldsymbol{\Lambda}+\lambda \boldsymbol{I}\right)^{-1}) \\
&\quad= \sum_{j=1}^{E}\sum_{s=1}^{S}r_{s,1}r_{s,j}a_{j,1} + \cdots + \sum_{j=1}^{E}\sum_{s=1}^{S}r_{s,E}r_{s,j}a_{j,E} \\
&\quad= \sum_{i=1}^{E}\sum_{j=1}^{E}\sum_{s=1}^{S}r_{s,i}r_{s,j}a_{j,i}  \\
&\quad= \sum_{s=1}^{S}(\sum_{i=1}^{E}\sum_{j=1}^{E}r_{s,i}r_{s,j}a_{j,i})  \\
&\quad= \sum_{s=1}^{S} \mathrm{tr}((\mathbf{R}^s)^\top\mathbf{R}^s\left(\boldsymbol{\Lambda}+\lambda \boldsymbol{I}\right)^{-1})\\
&\quad= \sum_{s=1}^{S} \mathrm{tr}(\mathbf{R}^s\left(\boldsymbol{\Lambda}+\lambda \boldsymbol{I}\right)^{-1}(\mathbf{R}^s)^\top)\\
&\quad= \sum_{s=1}^{S} \mathbf{R}^s\left(\boldsymbol{\Lambda}+\lambda \boldsymbol{I}\right)^{-1}(\mathbf{R}^s)^\top.
\end{align*}
\noindent Using notation $\mathbf{H} = \frac{\mathbf{R}^\top \mathbf{R}}{S}$, we have

\begin{align*}
&S\mathrm{tr}(\mathbf{H}\left(\boldsymbol{\Lambda}+\lambda \boldsymbol{I}\right)^{-1}) \\
&\quad=\mathrm{tr}(S\mathbf{H}\left(\boldsymbol{\Lambda}+\lambda \boldsymbol{I}\right)^{-1}) \\
&\quad=\mathrm{tr}(\mathbf{R}^\top\mathbf{R}\left(\boldsymbol{\Lambda}+\lambda \boldsymbol{I}\right)^{-1}) \\
&\quad= \sum_{s=1}^{S} \mathbf{R}^s\left(\boldsymbol{\Lambda}+\lambda \boldsymbol{I}\right)^{-1}(\mathbf{R}^s)^\top.
\end{align*}

\noindent Therefore, the joint posterior of $G, \boldsymbol{\rho}$ given data  $\boldsymbol{R}$ can  be represented as
\begin{align*}
&\exp\left\{-\frac{S}{2}\log \left(\det(\boldsymbol{\Lambda}+\lambda \boldsymbol{I})\right) \right. \\
&\qquad \left. - \frac{S}{2}\mathrm{tr}\left(\mathbf{H}\left(\boldsymbol{\Lambda}+\lambda \boldsymbol{I}\right)^{-1}\right)\right\} p(G)p(\boldsymbol{\rho}).
\end{align*}

\newpage
\section*{Appendix 2 }\label{append2}

\noindent {
\large{\bf{Derivation of the posterior distribution for  $\mathbf{\omega}$ and $\boldsymbol{\rho}$ }}}\\

\noindent The posterior distribution for the cluster assignment $\omega_i$ is proportional to the product of the likelihood and the prior probability,
\begin{align*}
&p(\omega_i=C_k \mid \mathbf{\omega}_{-i}, \boldsymbol{\rho}, \mathbf{R}) \\
&\propto p(\mathbf{R} \mid \omega_i=C_k, \mathbf{\omega}_{-i}, \boldsymbol{\rho})p(\omega_i=C_k)\\
&\propto p(\mathbf{R} \mid \mathbf{\Lambda})p(\omega_i=C_k).
\end{align*}
Here, we employ a prior derived from the degenerated Chinese Restaurant Process (CRP)~\cite{pitman2006combinatorial}. The prior probability that item $i$ is assigned to cluster $C_k$ is defined as $p(\omega_i=C_k)=\frac{m_{-ik}}{V-1}$, where $m_{-ik}$ denotes the number of items in cluster $C_k$, excluding item $i$.  Combining the prior and the likelihood yields the posterior distribution of $\omega_i$  conditional on all other parameters:

\begin{align*}
& p(\omega_i=C_k \mid \mathbf{\omega}_{-i}, \boldsymbol{\rho}, \mathbf{R}) \\
& \propto \exp\left\{ -\frac{S}{2}\log\left(\det\left(\mathbf{\Lambda}+\lambda \mathbf{I}\right)\right) \right. \\
& \phantom{\propto} \qquad \left. -\frac{S}{2}\mathrm{tr}\left(\mathbf{H}\left(\mathbf{\Lambda}+\lambda \mathbf{I}\right)^{-1}\right)\right\} \times \frac{m_{-ik}}{V-1}.
\end{align*}

\noindent The posterior distribution of $\rho_k$ ($k=0,1,\ldots,K$) is derived in an analogous manner.   
We can assume that the prior of  $\rho_k$ is a normal distribution with hyperparameters $\mu_k$ and $\tau_k^2$.   The posterior for $\rho_k$ is thus proportional to the product of the likelihood and the prior:
\begin{align*}
p(\rho_k\mid \boldsymbol{\omega},\mathbf{H},\boldsymbol{\rho}_{-k}) &\propto \exp\left\{ -\frac{S}{2}\log\left(\det\left(\mathbf{\Lambda}+\lambda \mathbf{I}\right)\right) \right. \\
& \phantom{\propto} \qquad \left. -\frac{S}{2}\mathrm{tr}\left(\mathbf{H}\left(\mathbf{\Lambda}+\lambda \mathbf{I}\right)^{-1}\right)\right\} \times p(\rho_k). 
\end{align*} 

\newpage
\section*{Appendix 3}\label{append3}
\begin{figure}[htbp]
    
    \centering
    \includegraphics[width=0.8\textwidth]{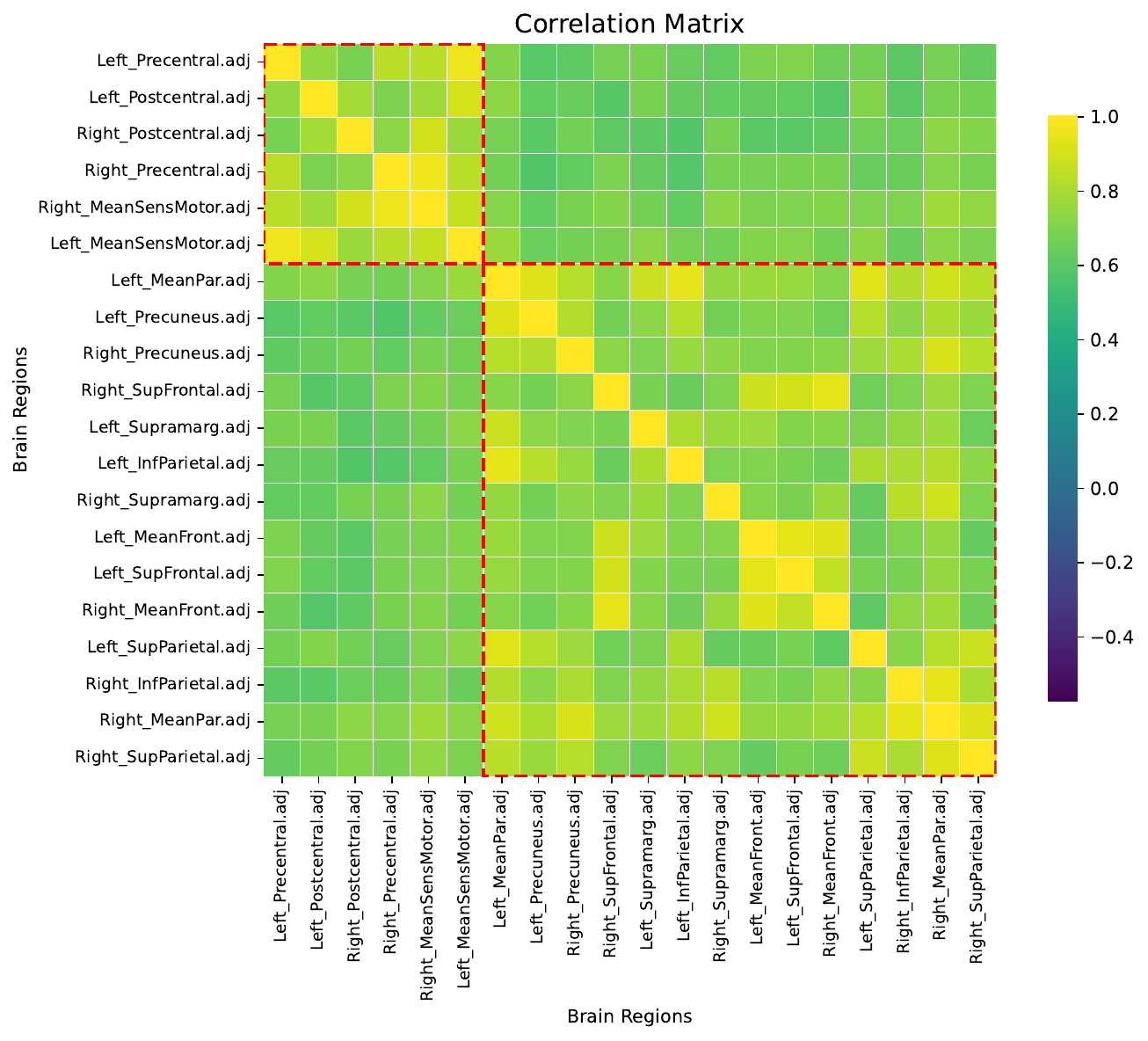}
    \caption{Covariance matrix of the 20 
selected ROIs. K-means clustering grouped the 20 ROIs into two major clusters, as highlighted by the red dashed boxes.}
    \label{fig:kmeans_matrix}
\end{figure}

\section*{Appendix 4}\label{append4}
To evaluate the feasibility and parameter estimation accuracy of the model for large-scale brain networks, we conducted simulations on networks with $V=100$ nodes.

When $V=100$, the network contains $4,950$ undirected edges, and the dimension of the covariance matrix to be estimated is $4950 \times 4950$. To construct the simulated data, the brain regions were uniformly divided into $10$ sub-networks, each containing $10$ nodes. The correlation strength parameters were sequentially set to $\exp(-\boldsymbol{\rho}) = [0.001, 0.1, 0.2, 0.3, 0.4, 0.5, 0.6, 0.7, 0.8, 0.9]$. The experiment investigated two different network structures, S1 and S2. Table \ref{tabel_V100} details the specific class allocation vectors $\boldsymbol{\omega}$ of the brain region nodes under the different network structures.

\begingroup
\renewcommand{\tabularxcolumn}[1]{m{#1}}
\begin{table}[htbp]
  \centering
  \captionsetup{justification=centering}
  \caption{Node allocation vectors when $V=100$ under different network structures.}
  \label{tabel_V100}
  \renewcommand{\arraystretch}{1.5} 
  \small  
  \begin{tabularx}{\textwidth}{>{\centering\arraybackslash}m{2cm} >{\centering\arraybackslash}m{2cm} >{\raggedright\arraybackslash}X} 
    \toprule
    \textbf{V} & \textbf{Structure} & \textbf{Node allocation vector} $\boldsymbol{\omega}$ \\
    \midrule
    \multirow{5}{*}{100} 
      & S1 & $(1, 1, 1, 1, 1, 1, 1, 1, 1, 1, 2, 2, 2, 2, 2, 2, 2, 2, 2, 2, $ \newline 
             $3, 3, 3, 3, 3, 3, 3, 3, 3, 3, 4, 4, 4, 4, 4, 4, 4, 4, 4, 4, $ \newline 
             $5, 5, 5, 5, 5, 5, 5, 5, 5, 5, 6, 6, 6, 6, 6, 6, 6, 6, 6, 6, $ \newline 
             $7, 7, 7, 7, 7, 7, 7, 7, 7, 7, 8, 8, 8, 8, 8, 8, 8, 8, 8, 8, $ \newline 
             $9, 9, 9, 9, 9, 9, 9, 9, 9, 9, 10, 10, 10, 10, 10, 10, 10, 10, 10, 10)$ \\
      \cmidrule{2-3}
      & S2 & $(5, 8, 7, 7, 1, 10, 7, 2, 1, 1, 10, 9, 6, 7, 8, 10, 6, 1, 8, 3, $ \newline
             $3, 5, 5, 5, 10, 3, 1, 2, 4, 5, 8, 10, 2, 9, 4, 3, 1, 6, 10, 6, $ \newline
             $9, 5, 7, 8, 4, 10, 1, 2, 3, 4, 9, 6, 1, 6, 2, 3, 9, 9, 4, 1, $ \newline
             $6, 7, 4, 6, 9, 7, 6, 4, 8, 6, 5, 4, 5, 2, 3, 8, 4, 10, 5, 8, $ \newline
             $2, 3, 1, 7, 2, 9, 3, 5, 8, 2, 3, 4, 2, 9, 7, 8, 10, 9, 10, 7)$ \\
    \bottomrule
  \end{tabularx}
\end{table}
\endgroup

The number of iterations for the MH algorithm was set to $2,500$. The initial values of the parameters $\boldsymbol{\rho}$ were uniformly drawn from the interval $[0, 0.5]$, and a random walk normal distribution with a step size of $0.01$ was used as the proposal distribution. For both S1 and S2 structures, the initial number of classes was preset to $K=12$.

The experimental results indicate that under both S1 and S2 structures, although the initially specified number of classes was $12$, during the iteration of the Markov chain, the two redundant classes eventually became empty because no nodes were assigned to them. The model successfully converged to the true number of classes, $10$.

For the S1 structure, the estimated strength parameters are as follows:
\begin{equation*}
\begin{aligned}
\exp(-\hat{\boldsymbol{\rho}}) = [&0.0010, 0.0979, 0.1877, 0.2869, 0.4690, \\
                                  &0.5541, 0.6744, 0.7131, 0.7798, 0.8908].
\end{aligned}
\end{equation*}

For the S2 structure, the corresponding estimated parameter values are:
\begin{equation*}
\begin{aligned}
\exp(-\hat{\boldsymbol{\rho}}) = [&0.0010, 0.0979, 0.1868, 0.2876, 0.4676, \\
                                  &0.5590, 0.6742, 0.7086, 0.7758, 0.8917].
\end{aligned}
\end{equation*}

These estimated values are highly consistent with the true parameter settings, and the inferred network topology is exactly the same as the actual preset structure. This indicates that the model can accurately identify the latent brain network structure even in high-dimensional scenarios. This experiment was conducted on a 13th Gen Intel(R) Core(TM) i7-13650HX (2.60 GHz) processor. For a total of $2,500$ iterations, the running time for the S1 structure was $86.1$ hours, and for the S2 structure, it was $87.9$ hours.

\begin{figure}[htbp]
    \centering
    % Row 1
    \begin{subfigure}[b]{0.48\textwidth}
        \centering
        \includegraphics[width=\textwidth]{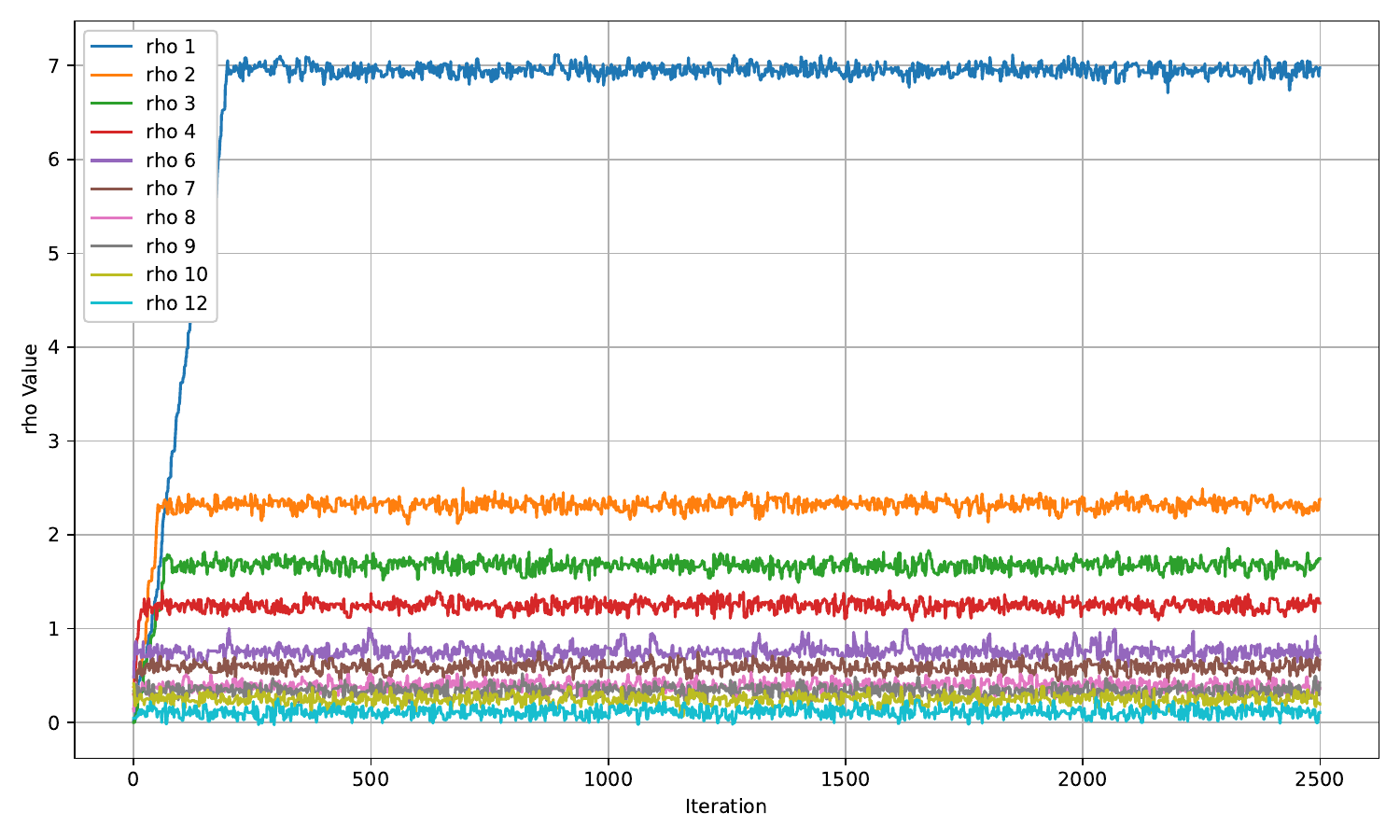}
        \caption{Trace plots}
    \end{subfigure}
    \hfill
    \begin{subfigure}[b]{0.48\textwidth}
        \centering
        \includegraphics[width=\textwidth]{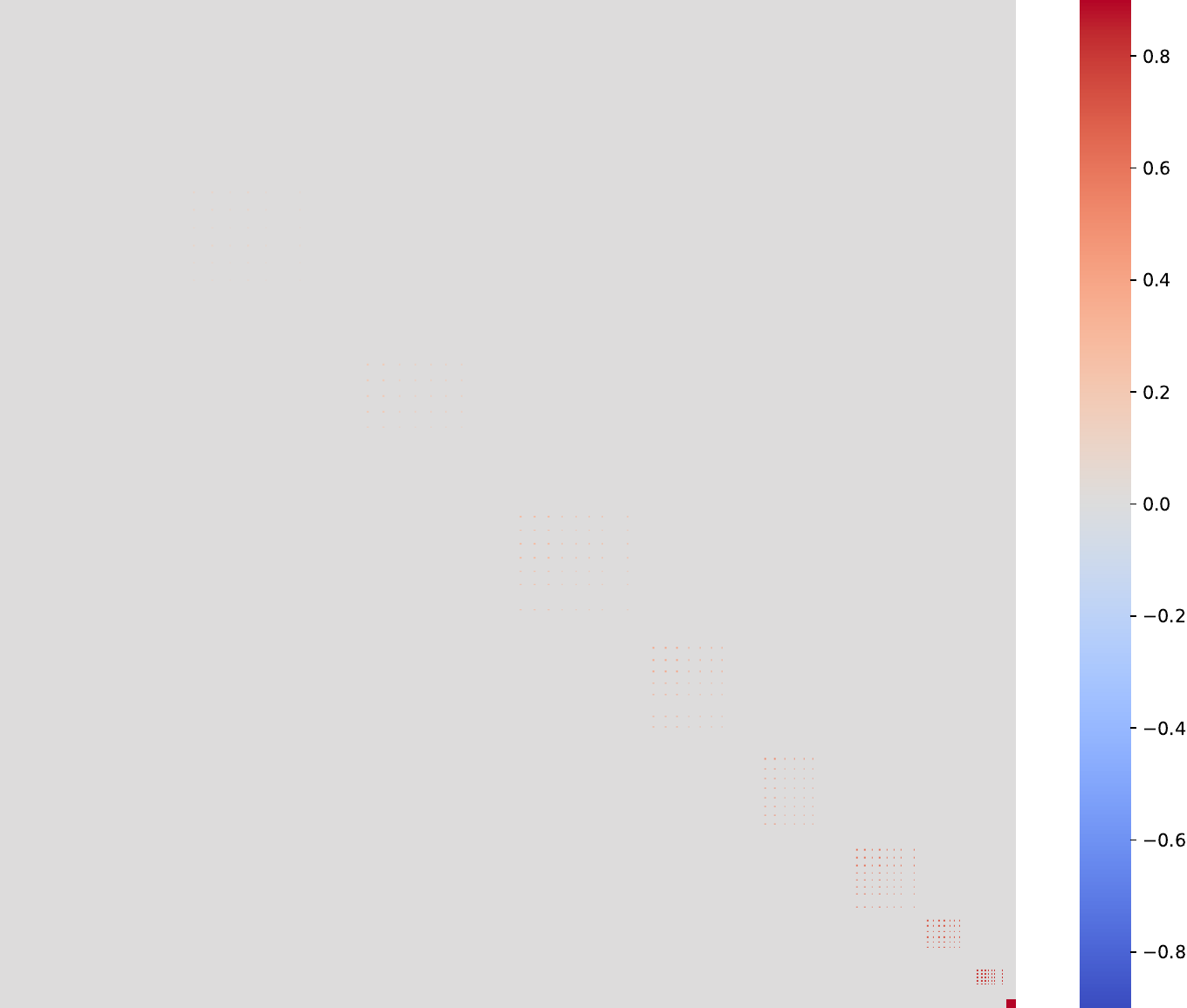}
        \caption{$\boldsymbol{\Lambda}$}
    \end{subfigure}

    \vspace{1em} 

    % Row 2
    \begin{subfigure}[b]{0.48\textwidth}
        \centering
        \includegraphics[width=\textwidth]{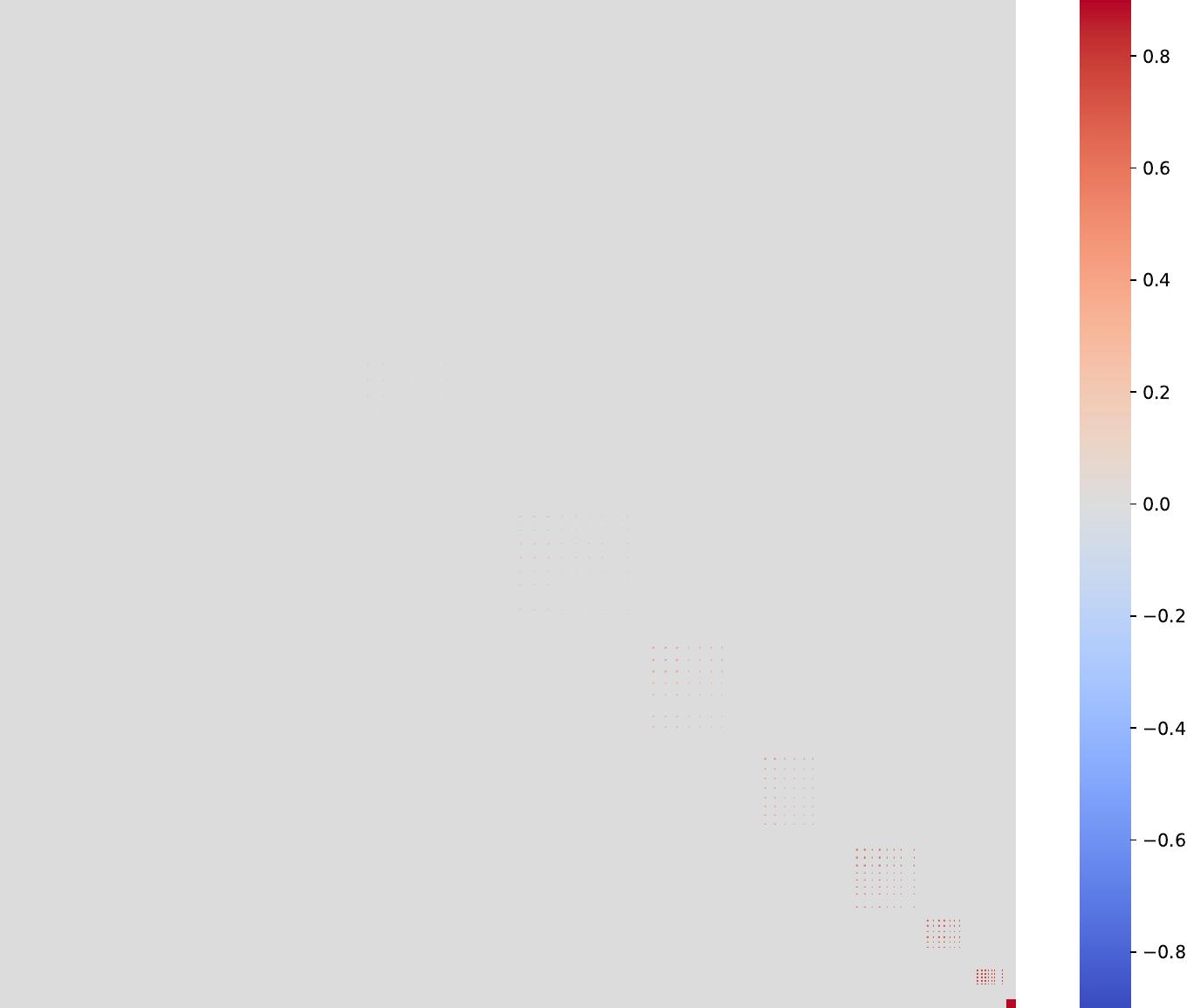}
        \caption{$\hat{\boldsymbol{\Lambda}}$}
    \end{subfigure}
    \hfill
    \begin{subfigure}[b]{0.48\textwidth}
        \centering
        \includegraphics[width=\textwidth]{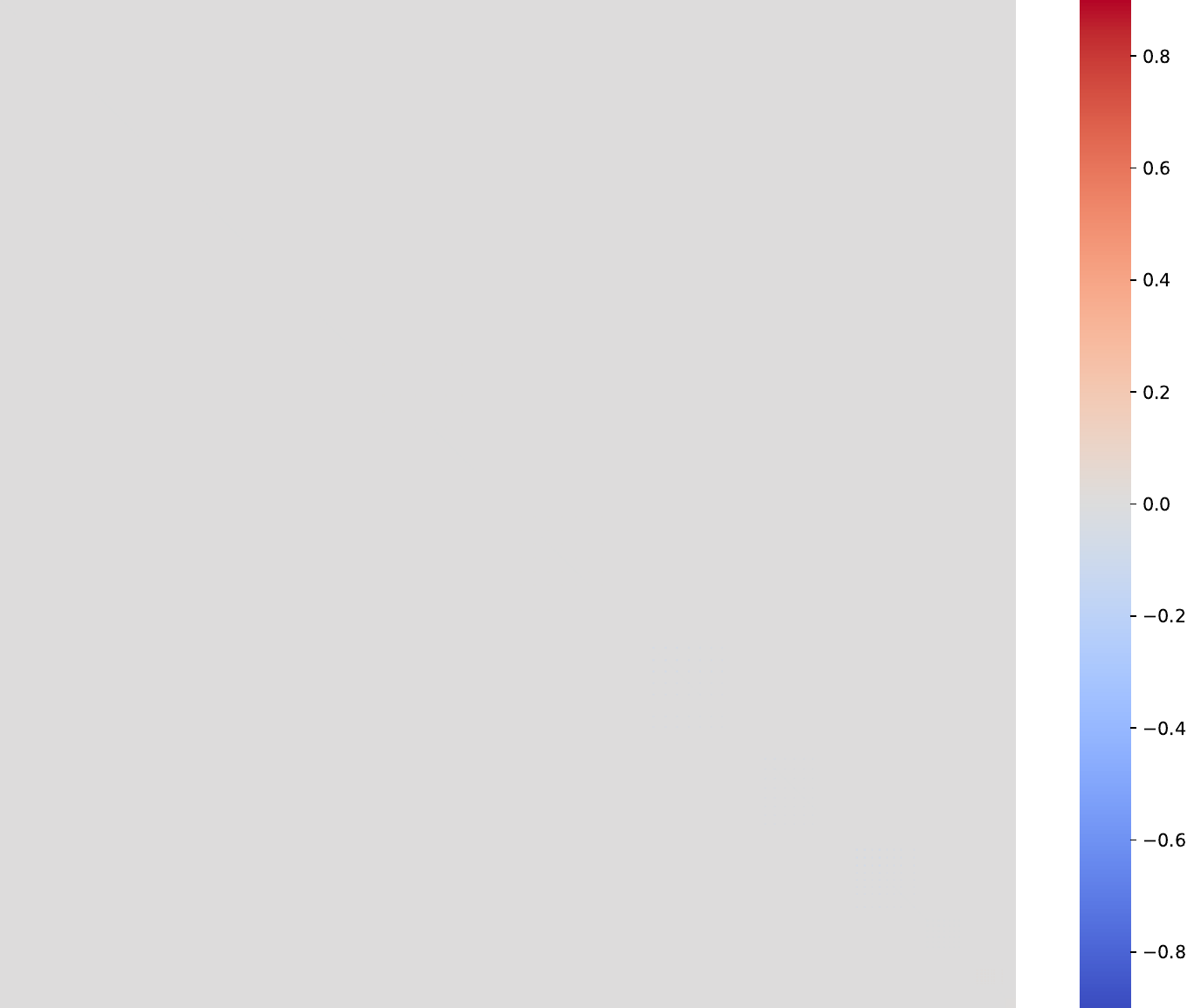}
        \caption{$\boldsymbol{\Lambda} - \hat{\boldsymbol{\Lambda}}$}
    \end{subfigure}

    \caption{Visualization of model estimation and convergence for the S1 structure in the $V=100$ simulation experiment. (a) Trace plots for convergence diagnostics; (b) True connectivity matrix ($\boldsymbol{\Lambda}$); (c) Estimated connectivity matrix ($\hat{\boldsymbol{\Lambda}}$); (d) Estimation error matrix ($\boldsymbol{\Lambda} - \hat{\boldsymbol{\Lambda}}$).}
    \label{fig:V=100_S1}
\end{figure}

\begin{figure}[htbp]
    \centering
    % Row 1
    \begin{subfigure}[b]{0.48\textwidth}
        \centering
        \includegraphics[width=\textwidth]{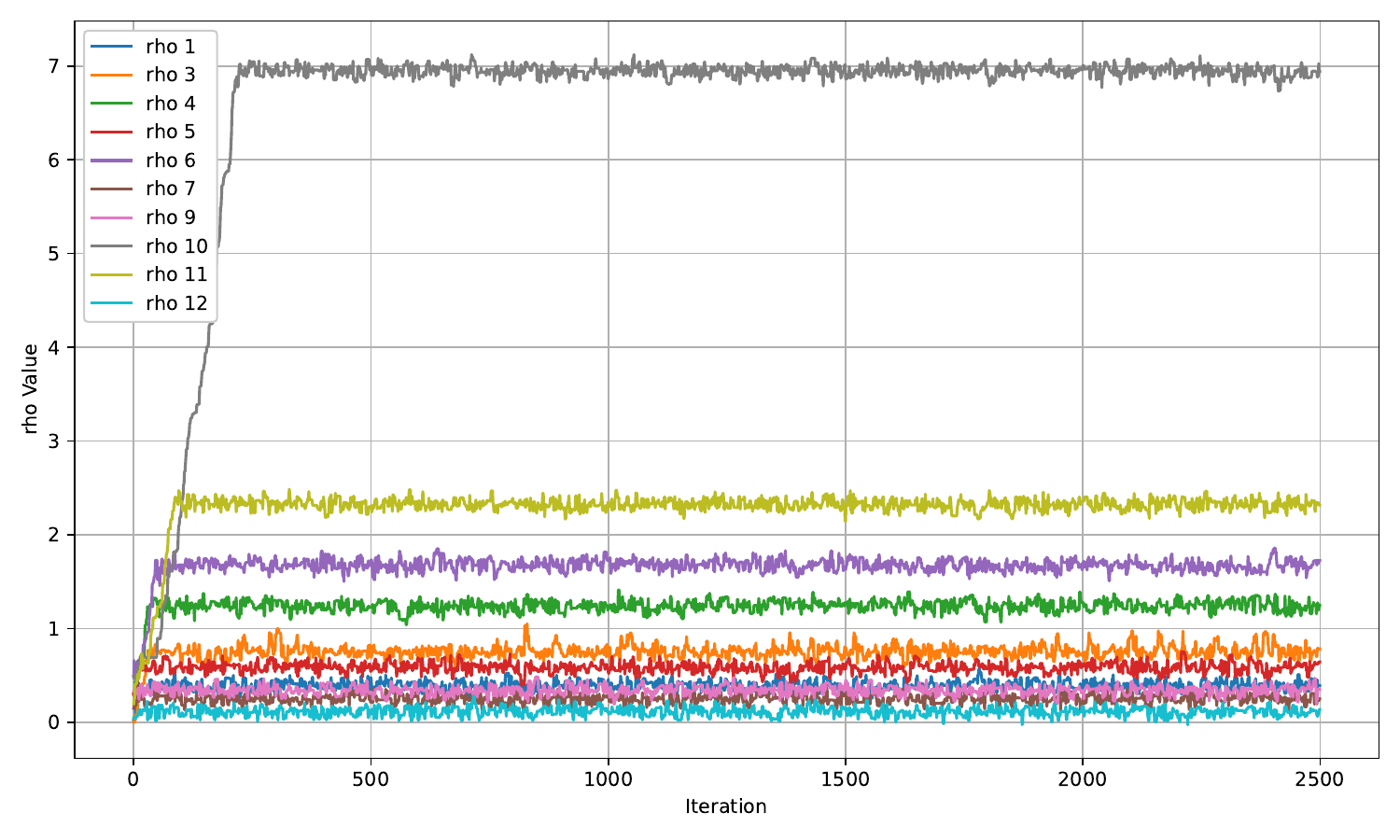}
        \caption{Trace plots}
    \end{subfigure}
    \hfill
    \begin{subfigure}[b]{0.48\textwidth}
        \centering
        \includegraphics[width=\textwidth]{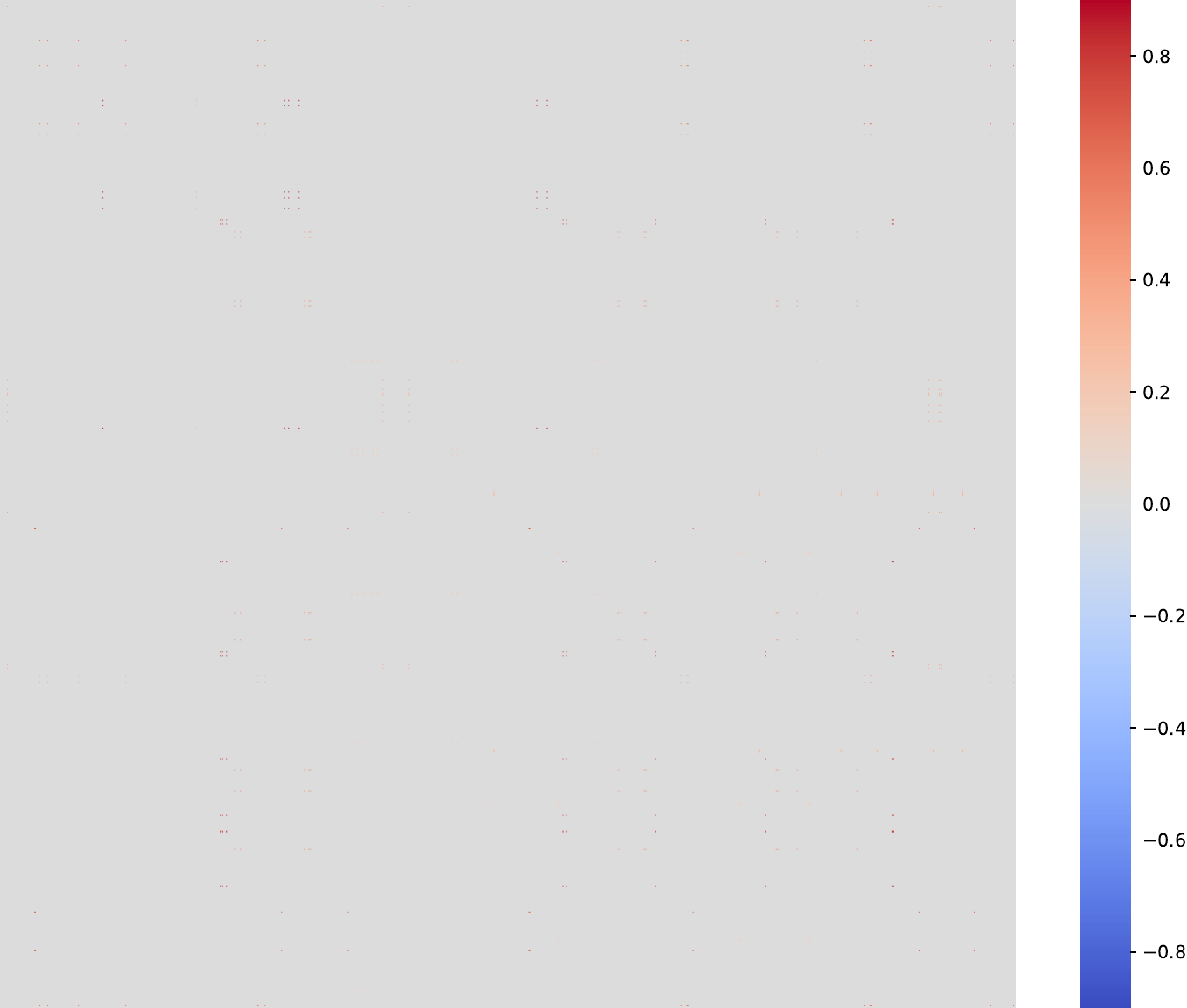}
        \caption{$\boldsymbol{\Lambda}$}
    \end{subfigure}

    \vspace{1em} 

    % Row 2
    \begin{subfigure}[b]{0.48\textwidth}
        \centering
        \includegraphics[width=\textwidth]{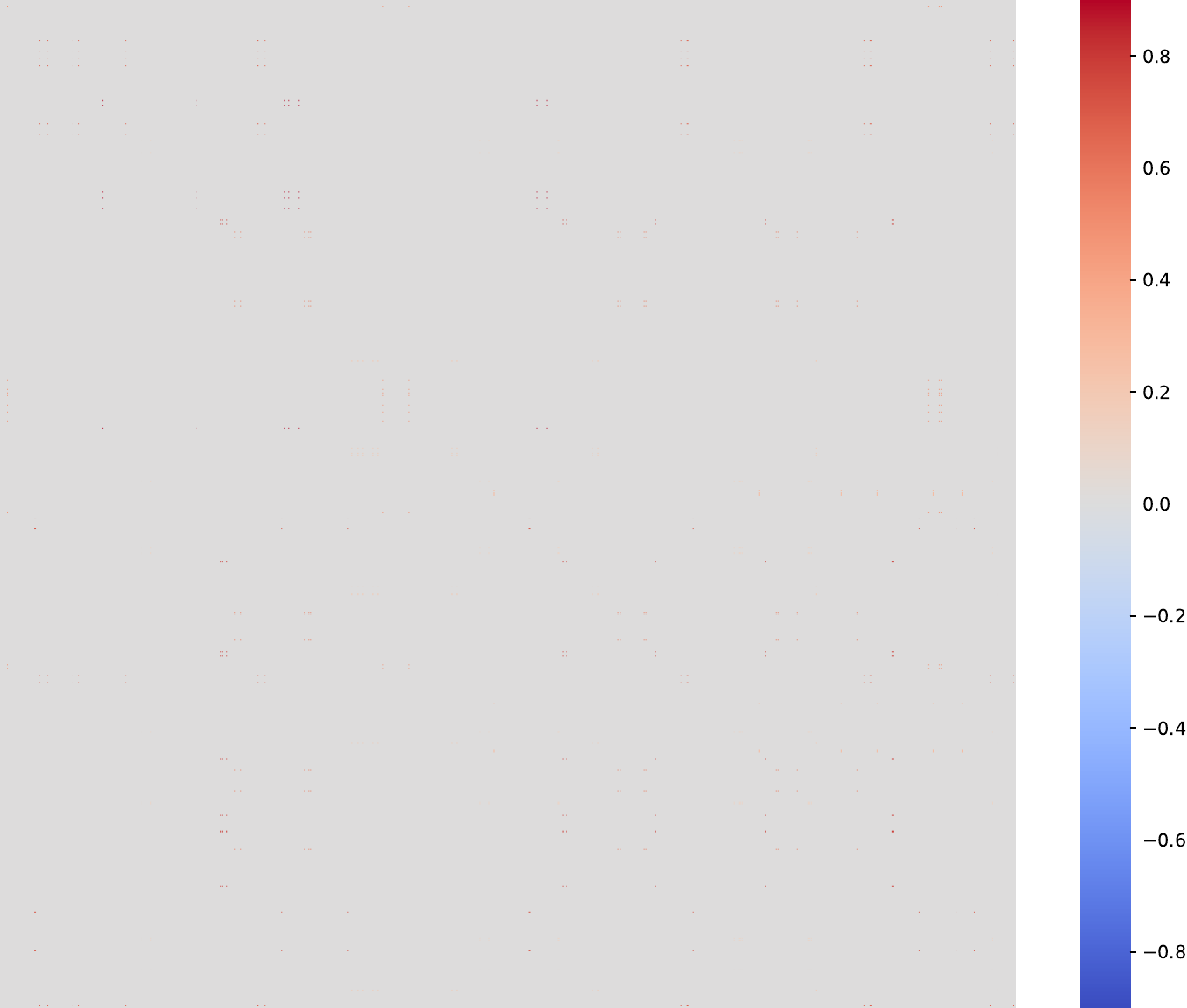}
        \caption{$\hat{\boldsymbol{\Lambda}}$}
    \end{subfigure}
    \hfill
    \begin{subfigure}[b]{0.48\textwidth}
        \centering
        \includegraphics[width=\textwidth]{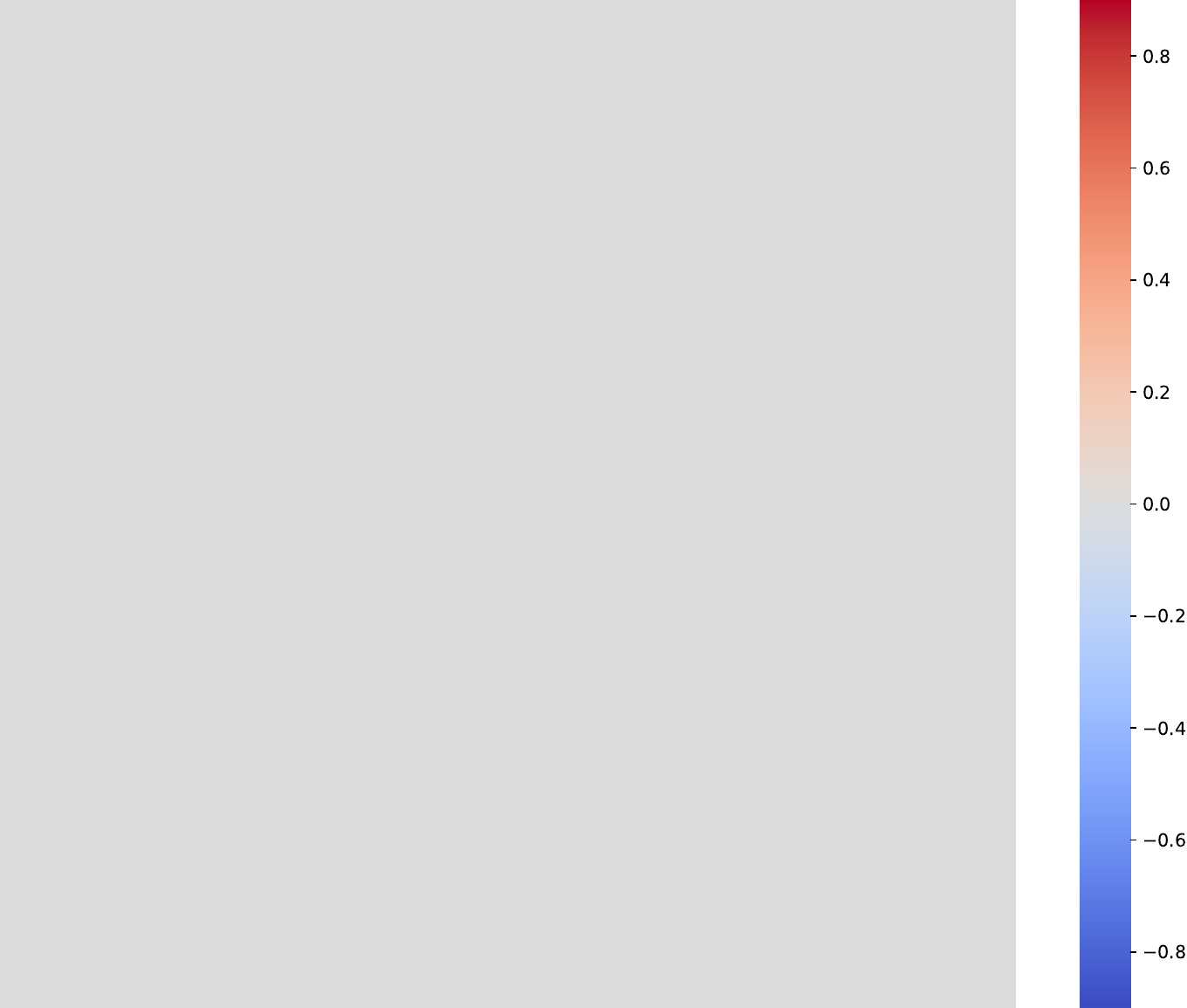}
        \caption{$\boldsymbol{\Lambda} - \hat{\boldsymbol{\Lambda}}$}
    \end{subfigure}

    \caption{Visualization of model estimation and convergence for the S2 structure in the $V=100$ simulation experiment. (a) Trace plots for convergence diagnostics; (b) True connectivity matrix ($\boldsymbol{\Lambda}$); (c) Estimated connectivity matrix ($\hat{\boldsymbol{\Lambda}}$); (d) Estimation error matrix ($\boldsymbol{\Lambda} - \hat{\boldsymbol{\Lambda}}$).}
    \label{fig:V=100_S2}
\end{figure}

Overall, the model maintains good estimation accuracy on large-scale networks. The computational optimization proposed in Section 3.2 effectively reduces the computational burden of whole-brain scale analysis, making it possible for the model to handle large-scale networks. The quantitative metrics in Table \ref{tab:result_V100} further demonstrate the model's accuracy in clustering and parameter estimation. The convergence diagnostics and matrix visualization results in Figure \ref{fig:V=100_S1} and Figure \ref{fig:V=100_S2} visually show the convergence state of the model and the accurate restoration of the true connectivity matrix.

\begin{table}[htbp]
\centering
\caption{Estimation accuracy under various metrics for large-scale networks when $V=100$.}
\label{tab:result_V100}
\begin{tabular}{p{3cm}>{\centering\arraybackslash}p{4.5cm}>{\centering\arraybackslash}p{4.5cm}}
\toprule
& \multicolumn{2}{c}{$V=100$} \\
\cmidrule(lr){2-3}
& \multicolumn{2}{c}{$G=(10,10,10,10,10,10,10,10,10,10)$} \\
& \multicolumn{2}{c}{$\exp(-\boldsymbol{\rho})=[0.001, 0.1, 0.2, 0.3, 0.4, 0.5, 0.6, 0.7, 0.8, 0.9]$} \\
\cmidrule(lr){2-3} 
& S1 & S2 \\
\midrule
normalized error & 0.00002  & 0.00002  \\
ARI              & 1.00     & 1.00     \\
sensitivity      & 100\%    & 100\%    \\
specificity      & 100\%    & 100\%    \\
Frobenius norm   & 0.00108  & 0.00109  \\
\bottomrule
\end{tabular}
\end{table}

\section*{Appendix 5}\label{append5}
This section examines whether the sequence of parameter sampling influences the model's estimation results. As described in Section 3.1, the inference process utilizes an alternating block-update strategy, where cluster labels $\omega_i$ are updated first via the MH algorithm, followed by the correlation parameters $\rho_k$ given the current assignments. 

To evaluate the effect of this sequence, we conducted a swap experiment where the correlation parameters $\rho_k$ are updated before the node labels $\omega_i$. The experiment was conducted using the S1 and S2 topological structures with $V=20$. In the original sequence, $K=6$ was used for both structures. In the swapped sequence, $K=6$ was maintained for S1, while S2 was adjusted to $K=7$. A comparison of the results is presented in Table~\ref{tab:comparison_order_V20}.

\begin{table}[htbp]
\centering
\caption{Comparison of estimation accuracy and convergence diagnostics for the BBeC model under different parameter update orders ($V=20$).}
\label{tab:comparison_order_V20}
\small
\begin{tabular*}{\linewidth}{@{\extracolsep{\fill}} l c c c c}
\toprule
\multicolumn{5}{c}{$V=20,~ G=(4,4,4,4,4),~ \exp(-\boldsymbol{\rho})=[0.2,0.4,0.6,0.8,0.9]$} \\
%\multicolumn{5}{c}{$\exp(-\boldsymbol{\rho})=[0.2,0.4,0.6,0.8,0.9], \quad G=(4,4,4,4,4)$} \\
\midrule
& \multicolumn{2}{c}{S1} & \multicolumn{2}{c}{S2} \\
\cmidrule(lr){2-3} \cmidrule(lr){4-5}
 & Original Order & Swapped Order & Original Order & Swapped Order \\
\midrule
normalized error                  & 0.0002  & 0.0002 & 0.0002  & 0.0003 \\
ARI                               & 1.00    & 1.00   & 1.00    & 1.00   \\
sensitivity                       & 100\%   & 100\%  & 100\%   & 100\%  \\
specificity                       & 100\%   & 100\%  & 100\%   & 100\%  \\
Frobenius Norm                    & 0.00251 & 0.00253 & 0.00256 & 0.00240 \\
Max $\hat{R}$ for $\boldsymbol{\omega}$ & 1.0000  & 1.0000 & 1.0000  & 1.0000 \\
Max $\hat{R}$ for $\boldsymbol{\rho}$   & 1.0120  & 1.0002 & 1.0248  & 1.0032 \\
\bottomrule
\end{tabular*}
\end{table}

As shown in Table~\ref{tab:comparison_order_V20}, with a suitable initial $K$, the sampling order has a negligible impact on the inference results, and consistent conclusions were obtained in both cases. Specifically, the model accurately recovered the topological structures of S1 and S2 regardless of the sequence. Regarding parameter estimation, the normalized error and Frobenius norm remained at very low levels with minimal differences between the two sequences, indicating high precision. Furthermore, convergence diagnostics showed that the maximum $\hat{R}$ values for all parameters were strictly below $1.05$, signifying that the model reaches a stable state under both update paths. Overall, the model demonstrates strong robustness to the choice of parameter sampling order.

To further examine whether the same conclusion holds for a larger network, we conducted an additional simulation study with $V=40$. The data-generation setting was kept the same as previously described, with
$$
G=(5,5,5,5,5,5,5,5),
$$
and
$$
\exp(-\boldsymbol{\rho})=[0.001,0.1,0.2,0.3,0.4,0.5,0.6,0.7].
$$
Both the ordered topology S1 and the randomized topology S2 were considered. In the original update order, the class labels $\boldsymbol{\omega}$ were updated before the correlation parameters $\boldsymbol{\rho}$. In the reversed update order, the correlation parameters $\boldsymbol{\rho}$ were updated before the class labels $\boldsymbol{\omega}$. For the original order, $K$ was set to $9$ for S1 and $10$ for S2. For the reversed order, $K$ was set to $10$ for both S1 and S2. The total number of MCMC iterations was $2500$. The results are summarized in Table~\ref{tab:comparison_order_V40}.

\begin{table}[htbp]
\centering
\caption{Comparison of estimation accuracy and convergence diagnostics under different parameter update orders for $V=40$. ``Original'' updates labels $\boldsymbol{\omega}$ before correlation parameters $\boldsymbol{\rho}$, while ``Reversed'' updates $\boldsymbol{\rho}$ before $\boldsymbol{\omega}$.}
\label{tab:comparison_order_V40}
\small
\begin{tabular*}{\linewidth}{@{\extracolsep{\fill}} l c c c c}
\toprule
\multicolumn{5}{c}{$V=40,~G=(5,5,5,5,5,5,5,5), ~ \exp(-\boldsymbol{\rho})=[0.001,0.1,0.2,0.3,0.4,0.5,0.6,0.7]$} \\
%\multicolumn{5}{c}{$\exp(-\boldsymbol{\rho})=[0.001,0.1,0.2,0.3,0.4,0.5,0.6,0.7]$} \\
%\multicolumn{5}{c}{$G=(5,5,5,5,5,5,5,5)$} \\
\midrule
& \multicolumn{2}{c}{S1} & \multicolumn{2}{c}{S2} \\
\cmidrule(lr){2-3} \cmidrule(lr){4-5}
& Original & Reversed & Original & Reversed \\
\midrule
normalized error                       & 0.00004 & 0.00003 & 0.00004 & 0.00003 \\
ARI                                    & 1.00    & 1.00    & 1.00    & 1.00    \\
sensitivity                            & 100\%   & 100\%   & 100\%   & 100\%   \\
specificity                            & 100\%   & 100\%   & 100\%   & 100\%   \\
Frobenius norm                         & 0.00164 & 0.00167 & 0.00167 & 0.00166 \\
Max $\hat{R}$ for $\boldsymbol{\omega}$ & 1.0000  & 1.0000 & 1.0000  & 1.0000 \\
Max $\hat{R}$ for $\boldsymbol{\rho}$ & 1.0200  & 1.0071  & 1.0155  & 1.0038  \\
\bottomrule
\end{tabular*}
\end{table}

As shown in Table~\ref{tab:comparison_order_V40}, the two update orders led to very similar results. The topological structure was correctly recovered in all settings, with both sensitivity and specificity equal to $100\%$. The normalized errors and Frobenius norms were also close across the two update orders. In addition, the maximum $\hat{R}$ values for the correlation parameters $\boldsymbol{\rho}$ were below $1.05$ in all settings, indicating stable convergence of the continuous parameters under both update orders.

\section*{Appendix 6}\label{append6}
%To specifically quantify the impact of model misspecification on performance, multivariate t-distributed noise and skew-normal distribution noise were introduced in this section.

To quantify the specific impact of model misspecification on performance, this section introduces multivariate t-distributed noise and skew-normal distribution noise.

First, the impact of multivariate $t$-distribution noise on network estimation accuracy was explored. The $t$-distribution is commonly used to model data with heavy-tailed characteristics. In this simulation experiment, the mixture proportion of outlier samples was denoted as $\eta$. For a dataset of size $S$, $(1-\eta)S$ samples were generated from a multivariate normal distribution, while the remaining $\eta S$ samples served as heavy-tailed noise, generated from a multivariate $t$-distribution with $\nu$ degrees of freedom. The specific data generation mechanism was as follows. Normal samples were sampled directly from a standard multivariate normal distribution, denoted as $D_N$. For the $t$-distribution noise samples, denoted as $D_t$, the generating process was achieved by scaling multivariate normal variables with a chi-square distribution. Specifically, normally distributed variables $\mathbf{y}_j$ with the same covariance structure were first generated, along with an independent random variable $u_j \sim \chi^2(\nu)$ following a chi-square distribution. Then, a scaling factor was calculated using $u_j$ to weight $\mathbf{y}_j$, resulting in $t$-distributed outlier samples with a mean of $\mathbf{0}$:
$$ \mathbf{x}_j = \mathbf{y}_j \sqrt{\frac{\nu}{u_j}}. $$
After generating the two sets of samples separately (i.e., $D_N, D_t$), they were vertically concatenated row-wise and randomly shuffled to ensure that the $t$-distributed samples were evenly distributed across the entire dataset.

This experiment utilized the S1 and S2 topological structures with $V=20$ and $V=40$. To demonstrate performance under different noise proportions, 5\%, 10\%, and 20\% $t$-distribution noise were introduced into the $V=20$ network ($E=190$). For a higher-dimensional $V=40$ network ($E=780$), given its relatively complex topology, only a 10\% noise level was tested due to computational constraints. Throughout the simulations, the degrees of freedom for generating $t$-distribution noise were uniformly set to $\nu = 4$. During the inference phase, the predefined number of categories for the $V=20$ and $V=40$ networks were initialized to $K=7$ and $K=11$, respectively. The network estimation results under the respective proportions of $t$-distribution noise are shown in Table~\ref{tab:comparison_noise_V20} and Table~\ref{tab:comparison_noise_V40}.

\begin{table}[htbp] 
\centering 
\caption{Network estimation accuracy of the BBeC model under different proportions of $t$-distribution noise when $V=20$.} 
\label{tab:comparison_noise_V20} 
\begin{tabular*}{\linewidth}{@{\extracolsep{\fill}} l c c c c c c} 
\toprule 
\multicolumn{7}{c}{$V=20,~G=(4,4,4,4,4),~ \exp(-\boldsymbol{\rho})=[0.2, 0.4, 0.6, 0.8, 0.9]$} \\ 
%\multicolumn{7}{c}{$G=(4,4,4,4,4), \quad \exp(-\boldsymbol{\rho})=[0.2, 0.4, 0.6, 0.8, 0.9]$} \\ 
\midrule 
& \multicolumn{3}{c}{S1} & \multicolumn{3}{c}{S2} \\ 
\cmidrule(lr){2-4} \cmidrule(lr){5-7} 
& 5\%  & 10\%  & 20\%  & 5\%  & 10\%  & 20\%  \\ 
\midrule 
normalized error       & 0.0004 & 0.0003 & 0.0008 & 0.0004 &  0.0003 & 0.0008 \\ 
ARI                    & 1.00 & 1.00 & 1.00 & 1.00 & 1.00 & 1.00 \\ 
sensitivity            & 100\% & 100\% & 100\% & 100\% & 100\% & 100\% \\ 
specificity            & 100\% & 100\% & 100\% & 100\% & 100\% & 100\% \\ 
Frobenius norm         & 0.00710 & 0.00609 & 0.01373 & 0.00715 & 0.00610 & 0.01376 \\ 
Max $\hat{R}$ for $\boldsymbol{\omega}$ & 1.0000 & 1.0000 & 1.0000 & 1.0000 & 1.0000 & 1.0000 \\ 
Max $\hat{R}$ for $\boldsymbol{\rho}$   & 1.0065 & 1.0066 & 1.0034 & 1.0088 & 1.0056 & 1.0084 \\ 
\bottomrule 
\end{tabular*} 
\end{table} 

\begin{table}[htbp] 
\centering 
\caption{Network estimation accuracy of the BBeC model under 10\% $t$-distribution noise when $V=40$.} 
\label{tab:comparison_noise_V40} 
\begin{tabular*}{\linewidth}{@{\extracolsep{\fill}} l c c } 
\toprule 
\multicolumn{3}{c}{$V=40,~G=(5,5,5,5,5,5,5,5), ~\exp(-\boldsymbol{\rho})=[0.001, 0.1, 0.2, 0.3,0.4, 0.5, 0.6, 0.7]$} \\ 
%\multicolumn{3}{c}{$G=(5,5,5,5,5,5,5,5)$} \\ 
%\multicolumn{3}{c}{\begin{tabular}{@{}c@{}}$\exp(-\boldsymbol{\rho})=[0.001, 0.1, 0.2, 0.3,0.4, 0.5, 0.6, 0.7]$\end{tabular}} \\ 
\midrule 
& S1 & S2 \\ 
\midrule 
normalized error       & 0.00008 & 0.00006 \\ 
ARI                    & 1.00 & 1.00 \\ 
sensitivity            & 100\% & 100\% \\ 
specificity            & 100\% & 100\% \\ 
Frobenius norm         & 0.00251 & 0.00254 \\ 
Max $\hat{R}$ for $\boldsymbol{\omega}$ & 1.0000 & 1.0000 \\ 
Max $\hat{R}$ for $\boldsymbol{\rho}$   & 1.0083 & 1.0050 \\ 
\bottomrule 
\end{tabular*} 
\end{table} 

As shown in Tables~\ref{tab:comparison_noise_V20}-\ref{tab:comparison_noise_V40}, the heavy-tailed noise introduced by the $t$-distribution has a limited impact on the overall inference, and the model accurately recovered the network topology without node misclassification. Concurrently, parameter estimation and algorithm convergence remained stable, indicating that the model possesses good resistance to heavy-tailed noise characteristics.

In addition, simulation experiments under a skew-normal distribution were conducted to explore the impact of asymmetric noise on model estimation.

In this experiment, the skewed noise samples were generated based on the construction of a multivariate skew-normal distribution \cite{azzalini1996multivariate}. Denote the true covariance matrix as $\mathbf{\Sigma}$. The specific data generation steps were as follows:

 First, all elements of the skewness vector $\boldsymbol{\gamma}$, which controls the degree of skewness, were set to 4.0. The standard deviation diagonal matrix $\boldsymbol{\Omega}$ was extracted from $\mathbf{\Sigma}$, and the corresponding correlation coefficient matrix $\mathbf{C} = \boldsymbol{\Omega}^{-1} \mathbf{\Sigma} \boldsymbol{\Omega}^{-1}$ was calculated.

Next, the augmented vector $\boldsymbol{\delta}$ was calculated using $\boldsymbol{\gamma}$ and $\mathbf{C}$, which was then used to construct the augmented covariance matrix $\mathbf{\Sigma}^*$ with one additional dimension:
$$ \boldsymbol{\delta} = \frac{\mathbf{C} \boldsymbol{\gamma}}{\sqrt{1 + \boldsymbol{\gamma}^T \mathbf{C} \boldsymbol{\gamma}}}, \quad \mathbf{\Sigma}^* = \begin{pmatrix} 1 & \boldsymbol{\delta}^T \\ \boldsymbol{\delta} & \mathbf{C} \end{pmatrix}. $$

Subsequently, a joint normal random vector $U = (U_0, \mathbf{U}_1)^T \sim \text{MVN}(\mathbf{0}, \mathbf{\Sigma}^*)$ with zero mean was generated from this augmented distribution. The sign of the auxiliary variable $U_0$ determined the orientation of the remaining variables $\mathbf{U}_1$, introducing an asymmetric skewed feature into the data:
$$ \mathbf{Z} = \begin{cases} \mathbf{U}_1, & U_0 > 0 \\ -\mathbf{U}_1, & U_0 \le 0 .\end{cases} $$

Finally, the variable $\mathbf{Z}$ was scaled using the standard deviation matrix $\boldsymbol{\Omega}$ to obtain the skewed noise sample $\mathbf{x}_j = \boldsymbol{\Omega} \mathbf{Z}$. After generating both the normal samples and the skewed noise samples, they were vertically concatenated and randomly shuffled.

When evaluating the impact of skewed noise, the settings for the network topology remained consistent with the previous experiments. For $V=20$ data, 5\%, 10\%, and 20\% skewed noise were introduced; for the high-dimensional $V=40$ data, a 10\% noise proportion was tested. In the initialization phase, the predefined number of categories for the $V=20$ network was initialized to $K=7$. For the $V=40$ network, the predefined categories for S1 and S2 structures were set to $K=12$ and $K=13$, respectively. The corresponding quantitative estimation results are summarized in Tables~\ref{tab:comparison_skew_noise_V20}-\ref{tab:comparison_skew_noise_V40}.

\begin{table}[htbp] 
\centering 
\caption{Network estimation accuracy of the BBeC model under different proportions of skewed distribution noise when $V=20$.} 
\label{tab:comparison_skew_noise_V20} 
\begin{tabular*}{\linewidth}{@{\extracolsep{\fill}} l c c c c c c} 
\toprule 
\multicolumn{7}{c}{$V=20,~G=(4,4,4,4,4), ~ \exp(-\boldsymbol{\rho})=[0.2, 0.4, 0.6, 0.8, 0.9]$} \\ 
%\multicolumn{7}{c}{$G=(4,4,4,4,4), \quad \exp(-\boldsymbol{\rho})=[0.2, 0.4, 0.6, 0.8, 0.9]$} \\ 
\midrule 
& \multicolumn{3}{c}{S1} & \multicolumn{3}{c}{S2} \\ 
\cmidrule(lr){2-4} \cmidrule(lr){5-7} 
& 5\%  & 10\%  & 20\%  & 5\%  & 10\%  & 20\%  \\ 
\midrule 
normalized error       & 0.00009 & 0.00004 & 0.0001 & 0.0001 & 0.0001 & 0.0001 \\ 
ARI                    & 1.00 & 1.00 & 1.00 & 1.00 & 1.00 & 1.00 \\ 
sensitivity            & 100\% & 100\% & 100\% & 100\% & 100\% & 100\% \\ 
specificity            & 100\% & 100\% & 100\% & 100\% & 100\% & 100\% \\ 
Frobenius norm         & 0.00225 & 0.00180 & 0.00166 & 0.00216 & 0.00172 & 0.00236 \\ 
Max $\hat{R}$ for $\boldsymbol{\omega}$ & 1.0000 & 1.0000 & 1.0000 & 1.0000 & 1.0000 & 1.0000 \\ 
Max $\hat{R}$ for $\boldsymbol{\rho}$   & 1.0017 & 1.0031 & 1.0051 & 1.0091 & 1.0181 & 1.0181 \\ 
\bottomrule 
\end{tabular*} 
\end{table} 

\begin{table}[htbp] 
\centering 
\caption{Network estimation accuracy of the BBeC model under 10\% skewed distribution noise when $V=40$.} 
\label{tab:comparison_skew_noise_V40} 
\begin{tabular*}{\linewidth}{@{\extracolsep{\fill}} l c c } 
\toprule 
\multicolumn{3}{c}{$V=40,~ G=(5,5,5,5,5,5,5,5),~\exp(-\boldsymbol{\rho})=[0.001, 0.1, 0.2, 0.3, 0.4, 0.5, 0.6, 0.7]$} \\ 
%\multicolumn{3}{c}{$G=(5,5,5,5,5,5,5,5)$} \\ 
%\multicolumn{3}{c}{\begin{tabular}{@{}c@{}}$\exp(-\boldsymbol{\rho})=[0.001, 0.1, 0.2, 0.3, 0.4, 0.5, 0.6, 0.7]$\end{tabular}} \\ 
\midrule 
& S1  & S2  \\ 
\midrule 
normalized error       & 0.00005 & 0.00004 \\ 
ARI                    & 1.00 & 1.00 \\ 
sensitivity            & 100\% & 100\% \\ 
specificity            & 100\% & 100\% \\ 
Frobenius norm         & 0.00154 & 0.00158 \\ 
Max $\hat{R}$ for $\boldsymbol{\omega}$ & 1.0000 & 1.0000 \\ 
Max $\hat{R}$ for $\boldsymbol{\rho}$   & 1.0071 & 1.0078 \\ 
\bottomrule 
\end{tabular*} 
\end{table} 

The results in Tables~\ref{tab:comparison_skew_noise_V20} and \ref{tab:comparison_skew_noise_V40} indicate that skew-normal distribution noise has a minimal impact on estimation accuracy. As the proportion of skewed noise increased, the model consistently identified the network topology accurately, and the estimation errors for the correlation parameters did not exhibit obvious fluctuations. This indicates that the model maintains stability in model inference and parameter estimation, even when perturbed by a certain proportion of asymmetric abnormal samples.

\section*{Appendix 7}\label{append7}

\textbf{Selection of the 20 ROIs} In this applciation study, the specifici 20 ROIs were used rather than standard Alzheimer's disease relevant sets. The selection was fundamentally driven by the mathematical assumptions of our current model framework. As described in Section 2, the proposed model employs a simplified parameterization where each group must contain at least two nodes, all edges within a specific group share a single similarity coefficient, and all edges between different groups share a global inter-group similarity coefficient. 

\begin{figure}[htbp]
\centering
\includegraphics[width=\textwidth]{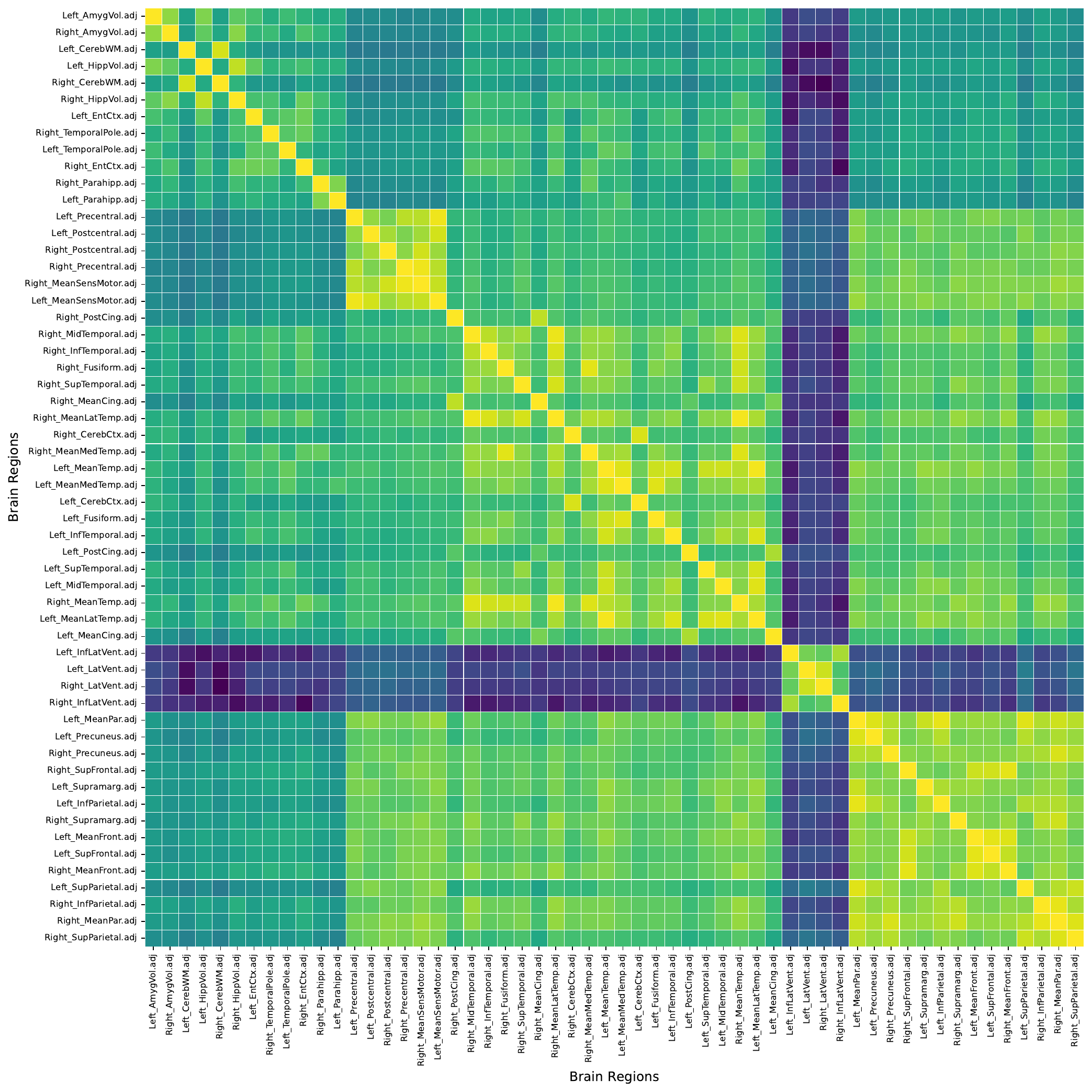}
\caption{Heatmap displaying the covariance matrix of the 56 ROIs. The arrangement of rows and columns reflects the group structure identified by K-means clustering.}
\label{fig:kmeans_pre}
\end{figure}

To ensure the input data satisfies these specific constraints, we performed a preliminary K-Means analysis on a broader set of 56 regions of interest from the ADNI dataset, as illustrated in Figure \ref{fig:kmeans_pre}. Based on the covariance matrix framework of the model, three core criteria were used for ROI selection. First, all correlation coefficients between communities (or functional modules) should be strictly positive. Second, the selected set of regions must exclude any isolated points—defined as nodes that fail to establish effective connections with any other region. Third, it is necessary to control the variation of correlations both within and between groups. Specifically, the variation in correlation coefficients among different nodes within the same community should be minimal, which aligns with the model's assumption that nodes within the same community are assigned identical correlation coefficients. Similarly, the variation in correlation coefficients between different communities should also be minimal, corresponding to the model's assumption that any two nodes from two different communities are assigned the same correlation coefficient. For both requirements, a relatively lenient threshold (variation $<$ $0.5$) was adopted. Based on these criteria, $20$ ROIs were ultimately selected from the candidate pool that  closely align with the model's fundamental assumptions.

%Stratified analysis results for the 20 selected ROIs. Estimated edge-based correlation matrices for the CN, LMCI, and AD groups (from left to right).

%To visually demonstrate the correlation patterns across different clinical stages, the estimated correlation matrices for the cognitively normal, late mild cognitive impairment, and Alzheimer's disease groups are presented below.

\begin{figure}[htbp]
    \centering
    \includegraphics[width=0.32\textwidth]{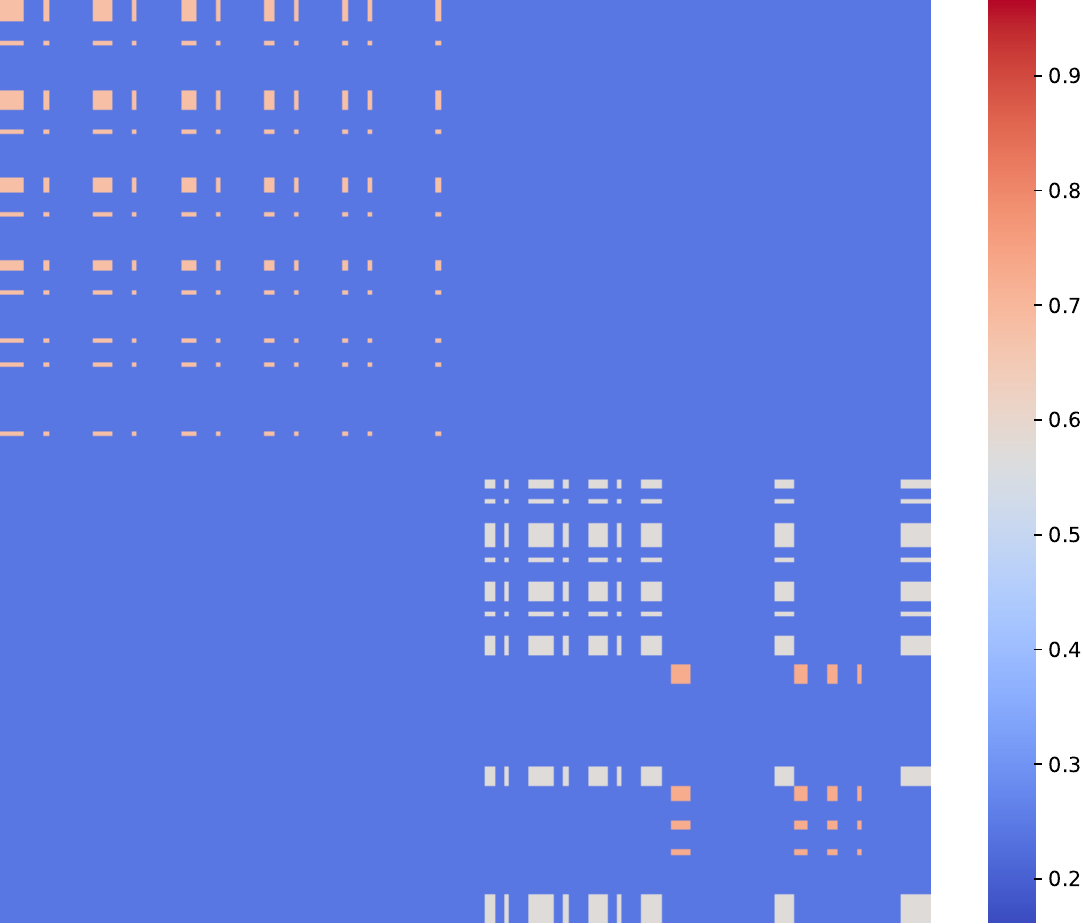}
    \includegraphics[width=0.32\textwidth]{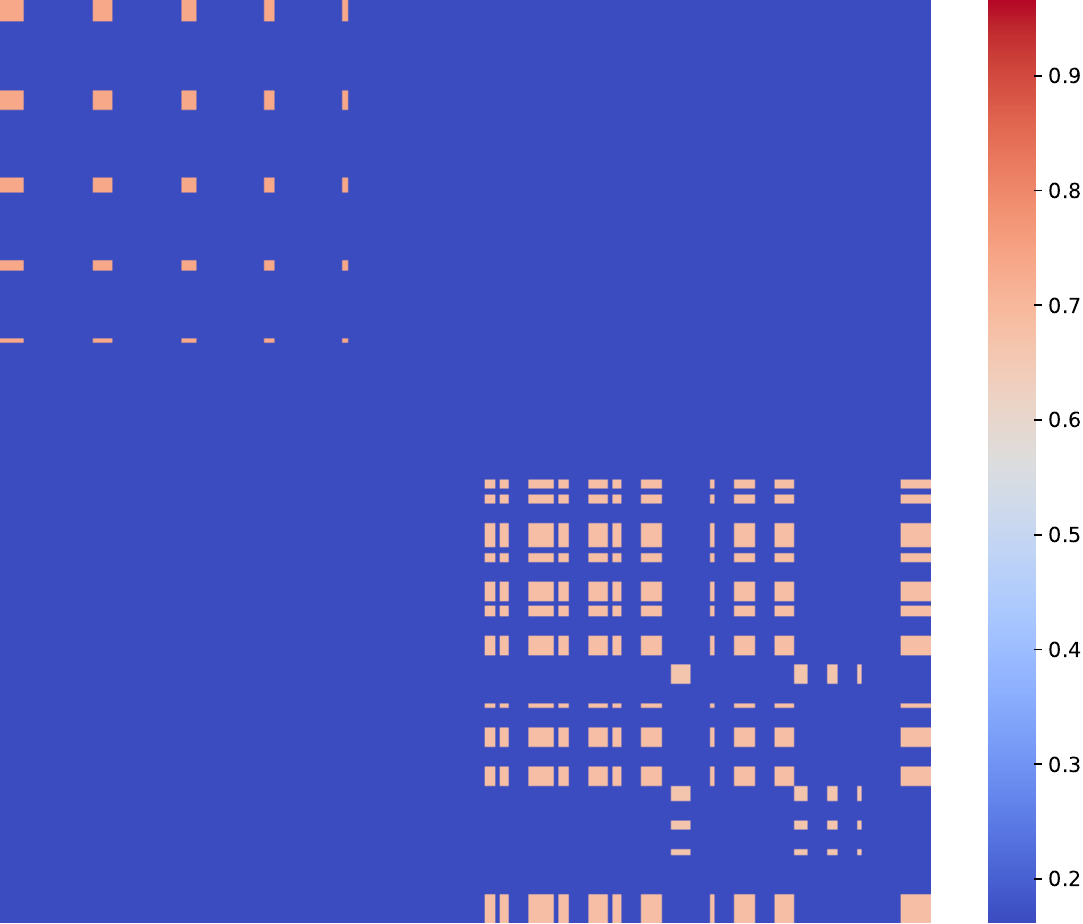}
    \includegraphics[width=0.32\textwidth]{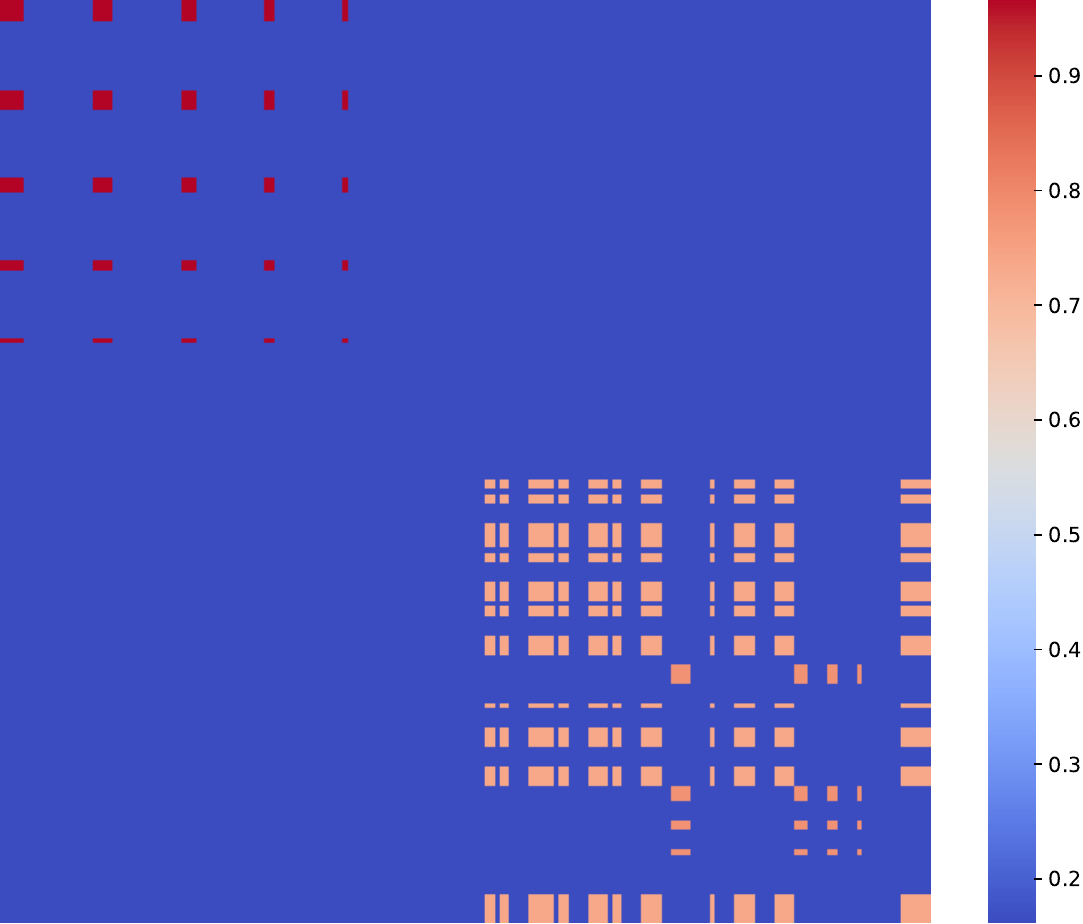}
    \caption{Stratified analysis results for the 20 selected ROIs. Estimated edge-based correlation matrices for the CN, LMCI, and AD groups (from left to right).}
    \label{fig:clinical_heatmaps}
\end{figure}

\textbf{Stratified analysis for ADNI}  
Figure \ref{fig:clinical_heatmaps} presents the estimated edge-based correlation matrices for the CN, LMCI, and AD groups (from left to right), based on the 20 selected regions of interest (ROIs).

\end{document}